\documentclass[letter]{aa}  

\usepackage{graphicx}
\usepackage{txfonts}
\usepackage{bm}
\usepackage[]{hyperref}%draft
\usepackage{xcolor}
\usepackage{array}
\DeclareMathOperator{\e}{e}

\newcommand{\ch}[1]{{\textcolor{black}{#1}}}
\newcommand{\cht}[1]{{\textcolor{black}{#1}}}
\newcolumntype{C}[1]{>{\centering\let\newline\\\arraybackslash\hspace{0pt}}m{#1}}

\begin{document} 
        
%\title{Disentangling stellar and planetary signals in spectroscopic measurements with individual spectral line velocities }
%\title{Velocity Extraction from Spectroscopic Measurements in the Presence of Stellar Noise}
\title{ The SOPHIE search for northern extrasolar planets. %in a chain of 3:2 MMR with SOPHIE radial velocities
        %The SOPHIE search for Neptunes and Super-Earths around bright, Solar-type stars: new analysis methods and detections of ....
}
        
        \subtitle{XVI. HD 158259: A compact planetary system in a near-3:2 mean motion resonance chain}
        
   \author{N. C. Hara
        \inst{\ref{i:geneve}}\thanks{CHEOPS fellow}
        \and
        F. Bouchy\inst{\ref{i:geneve}}
        \and
        M. Stalport\inst{\ref{i:geneve}}
        \and
        I. Boisse\inst{\ref{i:lam}}
        \and
        J. Rodrigues\inst{\ref{i:geneve}}
        \and
    J.-B. Delisle\inst{\ref{i:geneve}}
        \and
        A. Santerne\inst{\ref{i:lam}}
        \and
        G. W. Henry\inst{\ref{i:tennessee}}
        \and
        L. Arnold\inst{\ref{i:OHP}}
        \and
        N. Astudillo-Defru\inst{ \ref{i:chile}}
        \and
        S. Borgniet\inst{\ref{i:grenoble}}
        \and
        X.Bonfils\inst{\ref{i:grenoble}}
        \and
        V. Bourrier\inst{\ref{i:geneve}}
        \and
        B. Brugger\inst{\ref{i:lam}}
        \and
        B. Courcol\inst{\ref{i:lam}}
        \and
        S. Dalal\inst{\ref{i:paris}}
        \and
        M. Deleuil\inst{\ref{i:lam}}
        \and
    X. Delfosse\inst{\ref{i:grenoble}}
        \and
        O. Demangeon\inst{\ref{i:porto}}
        \and
        R. F. D\'iaz\inst{\ref{i:uba},\ref{i:conicet}}
        \and
        X. Dumusque\inst{\ref{i:geneve}}
        \and
        T. Forveille\inst{\ref{i:grenoble}}
        \and
        G. Hébrard\inst{\ref{i:paris}, \ref{i:OHP}}
        \and 
        M. J. Hobson\inst{\ref{i:chile2}}
        \and
        F. Kiefer\inst{\ref{i:paris}}
        \and
        T. Lopez\inst{\ref{i:lam}}
        \and
        L. Mignon\inst{\ref{i:grenoble}}
        \and
        O. Mousis\inst{\ref{i:lam}}
        \and
        C. Moutou\inst{\ref{i:lam},\ref{i:cfht}}
        \and
        F. Pepe\inst{\ref{i:geneve}}
        \and
        J. Rey\inst{\ref{i:geneve}}
        \and
        N. C. Santos\inst{\ref{i:porto},\ref{i:porto2}}
        \and
        D. Ségransan\inst{\ref{i:geneve}}
        \and
        S. Udry\inst{\ref{i:geneve}}
        \and
        P. A. Wilson\inst{\ref{i:warwick1},\ref{i:warwick2},\ref{i:paris}}
}

\institute{
                Observatoire Astronomique de l’Université de Genève, 51 Chemin des Maillettes, 1290 Versoix, Switzerland\label{i:geneve}  \email{nathan.hara@unige.ch}
        \and
        Aix Marseille Univ, CNRS, CNES, LAM, Marseille, France
\label{i:lam}
        \and
        Center of Excellence in Information Systems, Tennessee State University, Nashville, TN 37209, USA \label{i:tennessee}
        \and
   Observatoire de Haute-Provence, CNRS, Aix Marseille Université, Institut Pythéas UMS 3470, 04870 Saint-Michel-l’Observatoire, France\label{i:OHP}
        \and
        Departamento de Matem\'atica y F\'isica Aplicadas, Universidad Cat\'olica de la Sant\'isima Concepci\'on, Alonso de Rivera 2850, Concepci\'on, Chile\label{i:chile}
        \and
        Univ. Grenoble Alpes, CNRS, IPAG, 38000 Grenoble, France\label{i:grenoble}
        \and
Institut d’Astrophysique de Paris, UMR7095 CNRS, Université Pierre \& Marie Curie, 98bis boulevard Arago, 75014 Paris, France\label{i:paris}
        \and
Instituto de Astrofísica e Ciências do Espaço, Universidade do Porto, CAUP, Rua das Estrelas, 4150-762 Porto, Portugal\label{i:porto}
        \and
        Universidad de Buenos Aires, Facultad de Ciencias Exactas y Naturales. Buenos Aires, Argentina\label{i:uba}
        \and
        CONICET - Universidad de Buenos Aires. Instituto de Astronomía y Física del Espacio (IAFE). Buenos Aires, Argentina\label{i:conicet}        
        \and
   Instituto de Astrofísica, Pontificia Universidad Católica de Chile, Av. Vicuña Mackenna 4860, Macul, Santiago, Chile ; Millennium Institute for Astrophysics, Chilel\label{i:chile2}
        \and
Canada-France-Hawaii Telescope Corporation, 65-1238 Mamalahoa Hwy, Kamuela, HI 96743, USA\label{i:cfht}
        \and
        Departamento de Física e Astronomia, Faculdade de Ciências, Universidade do Porto, Rua do Campo Alegre, 4169-007 Porto, Portugal\label{i:porto2}
        %ASD/IMCCE, CNRS-UMR8028, Observatoire de Paris,  PSL, UPMC, 77 Avenue Denfert-Rochereau, 75014 Paris, France \label{i:ASD}
        \and
        Department of Physics, University of Warwick, Coventry CV4 7AL, UK\label{i:warwick1}
        \and
        Centre for Exoplanets and Habitability, University of Warwick, Coventry CV4 7AL, UK\label{i:warwick2}
}

%       \date{Received September 15, 1996; accepted March 16, 1997}
        
        % \abstract{}{}{}{}{} 
        % 5 {} token are mandatory
        
        \abstract
        % context heading (optional)
        {} %eave it empty if necessary  
        %{Since 2011, the SOPHIE spectrograph  has been used for a high precision ($\approx 1.5 - 2$ m/s precision) radial velocity survey of 124 targets in the northern hemisphere.}
        % aims heading (mandatory)
        {Since 2011, the SOPHIE spectrograph has been used to search for Neptunes and super-Earths in the Northern Hemisphere. As part of this observational program, 290 radial velocity measurements of  \ch{the 6.4 V magnitude star} HD 158259 were obtained. Additionally, TESS photometric measurements of this target are available. We present an analysis of the SOPHIE data and compare our results with the output of the TESS pipeline. 
    }
        % methods heading (mandatory)
        {The radial velocity data, ancillary spectroscopic indices, and ground-based photometric measurements were analyzed with classical and $\ell_1$ periodograms. The stellar activity was modeled as a correlated Gaussian noise and its impact on the planet detection was measured with a new technique.  }
        % results heading (mandatory)
        {The SOPHIE data support the detection of five planets,  each with $m \sin i \approx 6 M_\oplus$, orbiting HD 158259 in 3.4, 5.2, 7.9, 12, and 17.4 days. Though a planetary origin is strongly favored, the 17.4 d signal is classified as a planet candidate due to a slightly lower statistical significance and to its proximity to the expected stellar rotation period.  The data also present low frequency variations, most likely originating from a magnetic cycle and instrument systematics. Furthermore, the TESS pipeline reports a significant  signal at 2.17 days corresponding to a planet of radius $\approx 1.2 R_\oplus$. A compatible signal is seen in the radial velocities, which confirms the detection of an additional planet and yields a $\approx 2 M_\oplus$ mass estimate. 
         }
        {We find a system of five planets and a strong candidate near a 3:2 mean motion resonance chain orbiting HD 158259. %by radial velocity. %closely resembling the compact systems found by transit. 
        The planets are found to be outside of the two and three body resonances. }
        
        \maketitle
        %
        %-------------------------------------------------------------------
        
        \section{Introduction}
        \label{sec:intro}

Transit surveys have unveiled several multiplanetary systems where the planets are tightly spaced and close to low order mean motion resonances (MMRs). 
 For instance, Kepler-80~\citep{xie2013, lissauer2014, shallue2018}, Kepler-223~\citep{borucki2011, mills2016}, and  TRAPPIST-1~\citep{gillon2016,luger2017} present 5, 4, and 7 planets, respectively, in such configurations. These systems are often qualified as compact, in the sense that any two subsequent planets have a period ratio below 2. Compact, near resonant configurations could be the result of  a formation scenario where the planets encounter dissipation in the gas disk, are locked in resonance, and then migrate inwards before potentially leaving the resonance~\citep[e.g.,][]{terquem2007, macdonald2016, izidoro2017}.

%. More recently, the  system was detected with SPECULOOS.
 
   Near resonant, compact systems are detectable by radial velocity (RV), as demonstrated by follow up observations of transits~\citep{lopez2019}. However, such detections with only RV are rare: HD 40307~\citep{mayor2009} and HD 215152~\citep{delisle2018} both have three planets near 2:1 - 2:1 and 5:3 - 3:2 configurations, respectively.

        In the present work, we analyze the 290 SOPHIE radial velocity measurements of HD 158259. 
        We detect several signals, which are compatible with a chain of near resonant planets. The signals have an amplitude in the $1 - 3$ m/s range. At this level, in order to confirm their planetary origin, it is critical to consider whether these signals could be due to the star or to instrument systematics.
         To this end, we include the following data sets in our analysis: the bisector span and $\log {R'}_{HK}$ derived from the spectra as well as ground-based photometric data. The periodicity search is performed with a $\ell_1$ periodogram~\citep{hara2017} including a correlated noise model, selected with new techniques. The results are compared to those of a classical periodogram~\citep{baluev2008}. 
        
         Furthermore, HD 158259 has been observed in sector 17 of the TESS mission~\citep{ricker2014}. The results of the TESS reduction pipeline~\citep{jenkins2010, jenkins2016} are included in our analysis.

The data support the detection of six planets close to a 3:2 MMR chain, with a lower detection confidence for the outermost one.  The orbital stability of the resulting system is checked with numerical simulations, and we discuss whether the system is in or out of the two and three-body resonances.
         
         %\color{blue} We find that the period ratios of the planets are greater than 1.5, such that the system is not exactly within the isle of resonance, which is consistent with formation scenarii. \color{black}
         
         % including a covariance term as in~\cite{delisle2019a}, using the \texttt{spleaf} software package~\citep{delisle2019b}. 
        % The dynamical stability of the orbital configuration is checked with numerical simulations. 
        %and the stellar activity is modelled as a Gaussian process~\citep{haywood2014, rajpaul2015}
        
        The letter is structured as follows. The data and its analysis are presented in Sections~\ref{sec:data} and~\ref{sec:results}, respectively. The study of the system dynamics is presented in Section~\ref{sec:dynamics}, and we conclude in Section~\ref{sec:conc}.

                \section{Data}
                \label{sec:data}
                
\subsection{HD 158259}
\label{sec:star}
                        
        HD 158259 is a G0 type star in the Northern Hemisphere with  a $V$ magnitude of 6.4. The known stellar parameters are reported in Table~\ref{tab:stellarparam}. 
        The stellar rotation period is not known precisely, but it can be estimated.  The median $\log R'_{hK}$, which was obtained from SOPHIE measurements, is -4.8 $\pm 0.1$. With the empirical relationship of~\cite{mamajek2008}, this translates to an estimated rotation period of 18 $\pm$ 5 days. Additionally, the SOPHIE RV  data give $v\sin i = 2.9 \pm 1$  km/s~\citep[see][]{boisse2010}. Assuming $i = 90^\circ$ and taking the GAIA radius estimate of  1.21 $R_\odot$, the $v\sin i$ estimation yields a rotation period of $\approx$ 20 $\pm 7$ days. 
                
                  \begin{table}
                        \label{tab:stellarparam}
                        \centering
                        \caption{Known stellar parameters of HD 158259. Parallax, coordinates, proper motion, and radius are taken from \cite{gaiadr2}, spectral type is from  \cite{cannon1993}, and V magnitude is from ~\cite{hog2000}. Mass is from~\cite{chandler2016}.}
                        \begin{tabular}{p{3.5cm}|p{4.5cm}}
                                Parameter & Value \\    \hline \hline 
                                Right ascension (J2000) & 17h 25min 24.05s\\         
                                Declination (J2000) & +52.7906$^\circ$\\
                                Proper motion (mas/y) &         -91.047 $\pm$0.055, -49.639 $\pm$0.059\\
                                Parallax & 36.93 $\pm$ 0.029 mas\\
                                Spectral type & G0 \\
                                V magnitude & 6.46 \\
                                Radius & $1.21_{-0.08}^{+0.03}$ $R_{\odot}$ \\
                                Mass & 1.08 $\pm$ 0.1 $M_{\odot}$ \\
                                $v\sin i$ & 2.9 km/s\\
                                $\log R'_{hK}$  &  -4.8 \\
                        \end{tabular}
                \end{table}

                \subsection{SOPHIE radial velocities}
                
                \begin{figure} \centering
                        \includegraphics[width=10cm]{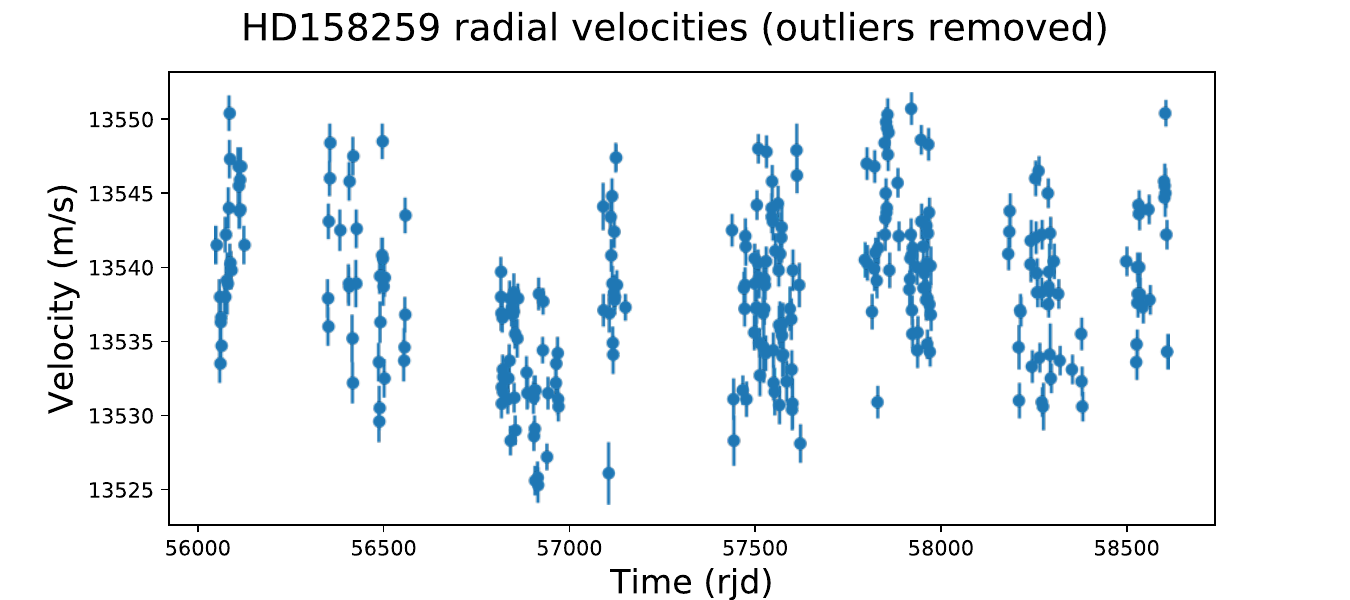}
                        \caption{SOPHIE radial velocity measurements of HD 158259 after outliers at BJD  2457941.5059, 2457944.4063, and 2457945.4585 have been removed. }
                                \label{fig:rv}
%\vspace{0.5cm}
        \end{figure}
        
        SOPHIE is an échelle spectrograph mounted on the 193cm telescope of the Haute-Provence Observatory~\citep{bouchy2011}. Several surveys have been conducted with SOPHIE, including a moderate precision survey (3.5 - 7 m/s), aimed at detecting Jupiter-mass companions~\citep[e.g.,][]{bouchy2009, moutou2014, hebrard2016}, as well as a search of smaller planets around M-dwarfs ~\citep[e.g.,][]{hobson2018, hobson2019, diaz2019}.
 
  Since 2011, SOPHIE has been used for a survey of bright solar-type stars, with the aim of  detecting Neptunes and super-Earths~\citep{bouchy2011}.  For all the observations performed in this survey, the instrumental drift was measured and corrected for by recording on the detector, close to the stellar spectrum, the spectrum of a reference lamp. This one is a thorium-argon lamp before barycentric Jullian date (BJD) 2458181 and a Fabry-Perot interferometer after this date.  The observations of HD 158259 were part of this program. Over the course of seven years, 290 measurements were obtained with an average error of 1.2 m/s. The data, which were corrected from instrumental drift and outliers (see Appendix~\ref{app:inst} and~\ref{app:dataselect}), are shown in Fig.~\ref{fig:rv}.
  
%   The raw SOPHIE radial velocities have a typical RMS of $2 - 3$ m/s, which is insufficient to detect Neptunes. This dispersion is partly due to the drift of the zero velocity point of the spectrograph, which has been monitored by the observations of very bright, quiet stars.  From these measurements, the drift is estimated with a method closely related to~\cite{courcol2015}, presented in detail in~\cite{hara2019}.  After this correction, the velocities can achieve a 1.5 m/s RMS. 
    
    %\color{blue}
    The RV is not the only data product that was extracted from the SOPHIE spectra. The SOPHIE pipeline also retrieves the bisector span~\citep{queloz2001} as well as the $\log R'_{hk}$~\citep{noyes1984}.%, which are leveraged in our analysis. 
  %Describe how the ancillary indicators are extracted
 % \color{black}
 % \subsection{Radial velocities }
 
   \subsection{APT Photometry }
   
   Photometry has been obtained with the T11 telescope at the Automatic Photoelectric Telescopes (APTs), located at Fairborn Observatory in southern Arizona. The data, covering four observation seasons, are presented in more detail in Appendix~\ref{app:apt}.%processed as described in~\cite{henry1999}. %In Fig.~\ref{fig:phot}, we show the magnitude of HD 158259 relative to three comparison stars,
  %We consider the photometric flux relative to three comparison stars, where an offset was added to the data of seasons 2, 3 and 4 so  their means equal the mean value of season 1.
  %The photometry does not present obvious features, which is also the case of its periodogram, presented in section~\ref{sec:ancillary}.

\subsection{TESS results }

HD 158259 was observed from  7 Oct to 2 Nov 2019 (sector 17). The TESS reduction pipeline~\citep{jenkins2010, jenkins2016} found evidence for a 2.1782 $\pm 0.0006$ d signal with a time of conjunction $T_c = 2458766.049072 \pm 003708$ BJD (TOI 1462.01).
The detection was made with a signal-to-noise ratio of 8.05, which is above the detection threshold of 7.3 adopted in~\cite{sullivan2015}.                  %\includegraphics[width=10cm]{HDx_photometry.pdf}
                %\caption{Photometric measurements obtained at the APT }
                %\label{fig:phot}

\section{Analysis of the data sets}
        \label{sec:results}

\subsection{Ground-based photometry and ancillary indicators}
\label{sec:ancillary}

If the bisector span, $\log R'_{HK}$, or photometry show signs of temporal correlation, in particular, periodic signatures, this might mean that the RVs are corrupted by stellar or instrumental effects~\citep[e.g.,][]{queloz2001}. To search for periodicities, we applied the generalized Lomb-Scargle periodogram iteratively~\citep{ferrazmello1981, zechmeister2009} as well as the $\ell_1$ periodogram for comparison purposes. The process is presented in detail in Appendix~\ref{app:periodsearch}.

The $\log R'_{Hk}$ periodogram presents a peak at 2900 d with a false alarm probability (FAP) of $4 \cdot 10^{-12}$. This  signal is the only clear feature of the ancillary indicators, and  it supports the presence of a magnetic cycle with a period  $\geqslant 1500 $ d. We note that both the APT photometry and bisector span present a peak around 11.6 d, though with a high FAP level. This periodicity as well as other periodicities found in the indicators were used to build candidate noise models for the analysis of the RVs, which is the object of the next section.

\subsection{Radial velocities analysis}
\label{sec:rv}
\label{sec:period}

\begin{figure} \centering
        \hspace{-0cm}
        \includegraphics[width=9.2cm]{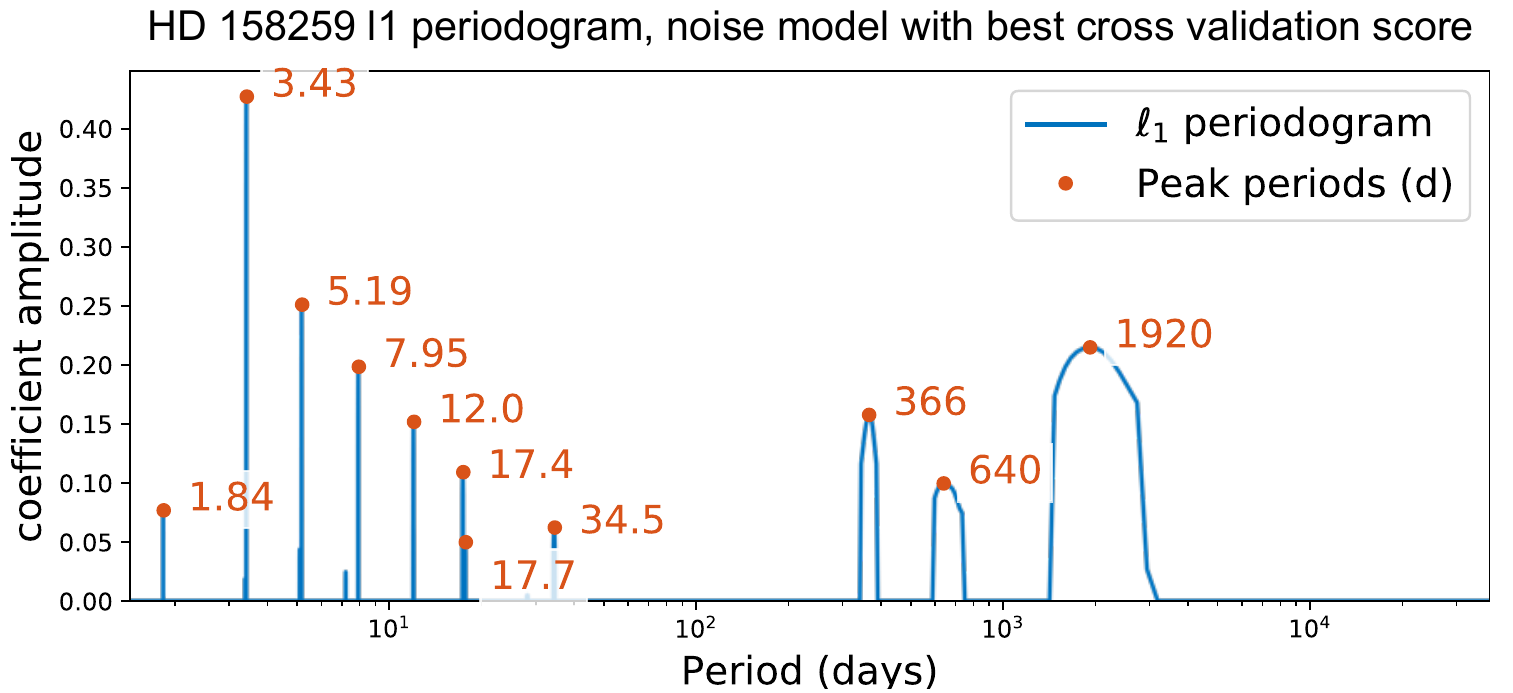}
        \caption{$\ell_1$ periodogram of the  SOPHIE radial velocities of HD 158259 corrected from outliers (in blue). The periods at which the main peaks occur are represented in red.  }
        \label{fig:l1perio}     
\end{figure}

\begin{table}
        \renewcommand{\arraystretch}{1.25}
        \caption{Periods appearing in the $\ell_1$ periodogram and their false alarm probabilities with their origin, the semi-amplitude of corresponding signals, and $M\sin i$ with $68.7\%$ intervals.}
        \label{tab:fapsbody}
        \centering
        \begin{tabular}{p{1.1cm}|p{1.3cm}|p{1.9cm}|p{1.1cm}|p{1.1cm}}%p{1.3cm}
                Peak period (d)  & FAP & Origin & $K$ (m/s) &  $M\sin i$ $(M_\oplus)$ \\ \hline \hline
                \multicolumn{5}{c}{Detected by SOPHIE RV} \\ \hline
                3.432 & $2\cdot10^{-5}$  & Planet $c$ &$2.2_{-0.2}^{+0.2}$ & $5.5^{+0.5}_{-0.6}$ \\
                5.198 & $8\cdot10^{-3}$ & Planet $d$& $1.9_{-0.2}^{+0.2}$ & $5.3^{+0.7}_{-0.7}$\\
                7.954&   $1.6\cdot10^{-3}$ & Planet  $e$&$1.8_{-0.3}^{+0.3}$& $6.0^{+0.9}_{-1.0}$ \\ 
                12.03&$2.8\cdot10^{-3}$ & Planet  $f$&$1.6_{-0.3}^{+0.4}$&  $6.1^{+1.2}_{-1.3}$\\
                17.39 &$2.3\cdot10^{-2}$  & Candidate  $g$ & $1.6_{-0.3}^{+0.5}$&  $6.8^{+1.8}_{-1.6}$ \\
                366 & $1.1\cdot10^{-6}$ & Systematic &$3.4_{-0.8}^{+0.7} $& 
                $\left(40^{+9}_{-10}\right)$ \\
                640 & $1.3\cdot10^{-2}$ & Activity & - & -  \\
                1920 & $2.1\cdot10^{-2}$  & Activity &$2.9_{-0.4}^{+0.4} $&       
                \\ \hline
                \multicolumn{5}{c}{Detected by TESS + confirmed by SOPHIE RV} \\ \hline
                2.177 & 0.44 &  Planet $b$ & $1.0^{+0.2}_{-0.2}$ & $2.2^{+0.4}_{-0.4}$ \\ 
                 \multicolumn{5}{l}{Alias of 1.84 d,  Radius: $ 1.2\pm1.3$ $R_\oplus$, Density: $1.09^{+0.23}_{-0.27}$ $\rho_\oplus$}  \\ \multicolumn{5}{c}{} \\ \hline 
                \multicolumn{5}{c}{Other signals}
                \\ \hline
                34.5 & 1.0 &  Candidate? & - & - \\
                17.7 & 0.5  & -& -  & -\\
                \hline
                %1.839 &  & -& -  & -\\
        \end{tabular}
\end{table}

%$39.^{+9.1}_{-9.7}$

The RV time series we analyze here was corrected for instrument drift and from outliers. The process is described in Appendix~\ref{app:inst} and~\ref{app:dataselect}. 
To search for potential periodicities, we computed the $\ell_1$ periodogram of the RV, as defined in~\cite{hara2017}. This tool is based on a sparse recovery technique called the basis pursuit algorithm~\citep{chen1998}. The $\ell_1$ periodogram takes in a frequency grid and an assumed covariance matrix of the noise as input. It aims to find a representation of the RV time series as a sum of a small number of sinusoids whose frequencies are in the input grid. It outputs a figure which has a similar aspect as a regular periodogram, but with fewer peaks due to aliasing. The peaks can be assigned a FAP, whose intrepretation is close to the FAP of a regular periodogram peak.

The signals found to be statistically significant might vary from one noise model to another. 
To explore this aspect, we considered several candidate noise models based on the periodicities found in the ancillary indicators. The noise models were ranked with cross-validation and~Bayesian evidence approximations (see Appendix~\ref{app:cv}). In Fig.~\ref{fig:l1perio}, we show the $\ell_1$ periodogram of the SOPHIE RVs corresponding to the noise with the best cross validation score on a grid of equispaced frequencies between 0 and 0.7 cycle/d. The FAPs of the peaks pointed by red markers in Fig.~\ref{fig:l1perio} are given in Table~\ref{tab:fapsbody}. They suggest the presence of signals, in decreasing strengths of detection, at 366, 3.43, 7.95, 12.0, 5.19, 1920, 640 , and 17.4  days. The significance of the signals at 1.84, 17.7, and 34.5 d is found to be marginal to null.

The FAPs reported in Table~\ref{tab:fapsbody} were computed with a certain noise model. 
In Appendix~\ref{app:cv}, we explore the sensitivity of the detections to the noise model choice.  We find the detection of signals at 3.43, 5.19, 7.95, and 12.0 d  to be robust. The detection of signals at 1920, 366, and 17.4 d is slightly less strong but still favored. There is evidence for a 640 d signal and hints of signals at 34.5 d and 1.84 d or 2.17 d. The latter two are aliases of one another. Indeed, the RV spectral window has a strong peak at the sidereal day (0.9972 d), which is common in RV time series~\citep[][]{dawsonfabricky2010}, and 1/1.840 + 1/2.177 = 1/0.9972 d\textsuperscript{-1}.

In Appendix~\ref{app:periodogram}, the results are compared to a classical periodogram approach with a white noise model, which gives similar results but fails to unveil the 1.84/2.17 d candidate in the signal. We also study whether the apparent signals could originate from aliases of the periods considered here. This possibility is found to be unlikely.

\begin{figure} \centering
        \begin{minipage}{0.5\textwidth}
                \includegraphics[width=8.2cm]{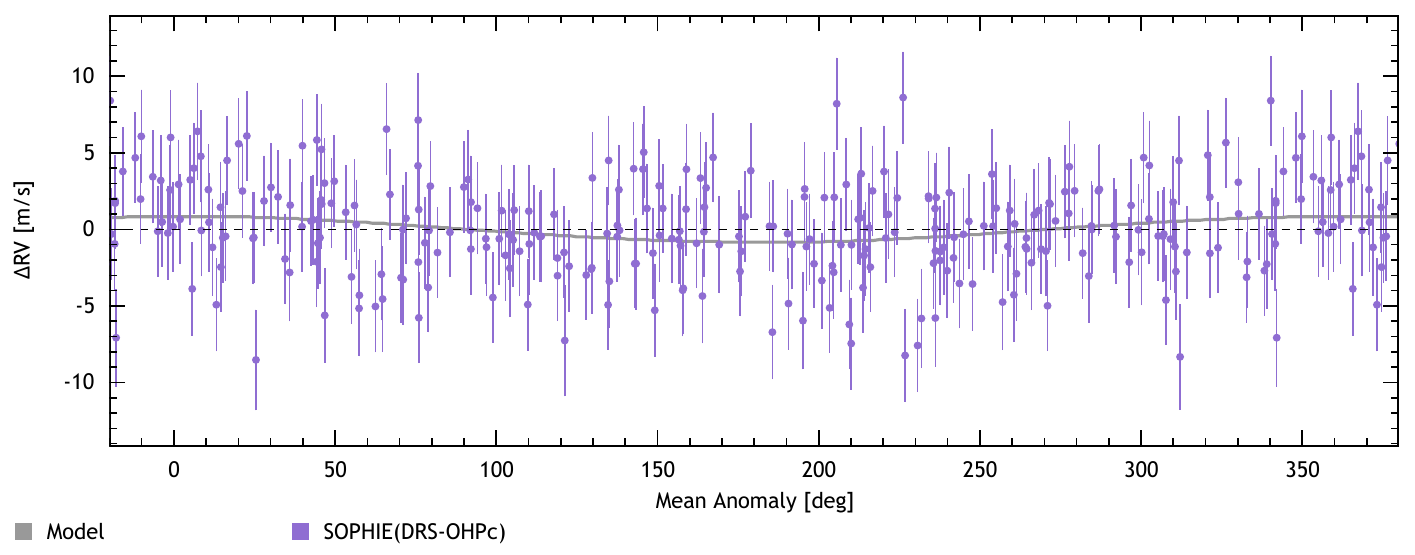}
                \includegraphics[width=8.2cm]{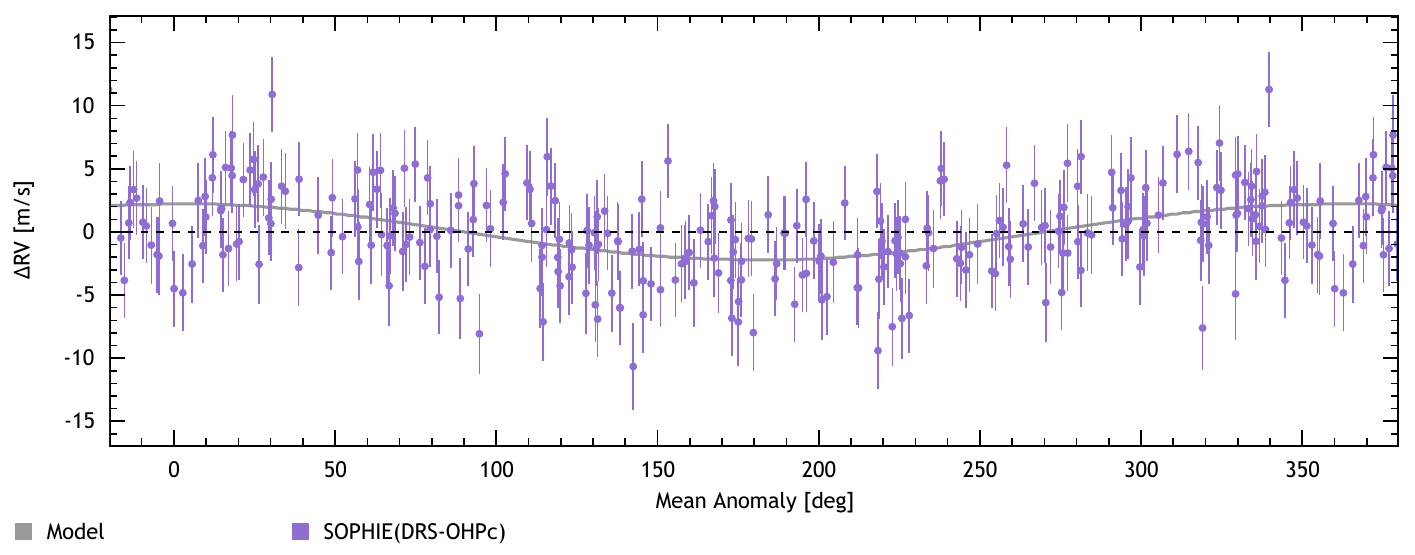}
                \includegraphics[width=8.2cm]{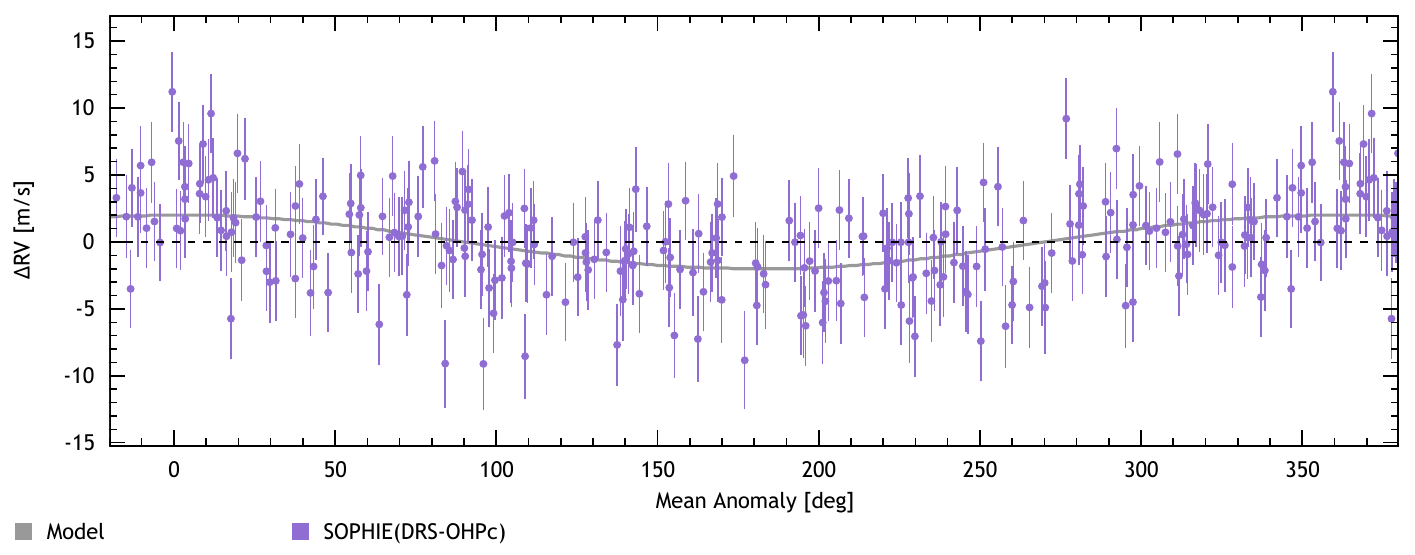}
                \includegraphics[width=8.2cm]{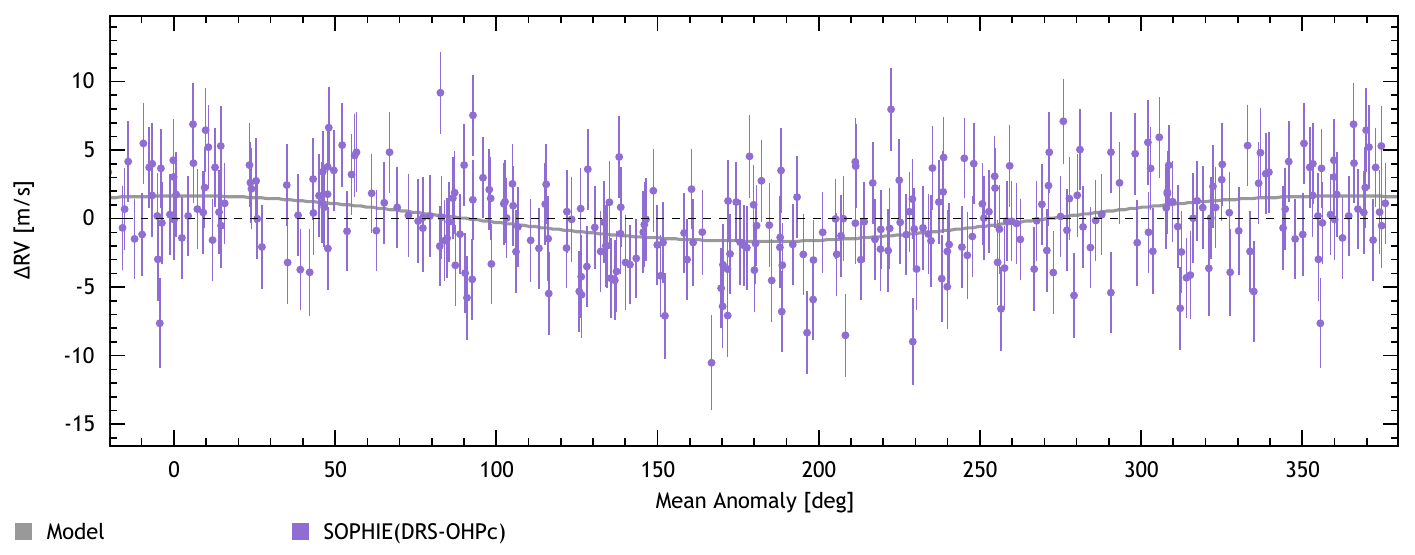}
                \includegraphics[width=8.2cm]{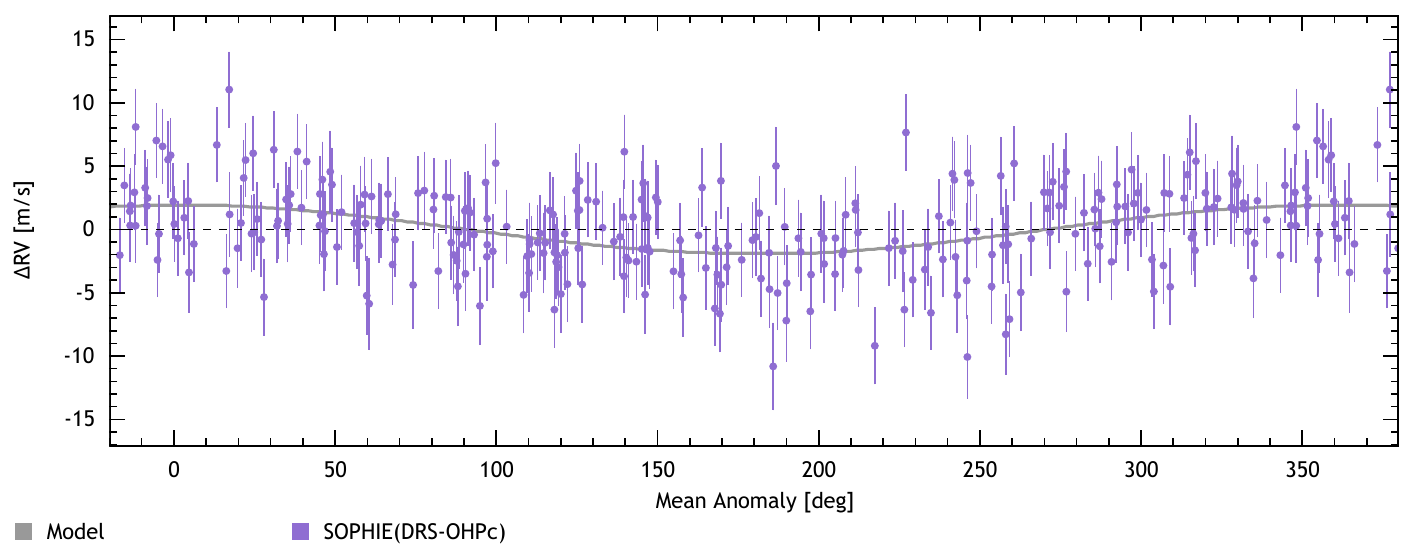}
                \includegraphics[width=8.2cm]{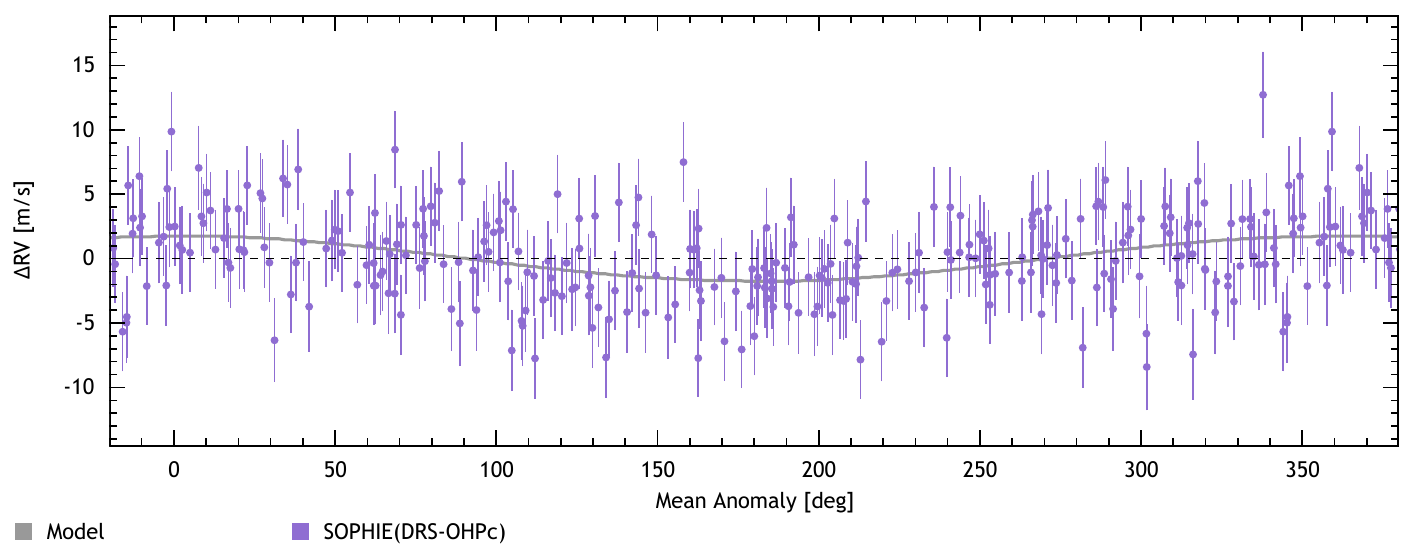}
        \end{minipage}
        \caption{Radial velocity phase-folded at the periods of the signals appearing in the period analysis. From top to bottom: 2.17, 3.43, 5.19, 7.95, 12.0, and 17.4 d. }
        \label{fig:phasefold}
\end{figure}

We fit a model with a free error jitter and nine sinusoidal functions initialized at the periods listed in the upper part of  Table~\ref{tab:fapsbody}. The data, which were phase-folded at the fitted periods, are shown in Fig.~\ref{fig:phasefold}.  The error bars correspond to the addition in quadrature of the  nominal uncertainties and the fitted jitter.

The residuals of the nine-signals model plus white noise fit have a root mean square (RMS) of 3.1 m/s, which is higher than the nominal uncertainties of the SOPHIE data (1.2 m/s). We studied the residuals with the methods of~\cite{hara2019ecc} and found that they are temporally correlated, which might corrupt the orbital element estimates. In Appendix~\ref{app:mcmc}, we show that a consistent model of the data can be obtained with a noise model, including white and correlated components.

\begin{figure} \centering
        \includegraphics[width=8.2cm]{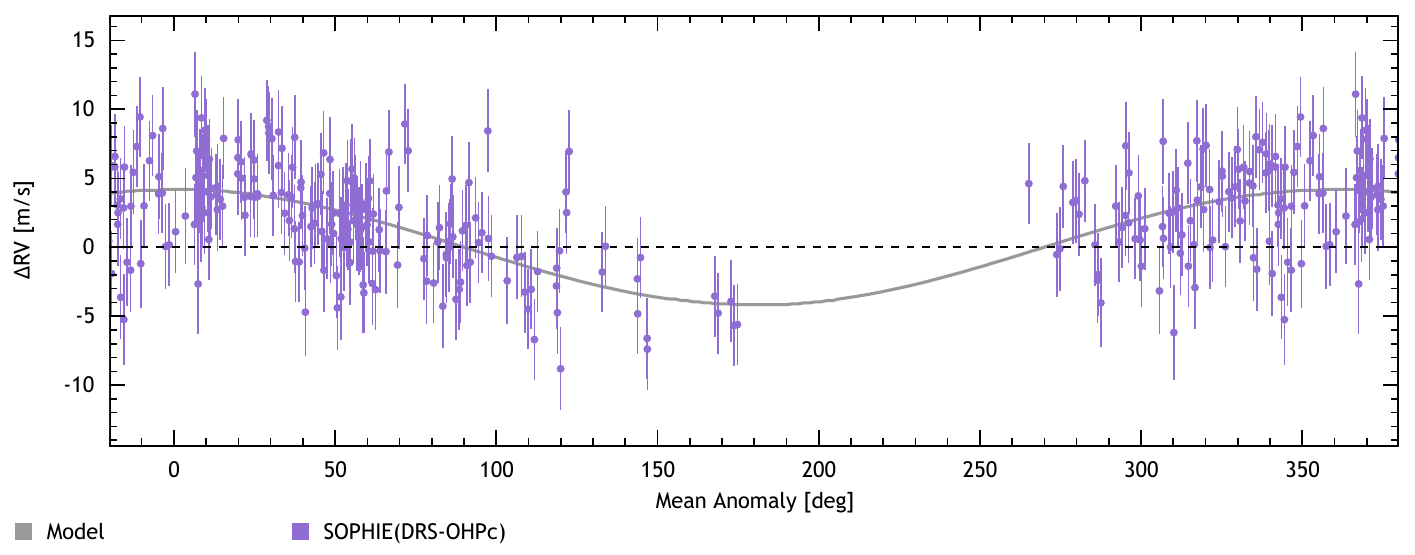}
        \includegraphics[width=8.2cm]{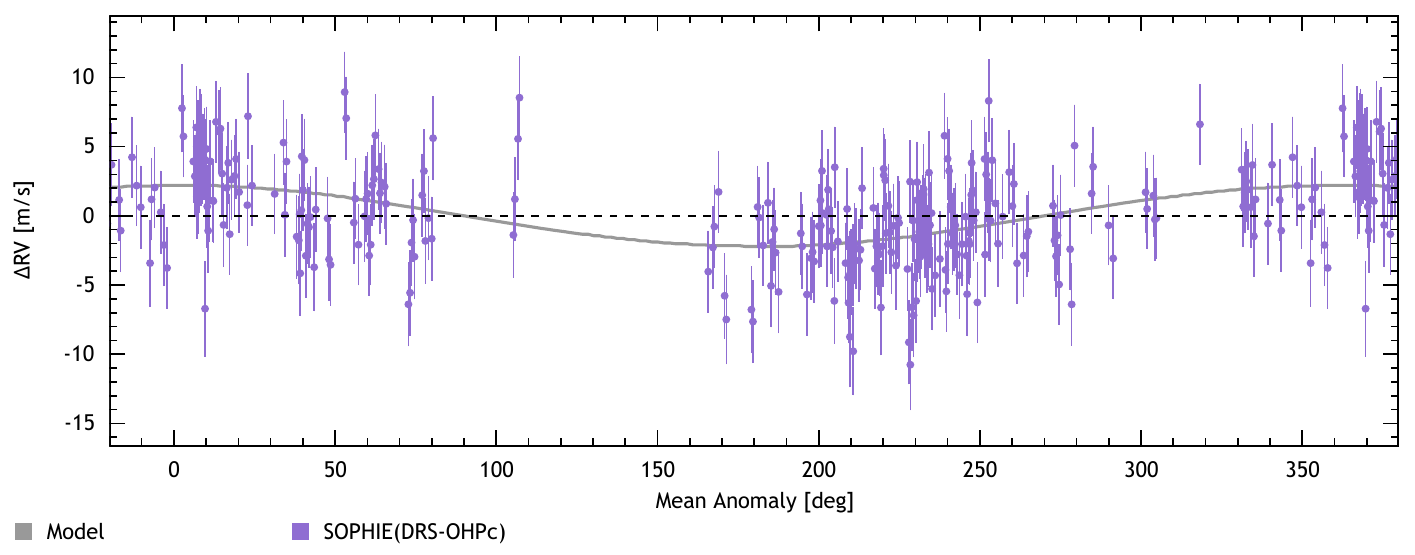}
        \includegraphics[width=8.2cm]{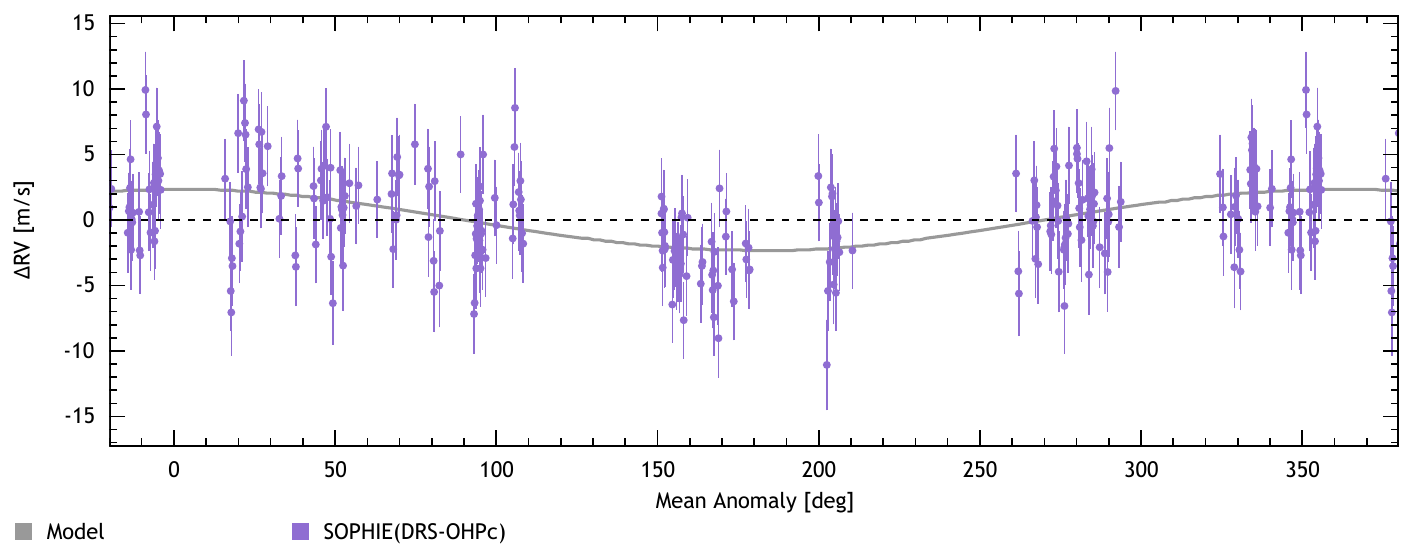}      
        \caption{Radial velocity phase-folded, from top to bottom, at: 360, 750, and 2000 d. }
        \label{fig:phasefold2}
\end{figure}

\subsection{Periodicity origin}
\label{sec:origin}

The origin of the 366, 640, and 1920  d signals is uncertain, and we do not claim planet detections at those periods.
The 366 d signal is fully compatible with a yearly signal. Instrument systematics, such as the stitching effect~\citep{dumusque2015}, could produce this signal, and they are deemed to be its most likely origin.  We fit a Gaussian process on the $\log R'_{HK}$ and used it as a linear predictor on the RVs, similarly to~\cite{haywood2014}. When computing the periodogram on the residuals of the fit, there is no trace of signals in the 1500 - 3000 d and 600 - 800 d regions, so they might stem from a magnetic cycle.
The signal at 34.5 d could be a faint trace of a planet near the 2:1  resonance, but its significance is too low to be conclusive.  

The periods at 3.43, 5.19, 7.95, 12.0, and 17.4 d, which are significant in the RV analysis, most likely stem from planets. Here, we list five arguments that support this claim. (i) None of these periods clearly appear  in the bisector span, $\log R'_{hk}$, or photometry. (ii) While eccentric planets can be mistaken for planet pairs near a 2:1 MMR, they are very unlikely to appear as planets near a 3:2 resonance~\citep{hara2019ecc}. (iii) The periods could be due to instrument systematics. We find it unlikely since the periods of the planet candidates do not consistently appear in the 123 other data sets of the survey HD 158259 is a part of~\citep{hara2019}.  (iv) All the signals are consistent in phase and amplitude (see appendix~\ref{app:phasecons}). 
(v) Most importantly, the period ratio of two subsequent planets is very close to 3:2, namely 1.51, 1.53, 1.51, and 1.44, and they have very similar estimated masses (see Section~\ref{sec:orbitalelts}). Pairs of planets close to the 3:2 period ratio are known to be common~\citep{lissauer_architecture_2011, steffen2015}, and it seems unlikely that the stellar features would mimic this specific spacing of periods. 

Signals at 12 and 17.4 d verify the five points listed above and are thus considered as planets. We, however, point out that the predicted rotation period of the star is 18 $\pm 5 $ d. This means that signals could be present in this period range, not necessarily at the stellar rotation period or its harmonic~\citep{nava2020}. Furthermore, in Appendix~\ref{app:cv}, we show that the 17.4 d signal is less significant and that certain noise models favor 17.7 d over 17.4 d. It appears that fitting signals at 17.4 or 17.7 d does not completely remove the other. This might point to differential rotation or dynamical effects, but it is most likely due to modeling uncertainties. Nonetheless, the nature and period of the 17.4 d signal are subject to a little more caution than the other planets.  It is thus conservatively classified as a strong planet candidate. 

The 2.17 d signal appearing in TESS has a counterpart in the RVs, which appear at 1.84 d (alias) here. The RV signal is marginally significant but compatible with the observation of a transit.  
        Indeed, the time of conjunction as measured by TESS is BJD 2458766.049072 $\pm$ 0.003708. At this epoch, the mean longitude of the innermost planet predicted from the RV with a prior on the period set from TESS data is ${95_{-23}^{+23}}^\circ$ (see Section~\ref{sec:orbitalelts} for details on the posterior calculation). On this basis, the 2.17 d signal is considered to stem from a planet. We note that no trace of the other planets is found in the TESS data.
        
\begin{figure}
        \centering
        \includegraphics[width=6.7cm]{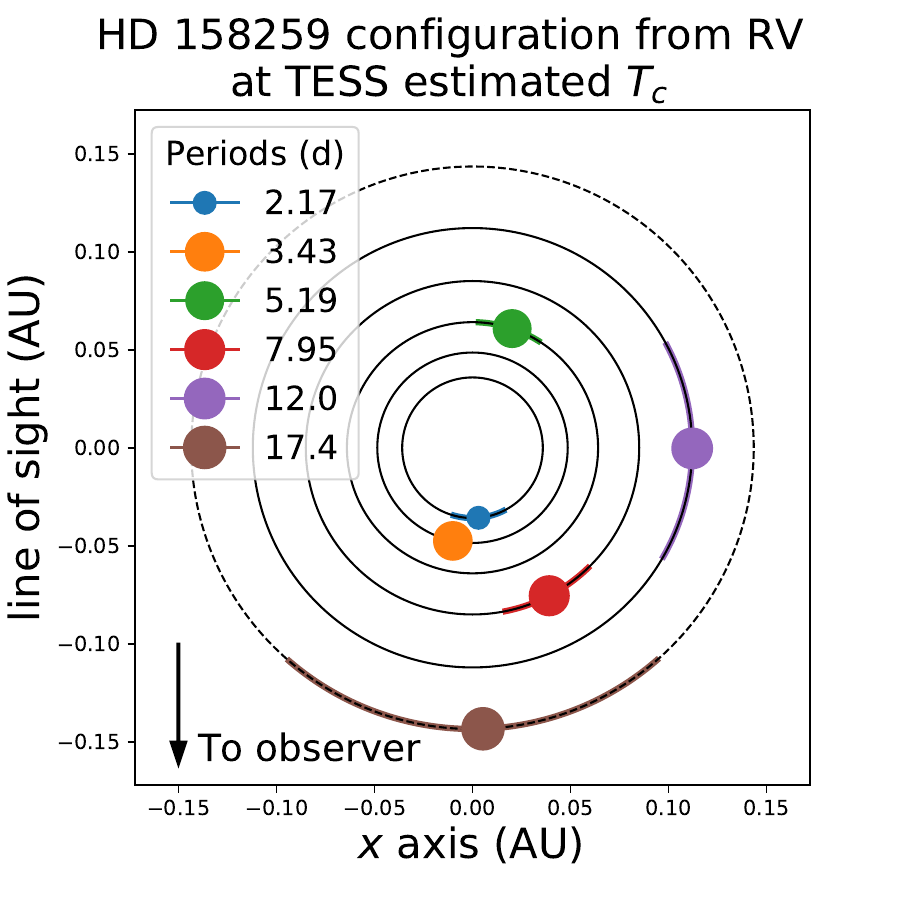}
        \caption{Configuration of the system estimated from the RV at the time of conjunction estimated by TESS (BJD 58766.049072). }
                \label{fig:config}
\end{figure}

        In Fig.~\ref{fig:config} we represent the configuration of the system at the time of conjunction given by the TESS pipeline. The markers represent the position of the planets corresponding to their posterior mean (semi major axis and mean longitude). The marker size is proportional to an estimated radius $\propto (m \sin i)^{0.55}$, following~\cite{bashi2017}. The plain colored lines correspond to 68\% credible intervals on the mean longitude.

In summary, we deem six of the nine significant signals to originate from planets: 2.17,  3.43, 5.19, 7.95, 12.0, and 17.4 d, with a slightly lower confidence in 17.4 d. From now on, they are referred to as planets $b, c, d, e, f$, and (strong) candidate $g$, respectively.

\subsection{Orbital elements}
\label{sec:orbitalelts}

To derive the uncertainties on the orbital elements of the planets, we computed their posterior distribution with a Monte Carlo Markov chain algorithm (MCMC). 
The model includes  signals at 2.17,  3.43, 5.19, 7.95, 12.0, and  17.4 d, a linear predictor fitted on the $\log R'_{HK}$ with a Gaussian process analysis,  
as well as a correlated noise model with an exponential decay. The prior on eccentricity was chosen to strongly disfavor $e>0.1$ (see Section~\ref{sec:dynamics} for justification). The details of the posterior calculations are presented in Appendix~\ref{app:mcmc} and the main features of the signals are reported in Table~\ref{tab:fapsbody}.
Planets $c,d,e,f$, and planet candidate $g$ exhibit $m\sin i \approx 6 M_\oplus$ and planet $b$ has $m \approx 2 M_\oplus$ and $R \approx 1.2 R_\oplus$, which is compatible with a terrestrial density.
The yearly signal has an amplitude of $\approx 3$ m/s. We find a significantly nonzero time-scale of the noise of $5.4_{-2.6}^{+0.8}$ d. A more detailed analysis shows hints of nonstationary noises (see Appendix~\ref{app:phasecons}).

The TESS photometry exhibits a transit signal from $b$, but not from $c,d,e,f$, or $g$. Given the masses of $c,d,e,f$, and  $g$, their transits should have been detected had they occurred. Assuming a coplanar system, this can be explained if the inclination departs from $90^\circ$ from at least ${7.08_{-0.43}^{+0.36}}^\circ$ (so that the transit of $c$ cannot be detected) to ${9.5_{-0.48}^{+0.57}}^\circ$ at most (the transit of $b$ must be seen). This would translate to a $\sin i$ between 0.985 and 0.993. Alternately, the planets could be relatively inclined.

\section{Dynamical analysis}
\label{sec:dynamics}

\subsection{Stability}

For the dynamical analysis of the system, we consider the planets b, c, d, e, f, and candidate g, whose period ratios are close to the 3:2 MMRs: $P_c/P_b$=1.57, $P_d/P_c$=1.51, $P_e/P_d$=1.53, $P_f/P_e$=1.51, and $P_g/P_f$=1.44. 
 To check the existence of stable solutions, we proceeded as follows. We performed the MCMC analysis with exactly the same priors as in Section~\ref{sec:orbitalelts} except with a looser prior on eccentricity. 
 Each MCMC sample was taken as an initial condition for the system. Its evolution was integrated 1 kyr in the future using the 15-th order N-body integrator IAS15~\citep{Rein2015} from the package REBOUND\footnote{The REBOUND code is freely available at \url{http://github.com/hannorein/rebound}.} \citep{Rein2012}. General relativity was included via REBOUNDx, using the model of~\citet{Anderson1975}. In total, 36 782 system configurations were integrated. We consider the system to be unstable if any two planets have an encounter below their mutual Hill radius. A total of 1700 samples lead to integrations where the stability condition was not violated. We computed their empirical probability distribution functions and found constraints on the eccentricities $\lesssim 0.1$.

\subsection{Resonances}

The planets of HD 158259 could be locked in 3:2 or three body resonances. This would translate to the following conditions. We consider three subsequent planets, indexed by $i=1,2,3$ from innermost to outermost. With $\lambda_i$, $P_i$, and $\varpi_i$, we denote their mean longitudes, periods, and arguments of periastron.
In case of two body resonances, the angle
\begin{align}
\label{eq:resangle}
\psi_{12} = 3\lambda_2 - 2\lambda_1 - \varpi_j, \;\; \; j=1,2
\end{align}
would librate. 
Following \citet{delisle_analytical_2017}, in case of three-body resonances, the so-called Laplace angle
\begin{align}
    \label{eq:laplaceangles}
        \phi_{123} = 3\lambda_3 + 2\lambda_1 - 5\lambda_2 
\end{align}
should librate around $\pi$. 
The posterior distributions of $\psi_{ij}$ and $\phi_{ijk}$ as well as their derivatives were computed for any doublet and triplet of planets from the MCMC samples, assuming that $\varpi_j$ is constant on the time-scale of the observations.  We conclude that the system is not locked in two and three body resonances. We note that the circulation of the angles occurs at different rates, $\phi_{cde}$ being the slowest.  

Nevertheless, period ratios so close to 3:2  are very unlikely to stem from pure randomness. It is therefore probable that the planets underwent migration in the protoplanetary disk, during which each consecutive pair of planets was locked in 3:2 MMR. The observed departure of the ratio of periods of two subsequent planets from exact commensurability might be explained by tidal dissipation, as was already proposed for similar Kepler systems~\citep[e.g.,][]{delisle_tidal_2014}. Stellar and planet mass changes have also been suggested as a possible cause of resonance breaking~\citep{matsumoto2020}. The reasons behind the absence of three-body resonances, which are seen in other resembling systems~\citep[e.g., Kepler-80, ][]{macdonald2016}, are to be explored.

\section{Conclusion}
\label{sec:conc}

In this work, we have analyzed 290 SOPHIE measurements of HD 158259. 
The analysis of the radial velocity data, including a correction of the instrument drift and over $7500$ correlated noise models, support the  detection of four planets ($c,d,e$, and $f$) and a strong candidate ($g$),  with respective periods of 3.43, 5.19, 7.95, 12.0, and 17.4 d. They all exhibit a $\approx 6 M_\oplus$ mass. There is substantial evidence for a planetary origin of the 17.4 d signal, and the remaining concerns on its nature should be cleared with a better estimate of the stellar period.  Furthermore, the TESS data exhibit a 2.17 d signal that is compatible with the RVs, which allows one to claim the detection of an additional planet ($b$) and to measure a density of $1.09^{+0.23}_{-0.27}$ $\rho_\oplus$. The TESS photometry does not show other transits.  There exist stable configurations  of the six-planet system that are compatible with the error bars.

While many compact near-resonance chains have been detected by transits, they have been rare in radial velocity surveys so far. The present analysis shows that they can be detected, provided there are enough data points and an appropriate accounting of correlated noises (instrumental and stellar).

HD 158259 $b,c,d,e,f$, and $g$ are such that subsequent planets have period ratios of 1.57, 1.51, 1.53, 1.51, and 1.44 with increasing period. Subsequent planet pairs and triplets are close to, but not within, 3:2 and three-body mean motion resonances. 
The period ratios are consistent with the distribution of period ratios of planet pairs found in Kepler, which exhibits a peak at 1.52~\citep[see][]{lissauer_architecture_2011,fabrycky_architecture_2014, steffen2015}. The similarity of the masses of the planets of the system is compatible with the hypothesis that planets within the same system have similar sizes~\citep{lissauer2011, ciardi2013, millholland2017, weiss2018}.
The configuration of the planets can be explained by existing formation scenarios~\citep[e.g.,][]{terquem2007,delisle_tidal_2014}. The proximity to 3:2 resonances of the HD 158259 system is reminiscent of Kepler-80~\citep{macdonald2016}. However, the latter presents three-body resonances, while HD 158259 does not. It would be interesting to investigate scenarios that can explain the differences between the two systems.

\begin{acknowledgements}
        We warmly thank the OHP staff for their support on the observations.
        N.C.H. and J.-B. D. acknowledge the financial support of the National Centre for Competence in Research PlanetS of the Swiss National Science Foundation (SNSF).
        This work was supported by the Programme National de Plan\'etologie (PNP) of CNRS/INSU, co-funded by CNE.
        G.W.H. acknowledges long-term support from NASA, NSF, Tennessee State University, and the State of Tennessee through its Centers of Excellence program.
        X.De., X.B., I.B., and T.F. received funding from the French Programme National de Physique Stellaire (PNPS) and the Programme National de Plan\'etologie (PNP) of CNRS (INSU).
        X.B. acknowledges funding from the European Research Council under the ERC Grant Agreement n. 337591-ExTrA. 
        This work has been supported by a grant from Labex OSUG@2020 (Investissements d'avenir – ANR10 LABX56).
        This work is also supported by the French National Research Agency in the framework of the Investissements d’Avenir program (ANR-15-IDEX-02), through the funding of the `Origin of Life' project of the Univ. Grenoble-Alpes.
        
        V.B. acknowledges support from the Swiss National Science Foundation (SNSF) in the frame of the National Centre for Competence in Research PlanetS, and has received funding from the European Research Council (ERC) under the European Unions Horizon 2020 research and innovation programme (project Four Aces; grant agreement No 724427). 
        
This work was supported by FCT - Fundação para a Ciência e a Tecnologia/MCTES through national funds (PIDDAC) by this grant UID/FIS/04434/2019, as well as through national funds (PTDC/FIS-AST/28953/2017 and PTDC/FIS-AST/32113/2017) and by FEDER - Fundo Europeu de Desenvolvimento Regional through COMPETE2020 - Programa Operacional Competitividade e Internacionalização (POCI-01-0145-FEDER-028953 and POCI-01-0145-FEDER-032113).

        N.A-D. acknowledges support from FONDECYT \#3180063.
        
        X.Du. is grateful to the Branco Weiss Fellowship—Society in Science for continuous support.
\end{acknowledgements}

        \bibliographystyle{aa}  
\bibliography{biblio.bib}

\begin{thebibliography}{74}
\expandafter\ifx\csname natexlab\endcsname\relax\def\natexlab#1{#1}\fi

\bibitem[{Akaike(1974)}]{akaike1974}
Akaike, H. 1974, IEEE Transactions on Automatic Control, 19, 716

\bibitem[{Amelunxen {et~al.}(2014)Amelunxen, Lotz, McCoy, \&
  Tropp}]{amelunxen2013}
Amelunxen, D., Lotz, M., McCoy, M.~B., \& Tropp, J.~A. 2014, Information and
  Inference: A Journal of the IMA, 3, 224

\bibitem[{{Anderson} {et~al.}(1975){Anderson}, {Esposito}, {Martin},
  {Thornton}, \& {Muhleman}}]{Anderson1975}
{Anderson}, J.~D., {Esposito}, P.~B., {Martin}, W., {Thornton}, C.~L., \&
  {Muhleman}, D.~O. 1975, \apj, 200, 221

\bibitem[{{Baluev}(2008)}]{baluev2008}
{Baluev}, R.~V. 2008, \mnras, 385, 1279

\bibitem[{{Bashi} {et~al.}(2017){Bashi}, {Helled}, {Zucker}, \&
  {Mordasini}}]{bashi2017}
{Bashi}, D., {Helled}, R., {Zucker}, S., \& {Mordasini}, C. 2017, \aap, 604,
  A83

\bibitem[{{Boisse} {et~al.}(2010){Boisse}, {Eggenberger}, {Santos}, {Lovis},
  {Bouchy}, {H{\'e}brard}, {Arnold}, {Bonfils}, {Delfosse}, {Desort},
  {D{\'\i}az}, {Ehrenreich}, {Forveille}, {Gallenne}, {Lagrange}, {Moutou},
  {Udry}, {Pepe}, {Perrier}, {Perruchot}, {Pont}, {Queloz}, {Santerne},
  {S{\'e}gransan}, \& {Vidal-Madjar}}]{boisse2010}
{Boisse}, I., {Eggenberger}, A., {Santos}, N.~C., {et~al.} 2010, \aap, 523, A88

\bibitem[{Borucki {et~al.}(2011)Borucki, Koch, Basri, Batalha, Brown, Bryson,
  Caldwell, Christensen-Dalsgaard, Cochran, DeVore, Dunham, Gautier, Geary,
  Gilliland, Gould, Howell, Jenkins, Latham, Lissauer, Marcy, Rowe, Sasselov,
  Boss, Charbonneau, Ciardi, Doyle, Dupree, Ford, Fortney, Holman, Seager,
  Steffen, Tarter, Welsh, Allen, Buchhave, Christiansen, Clarke, Das,
  D{\'{e}}sert, Endl, Fabrycky, Fressin, Haas, Horch, Howard, Isaacson,
  Kjeldsen, Kolodziejczak, Kulesa, Li, Lucas, Machalek, McCarthy, MacQueen,
  Meibom, Miquel, Prsa, Quinn, Quintana, Ragozzine, Sherry, Shporer, Tenenbaum,
  Torres, Twicken, Cleve, Walkowicz, Witteborn, \& Still}]{borucki2011}
Borucki, W.~J., Koch, D.~G., Basri, G., {et~al.} 2011, The Astrophysical
  Journal, 736, 19

\bibitem[{{Bouchy} {et~al.}(2011){Bouchy}, {H{\'e}brard}, {Delfosse}, {Udry},
  {Lagrange}, {Arnold}, {Boisse}, {Bonfils}, {Debondt}, {Diaz}, {Eggenberger},
  {Ehrenreich}, {Forveille}, {Lovis}, {Moutou}, {Pepe}, {Perrier}, {Queloz},
  {Santerne}, {Santos}, \& {S{\'e}gransan}}]{bouchy2011}
{Bouchy}, F., {H{\'e}brard}, G., {Delfosse}, X., {et~al.} 2011, in EPSC-DPS
  Joint Meeting 2011, Vol. 2011, 240

\bibitem[{{Bouchy} {et~al.}(2009){Bouchy}, {Hebrard}, {Udry}, {Delfosse}, \&
  {Boisse}}]{bouchy2009}
{Bouchy}, F., {Hebrard}, G., {Udry}, S., {Delfosse}, X., \& {Boisse}, I. 2009,
  arXiv e-prints, arXiv:0907.3559

\bibitem[{{Cannon} \& {Pickering}(1993)}]{cannon1993}
{Cannon}, A.~J. \& {Pickering}, E.~C. 1993, VizieR Online Data Catalog,
  III/135A

\bibitem[{{Chandler} {et~al.}(2016){Chandler}, {McDonald}, \&
  {Kane}}]{chandler2016}
{Chandler}, C.~O., {McDonald}, I., \& {Kane}, S.~R. 2016, VizieR Online Data
  Catalog, J/AJ/151/59

\bibitem[{Chen {et~al.}(1998)Chen, Donoho, \& Saunders}]{chen1998}
Chen, S.~S., Donoho, D.~L., \& Saunders, M.~A. 1998, SIAM JOURNAL ON SCIENTIFIC
  COMPUTING, 20, 33

\bibitem[{Ciardi {et~al.}(2013)Ciardi, Fabrycky, Ford, Gautier, Howell,
  Lissauer, Ragozzine, \& Rowe}]{ciardi2013}
Ciardi, D.~R., Fabrycky, D.~C., Ford, E.~B., {et~al.} 2013, The Astrophysical
  Journal, 763, 41

\bibitem[{{Courcol} {et~al.}(2015){Courcol}, {Bouchy}, {Pepe}, {Santerne},
  {Delfosse}, {Arnold}, {Astudillo-Defru}, {Boisse}, {Bonfils}, {Borgniet},
  {Bourrier}, {Cabrera}, {Deleuil}, {Demangeon}, {D{\'{\i}}az}, {Ehrenreich},
  {Forveille}, {H{\'e}brard}, {Lagrange}, {Montagnier}, {Moutou}, {Rey},
  {Santos}, {S{\'e}gransan}, {Udry}, \& {Wilson}}]{courcol2015}
{Courcol}, B., {Bouchy}, F., {Pepe}, F., {et~al.} 2015, \aap, 581, A38

\bibitem[{{Dawson} \& {Fabrycky}(2010)}]{dawsonfabricky2010}
{Dawson}, R.~I. \& {Fabrycky}, D.~C. 2010, \apj, 722, 937

\bibitem[{{Delisle}(2017)}]{delisle_analytical_2017}
{Delisle}, J.-B. 2017, \aap, 605, A96

\bibitem[{{Delisle} {et~al.}(2020){Delisle}, {Hara}, \&
  {S{\'e}gransan}}]{delisle2019a}
{Delisle}, J.~B., {Hara}, N., \& {S{\'e}gransan}, D. 2020, arXiv e-prints,
  arXiv:2001.10319

\bibitem[{{Delisle} {et~al.}(2019){Delisle}, {Hara}, \&
  {S\'egransan}}]{delisle2019b}
{Delisle}, J.~B., {Hara}, N.~C., \& {S\'egransan}, D. 2019, a\&A, submitted

\bibitem[{{Delisle} \& {Laskar}(2014)}]{delisle_tidal_2014}
{Delisle}, J.-B. \& {Laskar}, J. 2014, \aap, 570, L7

\bibitem[{{Delisle} {et~al.}(2018){Delisle}, {S{\'e}gransan}, {Dumusque},
  {Diaz}, {Bouchy}, {Lovis}, {Pepe}, {Udry}, {Alonso}, {Benz}, {Coffinet},
  {Collier Cameron}, {Deleuil}, {Figueira}, {Gillon}, {Lo Curto}, {Mayor},
  {Mordasini}, {Motalebi}, {Moutou}, {Pollacco}, {Pompei}, {Queloz}, {Santos},
  \& {Wyttenbach}}]{delisle2018}
{Delisle}, J.-B., {S{\'e}gransan}, D., {Dumusque}, X., {et~al.} 2018, \aap,
  614, A133

\bibitem[{{D{\'{\i}}az} {et~al.}(2019){D{\'{\i}}az}, {Delfosse}, {Hobson},
  {Boisse}, {Astudillo-Defru}, {Bonfils}, {Henry}, {Arnold}, {Bouchy},
  {Bourrier}, {Brugger}, {Dalal}, {Deleuil}, {Demangeon}, {Dolon}, {Dumusque},
  {Forveille}, {Hara}, {H{\'e}brard}, {Kiefer}, {Lopez}, {Mignon}, {Moreau},
  {Mousis}, {Moutou}, {Pepe}, {Perruchot}, {Richaud}, {Santerne}, {Santos},
  {Sottile}, {Stalport}, {S{\'e}gransan}, {Udry}, {Unger}, \&
  {Wilson}}]{diaz2019}
{D{\'{\i}}az}, R.~F., {Delfosse}, X., {Hobson}, M.~J., {et~al.} 2019, \aap,
  625, A17

\bibitem[{{Dumusque} {et~al.}(2015){Dumusque}, {Pepe}, {Lovis}, \&
  {Latham}}]{dumusque2015}
{Dumusque}, X., {Pepe}, F., {Lovis}, C., \& {Latham}, D.~W. 2015, \apj, 808,
  171

\bibitem[{{Fabrycky} {et~al.}(2014){Fabrycky}, {Lissauer}, {Ragozzine}, {Rowe},
  {Steffen}, {Agol}, {Barclay}, {Batalha}, {Borucki}, {Ciardi}, {Ford},
  {Gautier}, {Geary}, {Holman}, {Jenkins}, {Li}, {Morehead}, {Morris},
  {Shporer}, {Smith}, {Still}, \& {Van Cleve}}]{fabrycky_architecture_2014}
{Fabrycky}, D.~C., {Lissauer}, J.~J., {Ragozzine}, D., {et~al.} 2014, \apj,
  790, 146

\bibitem[{{Ferraz-Mello}(1981)}]{ferrazmello1981}
{Ferraz-Mello}, S. 1981, \aj, 86, 619

\bibitem[{{Foreman-Mackey} {et~al.}(2017){Foreman-Mackey}, {Agol},
  {Ambikasaran}, \& {Angus}}]{foremanmackey2017}
{Foreman-Mackey}, D., {Agol}, E., {Ambikasaran}, S., \& {Angus}, R. 2017, \aj,
  154, 220

\bibitem[{{Gaia Collaboration}(2018)}]{gaiadr2}
{Gaia Collaboration}. 2018, VizieR Online Data Catalog, I/345

\bibitem[{{Gillon} {et~al.}(2016){Gillon}, {Jehin}, {Lederer}, {Delrez}, {de
  Wit}, {Burdanov}, {Van Grootel}, {Burgasser}, {Triaud}, {Opitom}, {Demory},
  {Sahu}, {Bardalez Gagliuffi}, {Magain}, \& {Queloz}}]{gillon2016}
{Gillon}, M., {Jehin}, E., {Lederer}, S.~M., {et~al.} 2016, \nat, 533, 221

\bibitem[{{Hara} {et~al.}(2019{\natexlab{a}}){Hara}, {Bouchy}, {Boisse},
  {Stalport}, {Rodrigues}, {Delisle}, {Santerne}, \& {Delfosse}}]{hara2019}
{Hara}, N.~C., {Bouchy}, F., {Boisse}, I., {et~al.} 2019{\natexlab{a}}, in
  prep.

\bibitem[{{Hara} {et~al.}(2017){Hara}, {Bou{\'e}}, {Laskar}, \&
  {Correia}}]{hara2017}
{Hara}, N.~C., {Bou{\'e}}, G., {Laskar}, J., \& {Correia}, A.~C.~M. 2017,
  \mnras, 464, 1220

\bibitem[{{Hara} {et~al.}(2019{\natexlab{b}}){Hara}, {Bou{\'e}}, {Laskar},
  {Delisle}, \& {Unger}}]{hara2019ecc}
{Hara}, N.~C., {Bou{\'e}}, G., {Laskar}, J., {Delisle}, J.~B., \& {Unger}, N.
  2019{\natexlab{b}}, \mnras, 489, 738

\bibitem[{{Haywood} {et~al.}(2014){Haywood}, {Collier Cameron}, {Queloz},
  {Barros}, {Deleuil}, {Fares}, {Gillon}, {Lanza}, {Lovis}, {Moutou}, {Pepe},
  {Pollacco}, {Santerne}, {S{\'e}gransan}, \& {Unruh}}]{haywood2014}
{Haywood}, R.~D., {Collier Cameron}, A., {Queloz}, D., {et~al.} 2014, \mnras,
  443, 2517

\bibitem[{{H{\'e}brard} {et~al.}(2016){H{\'e}brard}, {Arnold}, {Forveille},
  {Correia}, {Laskar}, {Bonfils}, {Boisse}, {D{\'\i}az}, {Hagelberg},
  {Sahlmann}, {Santos}, {Astudillo-Defru}, {Borgniet}, {Bouchy}, {Bourrier},
  {Courcol}, {Delfosse}, {Deleuil}, {Demangeon}, {Ehrenreich}, {Gregorio},
  {Jovanovic}, {Labrevoir}, {Lagrange}, {Lovis}, {Lozi}, {Moutou},
  {Montagnier}, {Pepe}, {Rey}, {Santerne}, {S{\'e}gransan}, {Udry},
  {Vanhuysse}, {Vigan}, \& {Wilson}}]{hebrard2016}
{H{\'e}brard}, G., {Arnold}, L., {Forveille}, T., {et~al.} 2016, \aap, 588,
  A145

\bibitem[{{Henry}(1999)}]{henry1999}
{Henry}, G.~W. 1999, \pasp, 111, 845

\bibitem[{Hobson(2019)}]{hobsonthese}
Hobson, M. 2019, Exoplanet detection around M dwarfs with near infrared and
  visible spectroscopy

\bibitem[{{Hobson} {et~al.}(2019){Hobson}, {Delfosse}, {Astudillo-Defru},
  {Boisse}, {D{\'{\i}}az}, {Bouchy}, {Bonfils}, {Forveille}, {Arnold},
  {Borgniet}, {Bourrier}, {Brugger}, {Cabrera Salazar}, {Courcol}, {Dalal},
  {Deleuil}, {Demangeon}, {Dumusque}, {Hara}, {H{\'e}brard}, {Kiefer}, {Lopez},
  {Mignon}, {Montagnier}, {Mousis}, {Moutou}, {Pepe}, {Rey}, {Santerne},
  {Santos}, {Stalport}, {S{\'e}gransan}, {Udry}, \& {Wilson}}]{hobson2019}
{Hobson}, M.~J., {Delfosse}, X., {Astudillo-Defru}, N., {et~al.} 2019, \aap,
  625, A18

\bibitem[{{Hobson} {et~al.}(2018){Hobson}, {D{\'{\i}}az}, {Delfosse},
  {Astudillo-Defru}, {Boisse}, {Bouchy}, {Bonfils}, {Forveille}, {Hara},
  {Arnold}, {Borgniet}, {Bourrier}, {Brugger}, {Cabrera}, {Courcol}, {Dalal},
  {Deleuil}, {Demangeon}, {Dumusque}, {Ehrenreich}, {H{\'e}brard}, {Kiefer},
  {Lopez}, {Mignon}, {Montagnier}, {Mousis}, {Moutou}, {Pepe}, {Rey},
  {Santerne}, {Santos}, {Stalport}, {S{\'e}gransan}, {Udry}, \&
  {Wilson}}]{hobson2018}
{Hobson}, M.~J., {D{\'{\i}}az}, R.~F., {Delfosse}, X., {et~al.} 2018, \aap,
  618, A103

\bibitem[{{H{\o}g} {et~al.}(2000){H{\o}g}, {Fabricius}, {Makarov}, {Urban},
  {Corbin}, {Wycoff}, {Bastian}, {Schwekendiek}, \& {Wicenec}}]{hog2000}
{H{\o}g}, E., {Fabricius}, C., {Makarov}, V.~V., {et~al.} 2000, \aap, 355, L27

\bibitem[{{Izidoro} {et~al.}(2017){Izidoro}, {Ogihara}, {Raymond},
  {Morbidelli}, {Pierens}, {Bitsch}, {Cossou}, \& {Hersant}}]{izidoro2017}
{Izidoro}, A., {Ogihara}, M., {Raymond}, S.~N., {et~al.} 2017, \mnras, 470,
  1750

\bibitem[{{Jenkins} {et~al.}(2010){Jenkins}, {Chandrasekaran}, {McCauliff},
  {Caldwell}, {Tenenbaum}, {Li}, {Klaus}, {Cote}, \& {Middour}}]{jenkins2010}
{Jenkins}, J.~M., {Chandrasekaran}, H., {McCauliff}, S.~D., {et~al.} 2010, in
  \procspie, Vol. 7740, Proceedings of the SPIE, Volume 7740, id. 77400D
  (2010)., 77400D

\bibitem[{{Jenkins} {et~al.}(2016){Jenkins}, {Twicken}, {McCauliff},
  {Campbell}, {Sanderfer}, {Lung}, {Mansouri-Samani}, {Girouard}, {Tenenbaum},
  {Klaus}, {Smith}, {Caldwell}, {Chacon}, {Henze}, {Heiges}, {Latham},
  {Morgan}, {Swade}, {Rinehart}, \& {Vanderspek}}]{jenkins2016}
{Jenkins}, J.~M., {Twicken}, J.~D., {McCauliff}, S., {et~al.} 2016, in
  \procspie, Vol. 9913, Software and Cyberinfrastructure for Astronomy IV,
  99133E

\bibitem[{{Jones} {et~al.}(2017){Jones}, {Stenning}, {Ford}, {Wolpert},
  {Loredo}, \& {Dumusque}}]{jones2017}
{Jones}, D.~E., {Stenning}, D.~C., {Ford}, E.~B., {et~al.} 2017, ArXiv e-prints
  [\eprint[arXiv]{1711.01318}]

\bibitem[{Kass \& Raftery(1995)}]{kassraftery1995}
Kass, R.~E. \& Raftery, A.~E. 1995, Journal of the American Statistical
  Association, 90, 773

\bibitem[{{Lissauer} {et~al.}(2014){Lissauer}, {Marcy}, {Bryson}, {Rowe},
  {Jontof-Hutter}, {Agol}, {Borucki}, {Carter}, {Ford}, {Gilliland}, {Kolbl},
  {Star}, {Steffen}, \& {Torres}}]{lissauer2014}
{Lissauer}, J.~J., {Marcy}, G.~W., {Bryson}, S.~T., {et~al.} 2014, \apj, 784,
  44

\bibitem[{{Lissauer} {et~al.}(2011{\natexlab{a}}){Lissauer}, {Ragozzine},
  {Fabrycky}, {Steffen}, {Ford}, {Jenkins}, {Shporer}, {Holman}, {Rowe},
  {Quintana}, {Batalha}, {Borucki}, {Bryson}, {Caldwell}, {Carter}, {Ciardi},
  {Dunham}, {Fortney}, {Gautier}, {Howell}, {Koch}, {Latham}, {Marcy},
  {Morehead}, \& {Sasselov}}]{lissauer_architecture_2011}
{Lissauer}, J.~J., {Ragozzine}, D., {Fabrycky}, D.~C., {et~al.}
  2011{\natexlab{a}}, \apjs, 197, 8

\bibitem[{{Lissauer} {et~al.}(2011{\natexlab{b}}){Lissauer}, {Ragozzine},
  {Fabrycky}, {Steffen}, {Ford}, {Jenkins}, {Shporer}, {Holman}, {Rowe},
  {Quintana}, {Batalha}, {Borucki}, {Bryson}, {Caldwell}, {Carter}, {Ciardi},
  {Dunham}, {Fortney}, {Gautier}, {Howell}, {Koch}, {Latham}, {Marcy},
  {Morehead}, \& {Sasselov}}]{lissauer2011}
{Lissauer}, J.~J., {Ragozzine}, D., {Fabrycky}, D.~C., {et~al.}
  2011{\natexlab{b}}, \apjs, 197, 8

\bibitem[{{Lopez} {et~al.}(2019){Lopez}, {Barros}, {Santerne}, {Deleuil},
  {Adibekyan}, {Almenara}, {Armstrong}, {Brugger}, {Barrado}, {Bayliss},
  {Boisse}, {Bonomo}, {Bouchy}, {Brown}, {Carli}, {Demangeon}, {Dumusque},
  {D{\'{\i}}az}, {Faria}, {Figueira}, {Foxell}, {Giles}, {H{\'e}brard},
  {Hojjatpanah}, {Kirk}, {Lillo-Box}, {Lovis}, {Mousis}, {da N{\'o}brega},
  {Nielsen}, {Neal}, {Osborn}, {Pepe}, {Pollacco}, {Santos}, {Sousa}, {Udry},
  {Vigan}, \& {Wheatley}}]{lopez2019}
{Lopez}, T.~A., {Barros}, S.~C.~C., {Santerne}, A., {et~al.} 2019, arXiv
  e-prints [\eprint[arXiv]{1909.13527}]

\bibitem[{{Luger} {et~al.}(2017){Luger}, {Sestovic}, {Kruse}, {Grimm},
  {Demory}, {Agol}, {Bolmont}, {Fabrycky}, {Fernandes}, {Van Grootel},
  {Burgasser}, {Gillon}, {Ingalls}, {Jehin}, {Raymond}, {Selsis}, {Triaud},
  {Barclay}, {Barentsen}, {Howell}, {Delrez}, {de Wit}, {Foreman-Mackey},
  {Holdsworth}, {Leconte}, {Lederer}, {Turbet}, {Almleaky}, {Benkhaldoun},
  {Magain}, {Morris}, {Heng}, \& {Queloz}}]{luger2017}
{Luger}, R., {Sestovic}, M., {Kruse}, E., {et~al.} 2017, Nature Astronomy, 1,
  0129

\bibitem[{{MacDonald} {et~al.}(2016){MacDonald}, {Ragozzine}, {Fabrycky},
  {Ford}, {Holman}, {Isaacson}, {Lissauer}, {Lopez}, {Mazeh}, {Rogers}, {Rowe},
  {Steffen}, \& {Torres}}]{macdonald2016}
{MacDonald}, M.~G., {Ragozzine}, D., {Fabrycky}, D.~C., {et~al.} 2016, \aj,
  152, 105

\bibitem[{{Mamajek} \& {Hillenbrand}(2008)}]{mamajek2008}
{Mamajek}, E.~E. \& {Hillenbrand}, L.~A. 2008, \apj, 687, 1264

\bibitem[{{Matsumoto} \& {Ogihara}(2020)}]{matsumoto2020}
{Matsumoto}, Y. \& {Ogihara}, M. 2020, arXiv e-prints, arXiv:2003.01965

\bibitem[{{Mayor} {et~al.}(2009){Mayor}, {Udry}, {Lovis}, {Pepe}, {Queloz},
  {Benz}, {Bertaux}, {Bouchy}, {Mordasini}, \& {Segransan}}]{mayor2009}
{Mayor}, M., {Udry}, S., {Lovis}, C., {et~al.} 2009, \aap, 493, 639

\bibitem[{{Millholland} {et~al.}(2017){Millholland}, {Wang}, \&
  {Laughlin}}]{millholland2017}
{Millholland}, S., {Wang}, S., \& {Laughlin}, G. 2017, \apjl, 849, L33

\bibitem[{{Mills} {et~al.}(2016){Mills}, {Fabrycky}, {Migaszewski}, {Ford},
  {Petigura}, \& {Isaacson}}]{mills2016}
{Mills}, S.~M., {Fabrycky}, D.~C., {Migaszewski}, C., {et~al.} 2016, \nat, 533,
  509

\bibitem[{{Mortier} \& {Collier Cameron}(2017)}]{mortier2017}
{Mortier}, A. \& {Collier Cameron}, A. 2017, \aap, 601, A110

\bibitem[{{Moutou} {et~al.}(2014){Moutou}, {H{\'e}brard}, {Bouchy}, {Arnold},
  {Santos}, {Astudillo-Defru}, {Boisse}, {Bonfils}, {Borgniet}, {Delfosse},
  {D{\'\i}az}, {Ehrenreich}, {Forveille}, {Gregorio}, {Labrevoir}, {Lagrange},
  {Montagnier}, {Montalto}, {Pepe}, {Sahlmann}, {Santerne}, {S{\'e}gransan},
  {Udry}, \& {Vanhuysse}}]{moutou2014}
{Moutou}, C., {H{\'e}brard}, G., {Bouchy}, F., {et~al.} 2014, \aap, 563, A22

\bibitem[{{Nava} {et~al.}(2020){Nava}, {L{\'o}pez-Morales}, {Haywood}, \&
  {Giles}}]{nava2020}
{Nava}, C., {L{\'o}pez-Morales}, M., {Haywood}, R.~D., \& {Giles}, H. A.~C.
  2020, \aj, 159, 23

\bibitem[{{Nelson} {et~al.}(2018){Nelson}, {Ford}, {Buchner}, {Cloutier},
  {D{\'{\i}}az}, {Faria}, {Rajpaul}, \& {Rukdee}}]{nelson2018}
{Nelson}, B.~E., {Ford}, E.~B., {Buchner}, J., {et~al.} 2018, ArXiv e-prints
  [\eprint[arXiv]{1806.04683}]

\bibitem[{{Noyes}(1984)}]{noyes1984}
{Noyes}, R.~W. 1984, in Space Research in Stellar Activity and Variability, ed.
  A.~{Mangeney} \& F.~{Praderie}, 113

\bibitem[{{Queloz} {et~al.}(2001){Queloz}, {Henry}, {Sivan}, {Baliunas},
  {Beuzit}, {Donahue}, {Mayor}, {Naef}, {Perrier}, \& {Udry}}]{queloz2001}
{Queloz}, D., {Henry}, G.~W., {Sivan}, J.~P., {et~al.} 2001, \aap, 379, 279

\bibitem[{Rasmussen \& Williams(2005)}]{rasmussen2005}
Rasmussen, C.~E. \& Williams, C. K.~I. 2005, Gaussian Processes for Machine
  Learning (Adaptive Computation and Machine Learning) (The MIT Press)

\bibitem[{{Rein} \& {Liu}(2012)}]{Rein2012}
{Rein}, H. \& {Liu}, S.~F. 2012, \aap, 537, A128

\bibitem[{{Rein} \& {Spiegel}(2015)}]{Rein2015}
{Rein}, H. \& {Spiegel}, D.~S. 2015, \mnras, 446, 1424

\bibitem[{Ricker {et~al.}(2014)Ricker, Winn, Vanderspek, Latham, Bakos, Bean,
  Berta-Thompson, Brown, Buchhave, Butler, Butler, Chaplin, Charbonneau,
  Christensen-Dalsgaard, Clampin, Deming, Doty, Lee, Dressing, Dunham, Endl,
  Fressin, Ge, Henning, Holman, Howard, Ida, Jenkins, Jernigan, Johnson,
  Kaltenegger, Kawai, Kjeldsen, Laughlin, Levine, Lin, Lissauer, MacQueen,
  Marcy, McCullough, Morton, Narita, Paegert, Palle, Pepe, Pepper, Quirrenbach,
  Rinehart, Sasselov, Sato, Seager, Sozzetti, Stassun, Sullivan, Szentgyorgyi,
  Torres, Udry, \& Villasenor}]{ricker2014}
Ricker, G.~R., Winn, J.~N., Vanderspek, R., {et~al.} 2014, Journal of
  Astronomical Telescopes, Instruments, and Systems, 1, 1

\bibitem[{Schuster(1898)}]{schuster1898}
Schuster, A. 1898, Terrestrial Magnetism, 3, 13

\bibitem[{Schwarz(1978)}]{schwarz1978}
Schwarz, G. 1978, Ann. Statist., 6, 461

\bibitem[{{Shallue} \& {Vanderburg}(2018)}]{shallue2018}
{Shallue}, C.~J. \& {Vanderburg}, A. 2018, \aj, 155, 94

\bibitem[{Shapiro \& Wilk(1965)}]{shapirowilk1965}
Shapiro, S.~S. \& Wilk, M.~B. 1965, Biometrika, 52, 591

\bibitem[{{Steffen} \& {Hwang}(2015)}]{steffen2015}
{Steffen}, J.~H. \& {Hwang}, J.~A. 2015, \mnras, 448, 1956

\bibitem[{{Sullivan} {et~al.}(2015){Sullivan}, {Winn}, {Berta-Thompson},
  {Charbonneau}, {Deming}, {Dressing}, {Latham}, {Levine}, {McCullough},
  {Morton}, {Ricker}, {Vanderspek}, \& {Woods}}]{sullivan2015}
{Sullivan}, P.~W., {Winn}, J.~N., {Berta-Thompson}, Z.~K., {et~al.} 2015, \apj,
  809, 77

\bibitem[{{Terquem} \& {Papaloizou}(2007)}]{terquem2007}
{Terquem}, C. \& {Papaloizou}, J. C.~B. 2007, \apj, 654, 1110

\bibitem[{Tibshirani(1994)}]{tibshirani1994}
Tibshirani, R. 1994, Journal of the Royal Statistical Society, Series B, 58,
  267

\bibitem[{{Weiss} {et~al.}(2018){Weiss}, {Marcy}, {Petigura}, {Fulton},
  {Howard}, {Winn}, {Isaacson}, {Morton}, {Hirsch}, {Sinukoff}, {Cumming},
  {Hebb}, \& {Cargile}}]{weiss2018}
{Weiss}, L.~M., {Marcy}, G.~W., {Petigura}, E.~A., {et~al.} 2018, \aj, 155, 48

\bibitem[{{Xie}(2013)}]{xie2013}
{Xie}, J.-W. 2013, \apjs, 208, 22

\bibitem[{{Zechmeister} \& {K{\"u}rster}(2009)}]{zechmeister2009}
{Zechmeister}, M. \& {K{\"u}rster}, M. 2009, \aap, 496, 577

\end{thebibliography}

        \appendix

        \section{Complementary analysis of the time series }
        \label{app:compfig}
        
        \subsection{Accounting for instrumental effects in RV}
\label{app:inst}
SOPHIE experiences a drift of the zero velocity point due to several factors, as follows: a change of fiber, calibration lamp aging, and other systematic effects.
A drift estimate was obtained by observing reference stars, which are deemed to have a nearly constant velocity,  each night of observations. The reference stars' velocities  were combined and interpolated to create an estimate of the drift as a function of time. This one was then subtracted from all the time series of the observation program. The estimation procedure, which is similar to that of~\cite{courcol2015}, is presented in detail in~\cite{hara2019}. We performed the time-series analysis of the data obtained with the original correction of~\cite{courcol2015}. The results are very similar; the only notable difference is a lower statistical significance of the 12.0 days signal.

\subsection{Data selection}
\label{app:dataselect}

%In the obtained time series, there might remain outliers, most likely due to bad observation conditions. \ch{Our outlier selection criterion, based on the mean absolute deviation, leads to remove three measurements at barycentric Jullian day (BJD) 2457941.5059, 2457944.4063, 2457945.4585.} The selection procedure is detailed in appendix~\ref{app:dataselect}.

Some of the data points of the radial velocity and  $\log R'_{Hk}$  time series are excluded from the analysis. In this section, we present the methodology adopted. 

We removed outliers from the time series with a criterion based on the median absolute deviation (MAD). We computed $\sigma = 1.48$ MAD, which is the relation between the standard deviation $\sigma$ and the MAD of a Gaussian distribution. We excluded the data points if their absolute difference to the median is greater than $k \times \sigma$ with $k=4$.

\begin{figure} \centering
        \centering              
        \includegraphics[width=9cm]{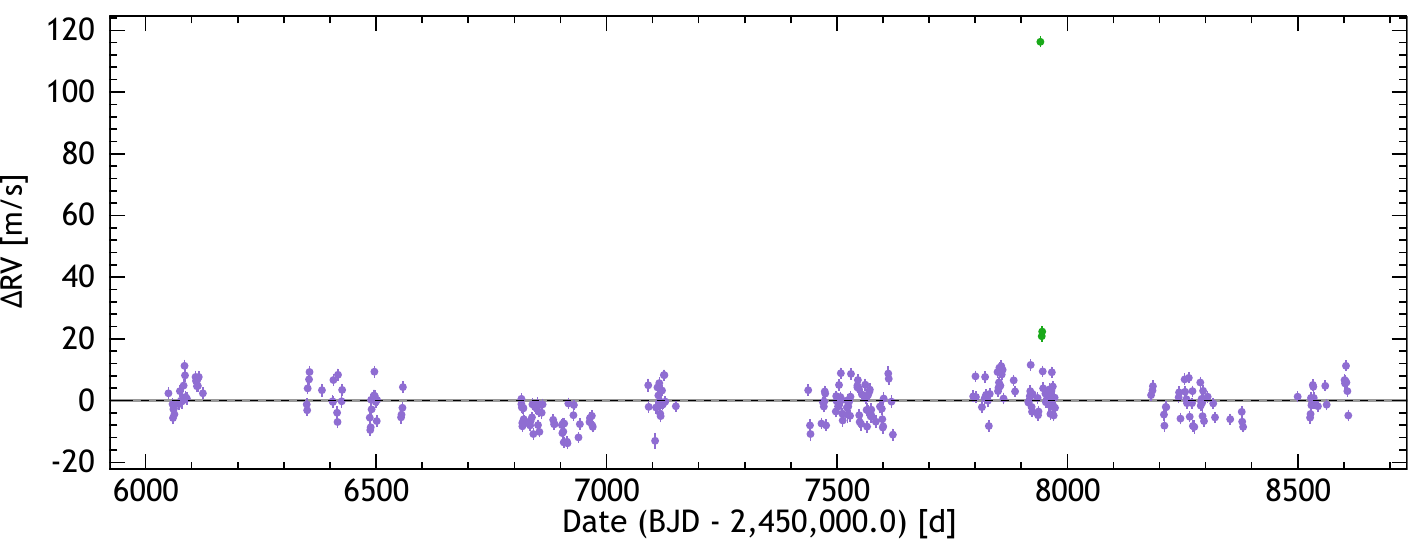}
        \caption{ SOPHIE radial velocities and nominal $1\sigma$ error bars, the three points which were removed are shown in green. }
        \label{fig:rvoutliers}
\end{figure}

In Fig.~\ref{fig:rvoutliers}, we show the radial velocity data points and their error bars based on the nominal uncertainties of the measurements.    The three points excluded from the analysis based on the MAD criterion are represented  in green. There is one point with a very clear deviation and two  outliers with a lower deviation. The farthest outlier prevents finding any significant signal in the RVs besides a signal with a time-scale of 2000 d. Including the two other outliers in  the RV analysis yields minor changes in the analysis, and it does not challenge the detection of the planets. 
The $\log R'_{Hk}$ and bisector measurements at the dates of the three outliers are also excluded from the analysis. 

\begin{figure} \centering
        \hspace{-0.2cm}
        \includegraphics[width=10cm]{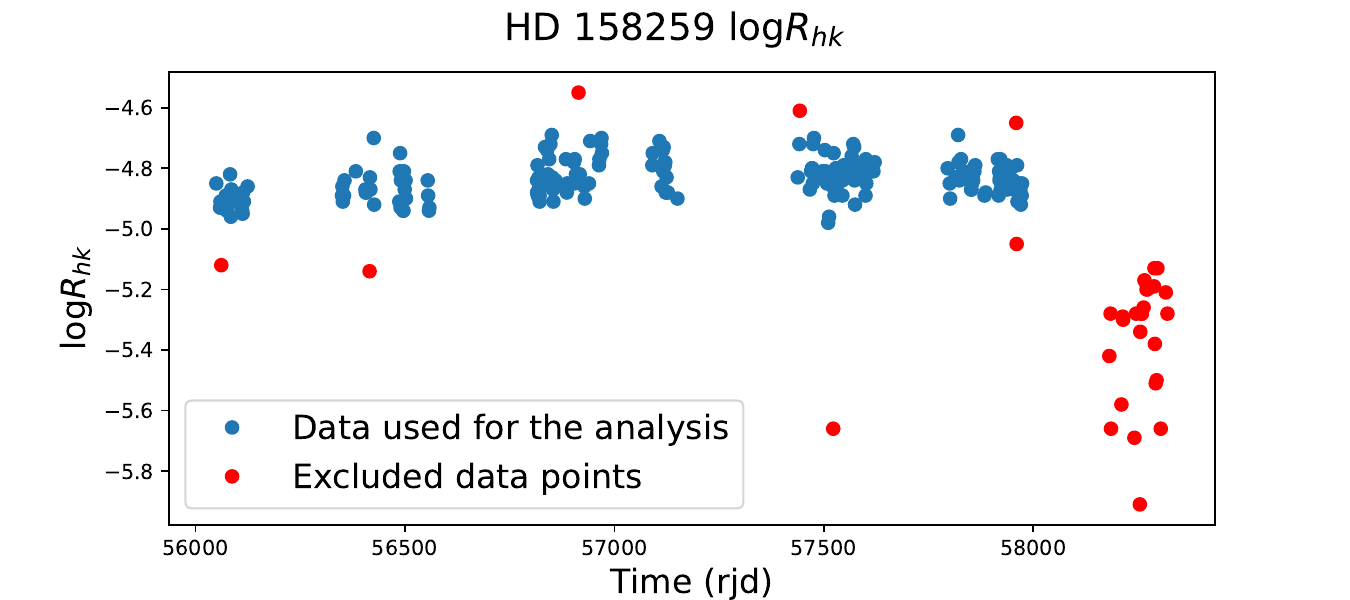}
        \caption{ $\log R'_{HK}$ measurements of HD 158259. Points in blue are conserved for the analysis; the points in red are excluded from the analysis.}
        \label{fig:rhk}
\end{figure}

In the case of the $\log R'_{Hk}$, besides the removal criterion based on the MAD,  the  $\log R'_{HK}$ measured after BJD 2458000 are excluded. After this date, the values of the $\log R'_{HK}$ are not reliable. \ch{ This is due to the change of the calibration lamp from thorium-argon to Fabry-Perot, which leaks on the stellar spectrum as a continuum background. No effects are seen in the RV of the constant stars, which were observed each night, but the leaking affects the measurements of the $\log R'_{Hk}$. This issue is described in greater detail in~\cite{hobsonthese}, p. 92. We
note that excluding some of the $\log R'_{HK}$ points has a very limited impact on our results, which do not depend on the exact modeling of low frequency components in the RV signal. Furthermore, excluding the RV  points after the calibration change does not affect our results.} The points selected for the analysis are presented in Fig.~\ref{fig:rhk} in blue, while the points excluded are represented in red. \ch{In our analysis, we do not exclude the RV points at the dates of the points excluded from the $\log R'_{HK}$. }
The application of the MAD criterion on the bisector span and photometry does not lead to the removal of any point.

        \subsection{APT Photometry}
\label{app:apt}

\ch{
We acquired 320 observations of HD~158259 during the 2002, 2003, 2004, and
2005 observing seasons with the T11 0.80~m Automatic Photoelectric Telescope 
(APT) at Fairborn Observatory in southern Arizona.  T11 is equipped with a 
two-channel precision photometer that uses a dichroic mirror and two 
EMI 9124QB bi-alkali photomultiplier tubes to measure the 
Str\"omgren $b$ and $y$ passbands simultaneously.  We observed the program star with respect 
to three comparison stars and computed the differential magnitudes as the 
difference in brightness between the program star and the mean brightness of 
the two best comparison stars.  To improve the photometric precision, we 
combined the differential $b$ and $y$ magnitudes into a single $(b+y)/2$ 
passband.  The precision of a single observation is typically around 
~0.0015~mag.  The T11 APT is functionally identical to our T8 0.80~m APT 
described in \citet{henry1999}, where further details of the telescope, precision 
photometer, and observing and data reduction procedures can be found. }

\ch{
Table~\ref{tab:photo} gives a summary of the photometric results.  No significant variability is found within any observing season.  The seasonal means exhibit a range of 0.0018~mag, which is probably due to similar variability seen in the mean magnitudes of the comparison stars.  The lack of observed spot activity is 
consistent with the star's low value of $\log R^{\prime}_{\rm HK}$ = -4.8.  For further analysis, the four observing seasons of APT photometry were normalized such that the last three seasons have the same mean magnitude as the first (see next section).}

\begin{table}
        \caption{Summary of APT photometric observations for HD 15825.}
        \label{tab:photo}
        \begin{tabular}{p{0.7cm}p{0.4cm}p{2cm}p{0.9cm}p{2.75cm}}
                Obs. season & $N_{\mathrm{obs}}$ & Date range ($HJD$-2,400,000) & Sigma (mag)& Seasonal mean (mag) \\ \hline \hline
                2002   &  82 & 52289--52584 & 0.0013 & $-1.22669\pm0.00015$   \\
                2003   &  71 & 52653--52948 & 0.0010 & $-1.22694\pm0.00011$   \\
                2004   & 100 & 53024--53308 & 0.0013 & $-1.22677\pm0.00013$   \\
                2005   &  67 & 53384--53560 & 0.0010 & $-1.22510\pm0.00012$  \\
        \end{tabular}
\end{table}

\subsection{Period search}
\label{app:periodsearch}

The presence of signals in the ancillary indicators might give hints as to the instrumental and stellar features in the radial velocity. We are particularly interested in periodic signals, which could also appear in the RV and mimic a planetary signal. To search for periodic signals, we  analyzed the photometry and ancillary indicators with classical and $\ell_1$ periodograms. The periods found are used to build candidate  quasi-periodic noise models  in  the RV. 

We first present the classical periodogram analysis. We computed the generalized Lomb-Scargle periodogram~\citep{ferrazmello1981, zechmeister2009} of the ancillary indicators on a grid of frequencies spanning from 0 to 1.5 cycles/day. We report the strongest periodic signatures. The false alarm probability (FAP) of the highest peaks of the periodograms were computed using the~\cite{baluev2008} analytical formula. After computing the periodogram and checking that no significant high frequency signal is found,  we performed the search on a grid of frequencies from 0 to 0.95 cycle/day to avoid aliases in the one day region. We  subtracted signals at the periods found iteratively and computed the periodograms of the residuals. We also applied $\ell_1$ periodograms for comparison purposes. 

\begin{figure} \centering
        \hspace{-0.2cm}
        \includegraphics[width=10cm]{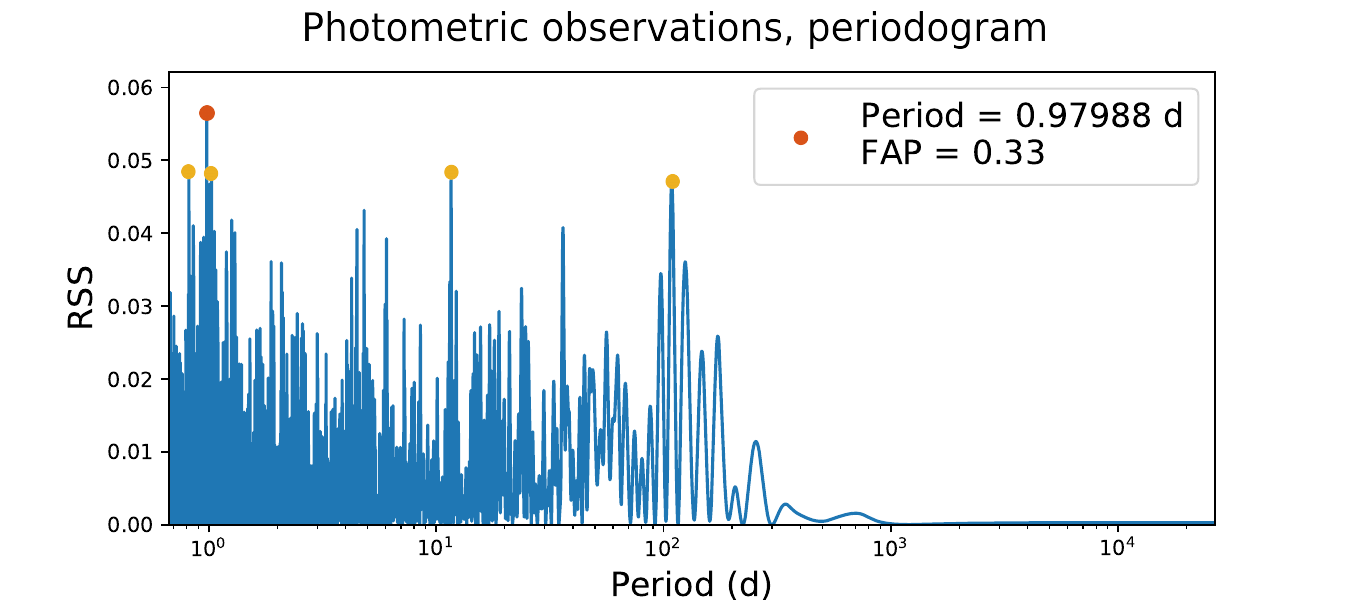}
        \caption{Periodogram of the photometric data computed between on an equispaced frequency grid between 0 and 1.5 cycle/d. The maximum peak is attained at 0.979 d (in red). The four subsequent tallest peaks are, in decreasing order, at 0.817  11.636,   1.028, and 108 d (in yellow).  }
        \label{fig:photperio}
        \vspace{0.5cm}
        
        \hspace{-0.2cm}
        \includegraphics[width=10cm]{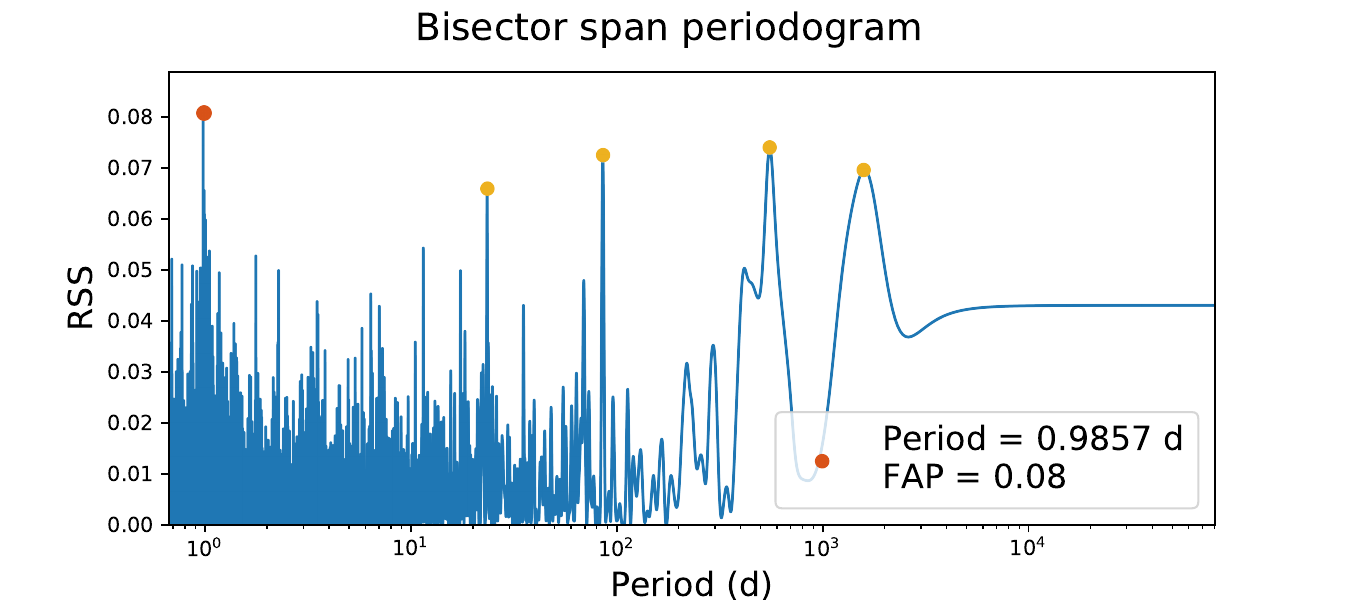}
        \caption{Periodogram of the bisector span  computed on an equispaced frequency grid between 0 and 1.5 cycle/d. The maximum peak is attained at 0.985 d (in red). The four subsequent tallest peaks are, in decreasing order, at 552, 85.5, 1576, and 23.48 d (in yellow).}
        \label{fig:bistperio}   
        
        \hspace{-0.2cm}
        \includegraphics[width=10cm]{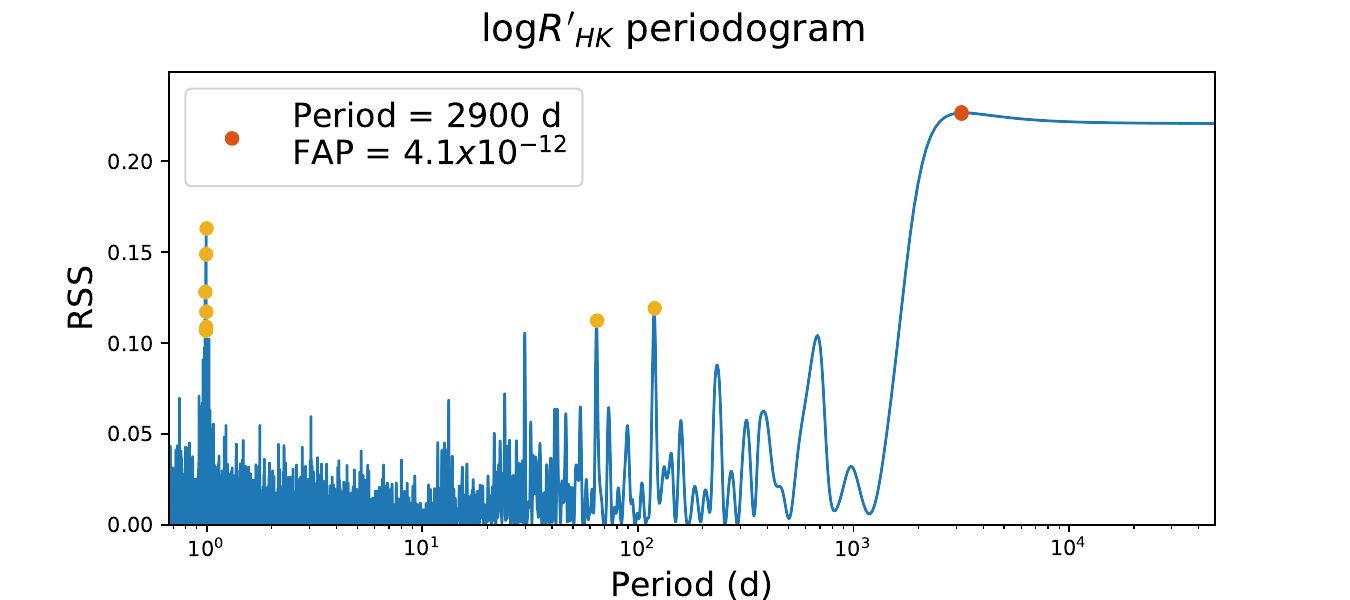}
        \caption{Periodogram of the $\log R'_{Hk}$  computed  on an equispaced frequency grid between 0 and 1.5 cycle/d. The maximum peak is attained at 2900 d (in red). Besides aliases in the one day region, there is a peak at 119 and 64 days. }
        \label{fig:rhkperio}    
\end{figure}

The $\log R'_{Hk}$ analysis was performed after the data selection, which is presented in Section~\ref{app:dataselect}.
The  periodogram (see Fig.~\ref{fig:rhkperio}) presents a clear long-term trend and, besides peaks in the one day region, it peaks at 119 and 64 d. The  iterative period search on a frequency grid from 0 to 0.95 cycle/day gives signals at 3500, 119, and 32 d, with FAPs $4\cdot 10^{-12}$, 0.18 and 0.14. The long-term signal in the $\log R'_{Hk}$ might correspond to a magnetic cycle. In Section~\ref{app:cv2}, we fit a Gaussian process model to the $\log R'_{Hk}$ data (see Fig.~\ref{fig:rhkgp}), which was used in the RV processing.  

The bisector span periodogram is presented in Fig.~\ref{fig:bistperio}. It presents peaks, in order of decreasing amplitude at 
0.9857, 552, 85,  1576, and 23.5 days, where 0.9857 is an alias of 85 days. The FAP of the highest peak is 0.08, which indicates \cht{marginal} evidence against the hypothesis that the bisector behaves like white noise. The iterative period search from 0 to 0.95 cycle/day points to signals at 550, 85, and 11.5 d with FAPs of 0.13, 0.21, and 0.25.

In Fig.~\ref{fig:photperio}, we represent the periodogram of the photometric data. The maximum peak occurs at a period of 0.979 days, which is an alias of 11.6 days, and has a FAP of 0.33. 
The iterative search yields 11.63 and 108 days as dominant periods with FAPs of 0.75 and 0.43.  The fact that 11.5 days both appear in the bisector and photometry might indicate that there is a weak stellar feature at this period, which is possibly due to the rotation period or one of its harmonics. \cht{The fitted amplitudes of sine functions initialized at the orbital periods are given in Table~\ref{tab:tranist}.  In each case, the peak-to-peak amplitude is very small (a fraction of a millimagnitude) and of the same order as its uncertainty.  %This isfirm evidence that the observed Doppler shifts are due to planetary reflex motion and not intrinsic to the star itself (spots or pulsations).
}

\begin{table}
        \caption{Photometric amplitudes on the planetary orbital periods}
        \label{tab:tranist}
        \begin{tabular}{cc}
                Orbital period (d) & Peak-to-Peak Amplitude (mag) \\ \hline \hline
                2.177   &   $0.00019\pm0.00021$ \\
                3.432   &   $0.00010\pm0.00022$ \\
                5.198   &   $0.00029\pm0.00022$ \\
                7.954   &   $0.00024\pm0.00021$ \\
                12.03    &   $0.00056\pm0.00021$ \\
                17.39    &   $0.00038\pm0.00021$ \\
        \end{tabular}
\end{table}

\begin{figure} \centering
        \includegraphics[width=10cm]{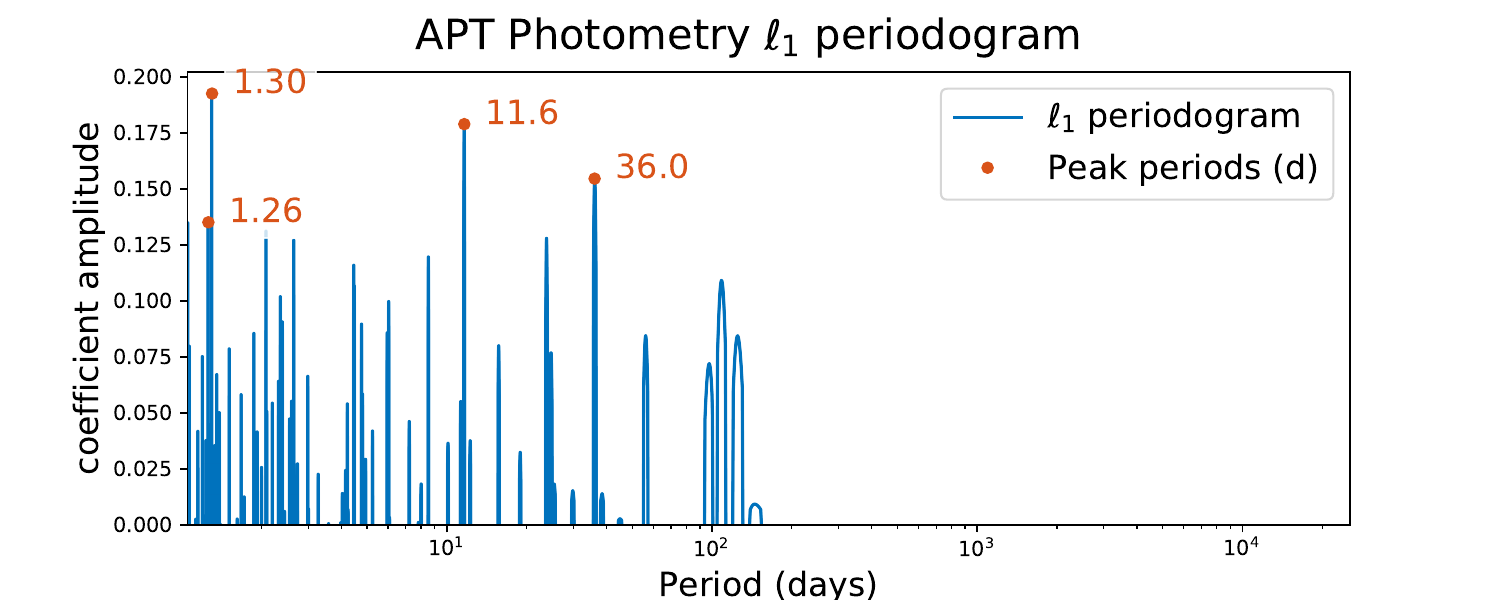}
        \caption{ $\ell_1$ periodogram of the APT photometric measurements.  }
        \label{fig:l1phot}
\end{figure}

\begin{figure} \centering
        \includegraphics[width=10cm]{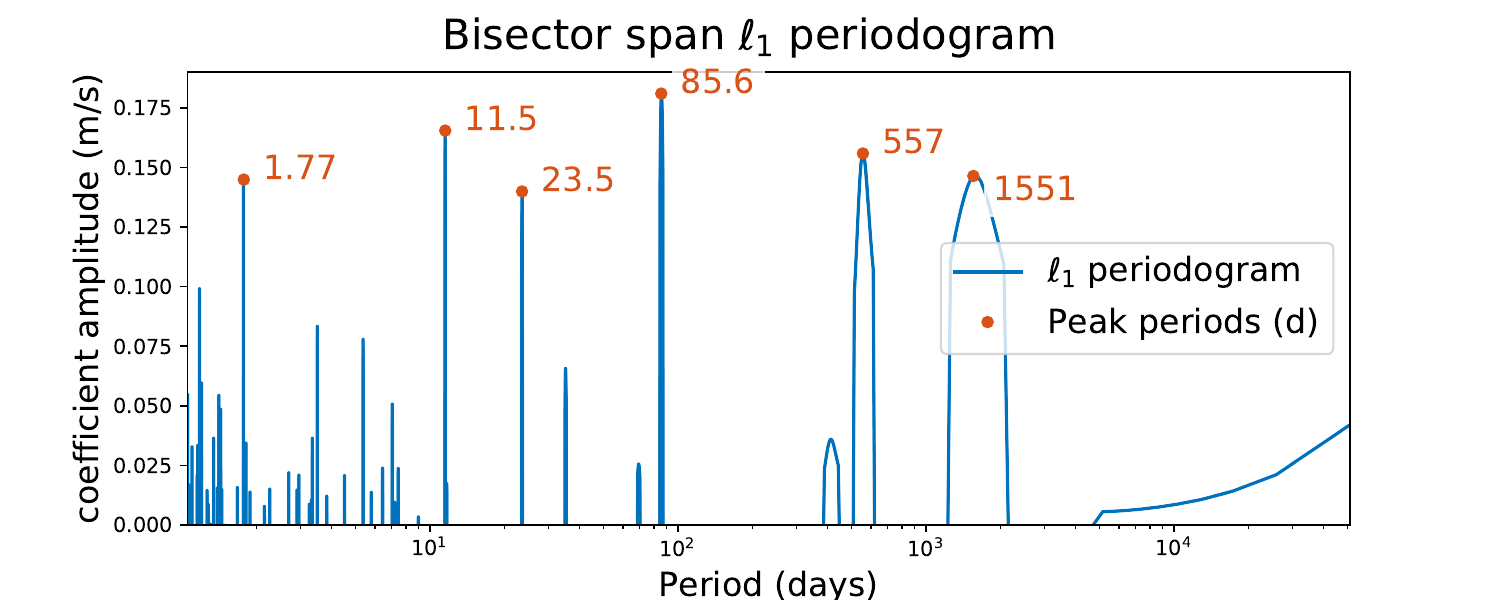}
        \caption{ $\ell_1$ periodogram of the bisector span.  }
        \label{fig:l1bis}
\end{figure}

\begin{figure} \centering
        \includegraphics[width=10cm]{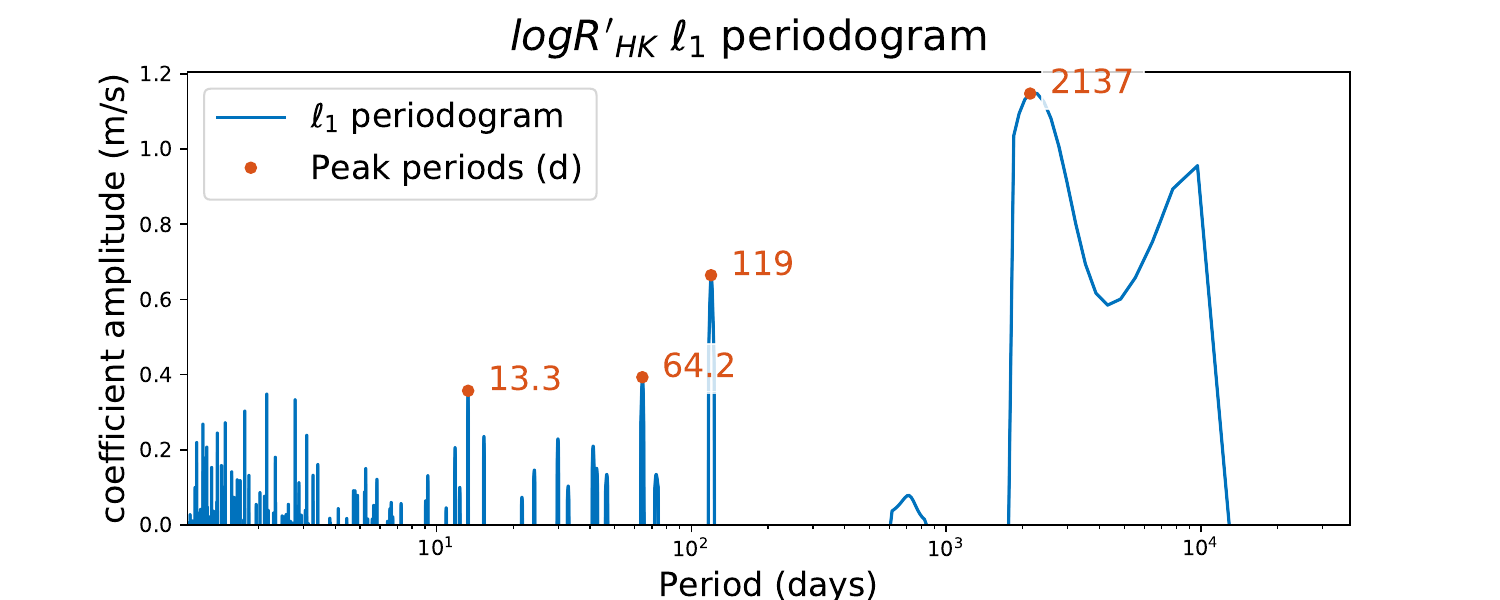}
        \caption{ $\ell_1$ periodogram of the $\log R'_{HK}$. }
        \label{fig:l1rhk}
\end{figure}

\ch{The $\ell_1$ periodograms of the photometry, bisector span, and the $\log R'_{HK}$ (resp. Fig.~\ref{fig:l1phot} Fig.~\ref{fig:l1bis} and Fig.~\ref{fig:l1rhk}) were computed on a frequency grid from 0 to 0.95 cycles/day to avoid the one day region. The $\ell_1$ periodogram aims to express the signal as a sum of a small number of sinusoidal components. Here, the $\ell_1$ periodograms present numerous peaks, which is characteristic of noisy signals,  as they do not have a sparse representation in frequency.  The periods appearing in $\ell_1$ periodograms are consistent with those appearing in the classical analysis. We simply note the addition of a signal at 23.5 d in the bisector span $\approx 11.6 \times 2$. }

In conclusion, the analysis of the photometry and ancillary indicators supports the existence of a long-term magnetic cycle, appearing in the $\log R'_{Hk}$ periodogram, with a period  $\geqslant 1500 $ d. The period cannot be resolved due to the time-span of the SOPHIE observations (2560 days). The presence of several marginally significant periods in the bisector span might indicate the presence of correlated noise in the bisector, and possibly in the RV.

% There is marginal evidence for a signal at 119 d in the $\log R'_{Hk}$  data, which  does not necessarily indicate a periodic structure in the noise, and might result only from correlated noise structures without a periodic covariance.  Let us note that a peak at 108 days appears in the photometry, which could have a common origin. None of the periods appearing in the bisector span are very statistically significant.

%We simply note the possibility that a 23.5 and 85.5 days might be due to the stellar rotation. 

        \section{Impact of the noise model on the planet detection}
\label{app:cv}

\subsection{Selection with cross validation}
\label{app:cv1}

In this section, we present the noise model chosen and the sensitivity of the detection to the noise model. The various sources of noise in the RV were modeled as in~\cite{haywood2014} by a correlated Gaussian noise model. In practice, one chooses a parametrization of the covariance matrix of the noise $V(\theta)$, where the element of $V$ at index $k,l$  depends on $|t_k-t_l|$ and a vector of parameters $\theta$. In the following analysis, the parametrization chosen for $V$ is such that its element at index $k,l$ is
\begin{align}
\begin{split}
V_{kl}(\theta ) & =  \delta_{k,l} (\sigma_{k}^2 + \sigma_{W}^2) + \sigma_{C}^2 c(k,l) \\  &+   \sigma_{R}^2 \e^{-\frac{|t_k-t_l|}{\tau_R}} +
\sigma_{QP}^2\e^{-\frac{|t_k-t_l|}{\tau_{QP}}} \frac{1}{2} \left(1 + \cos\left(\frac{2\pi(t_k-t_l)}{P_\mathrm{act}} \right)  \right)
\end{split}
\label{eq:kernel}
\end{align}
where $\sigma_k$ is the nominal measurement uncertainty, $\sigma_W$ is an additional white noise, $\sigma_{C}$ is a calibration noise, and where $c$ equals one if measurements $k$ and $l$ are taken within the same night and zero otherwise. The quantities $\sigma_{R}$ and $\tau_R$ parametrize a correlated noise, which might originate from the star or the instrument. Additionally,  $\sigma_{QP}, \tau_{QP} $, and $P_\mathrm{act}$ parametrize a quasi-periodic covariance, potentially resulting from spots or faculae. The form of this covariance is compatible with the \texttt{spleaf} software~\citep{delisle2019b} and, except for the calibration noise, the \textsc{CELERITE}  model~\citep{foremanmackey2017}.

To study the sensitivity of the detection to the noise model $\theta = (\sigma_W$ , $\sigma_{C}$, $\sigma_{R}$ $\tau_R$ , $\sigma_{QP}, \tau_{QP} $,  $P_\mathrm{act})$, we proceeded as follows.
%We here give a brief outline of the procedure which is described in detail in~\cite{hara2019}. 
We first considered a grid of possible values for each component of $\theta$. For instance, $\sigma_R = 0, 1,2,3,4 $ m/s, $\tau_{R} = 0,1,6$ d etc. The $\theta$ with all the possible combinations of the values of its components were generated, and the corresponding covariance matrices were created according to Eq.~\eqref{eq:kernel}. 

The particular values taken in the present analysis are reported in Table~\ref{tab:cv}. The decay time scales of the red and quasi periodic components (subscripts R and QP) include 0 in which case they are white noise jitters. 
The $\sigma_R$ and $\sigma_{QP}$ were chosen such that when $\tau_{R} = \tau_{QP} = 0,$ there exists a value of $\sigma_W^2 + \sigma_C^2 + \sigma_R^2 + \sigma_{QP}^2$ that is greater than the total variance of the data, and we subdivided the possible values of $\sigma_R , \sigma_{QP}$ in smaller steps. The correlation time-scales of $\tau_R = 0, 1, $ and $6$ d correspond to no correlation, daily correlation, and noise correlated on a whole run of observations, which is typically 6 days. The $P_{\mathrm{act}}$ candidate corresponds to peaks that appear in the period analysis of the radial velocity or ancillary indicators (section~\ref{app:periodsearch}). We note that 2000 d was not considered as it is degenerate with an exponential decay due to the observation time span.

For each matrix in the list of candidate noise models, the $\ell_1$ periodogram was computed, and the frequencies that have a peak with FAP < 0.05 were selected. \ch{We then attributed a score to the couple (noise model, planets) with two different metrics.} 

\ch{The first metric is based on cross validation}. We selected 70\% of the data points randomly -- the training set -- and fit a sinusoidal model at the selected frequencies on this point. On the remaining 30\% -- the test set -- we computed the likelihood  of the data knowing the model fitted. The operation of selecting of training set randomly and evaluating the likelihood on the test set was repeated 200 times. We took the median of the 200 values of the likelihood as the cross validation score of the noise model. As a result of this procedure, each noise model has a cross validation score (CV score). 

\ch{The second metric is the approximation of the Bayesian evidence of the (noise model, planets) couple. For a given covariance matrix, we fit a sinusoidal model initialized at the periods selected by the FAP criterion. We then made a second order Taylor expansion of the posterior (with priors \cht{set to one} in all the variables) and estimated the evidence of the model, which is also called the Laplace approximation~\citep{kassraftery1995}. The procedure is explained in greater detail in~\cite{nelson2018}, Appendix A.4. }

\begin{figure} \centering
        %\begin{minipage}[l]{0.48\textwidth} 
        \hspace{-0.2cm}
        \includegraphics[width=10cm]{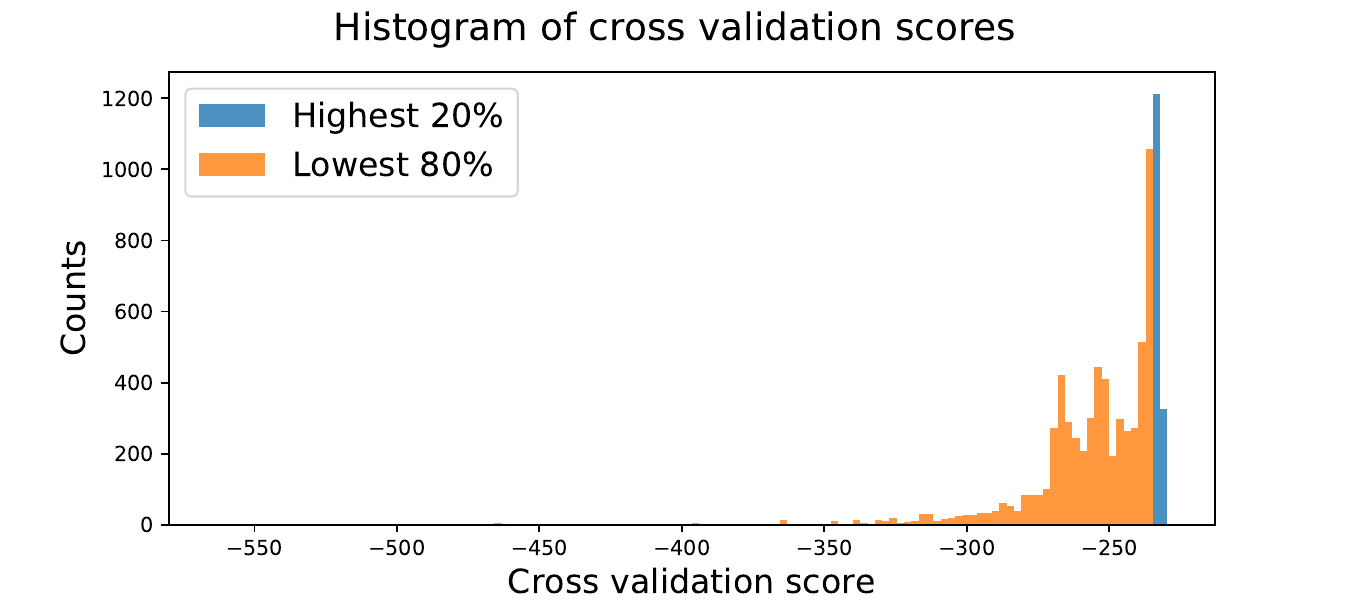}
        \caption{Histogram of the values of the cross validation score. Best 20\% and lowest 80\% are represented in blue and orange, respectively. }
        \label{fig:histogram}   
        %\end{minipage}
\end{figure}

 In Fig.~\ref{fig:histogram}, we represent the histogram of the values of the CV score for all the noise models considered. The models with the 20\% highest CV score are represented in blue, and they present similar values of the CV score. We call the set of these models $\mathrm{CV}_{20}$. 

\begin{figure} \centering
        \hspace{-0.2cm}
        \includegraphics[width=9 cm]{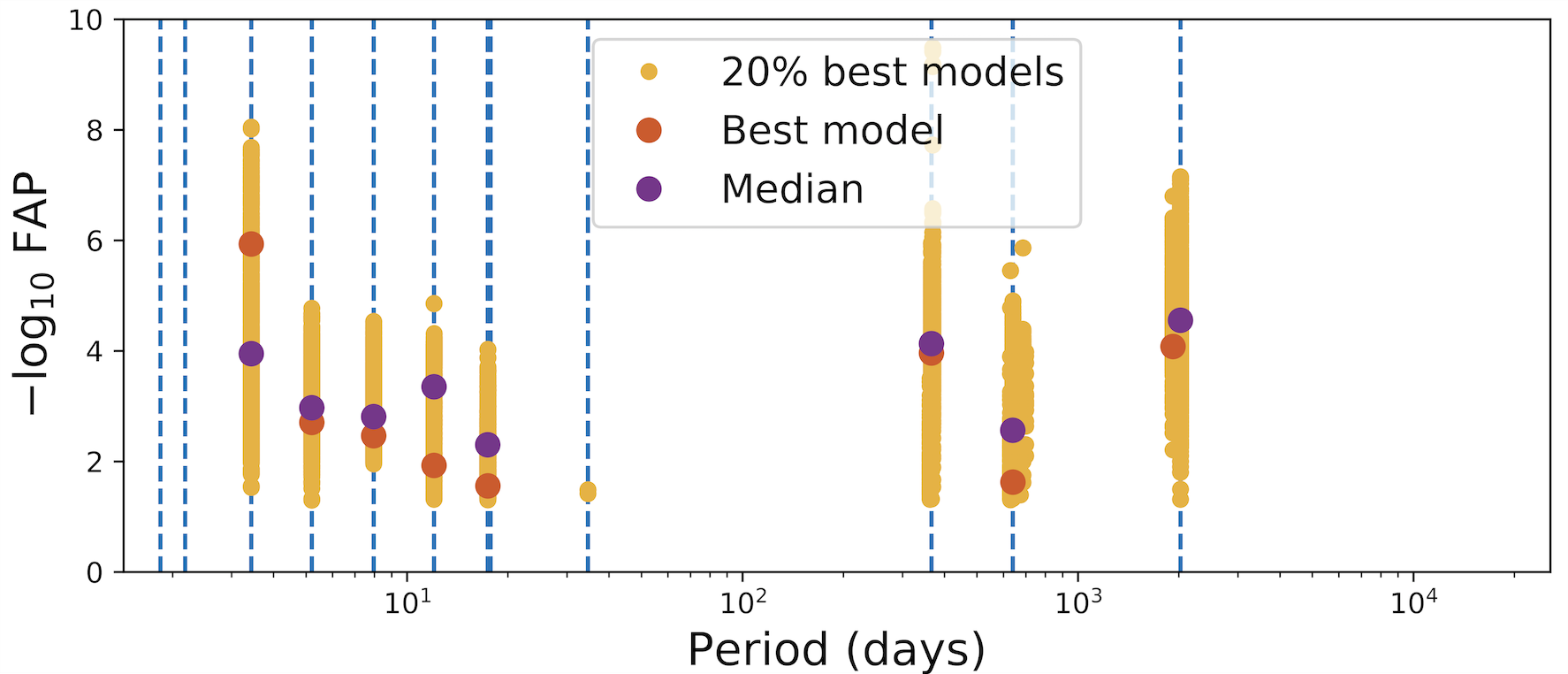}
        \caption{ FAPs of the peaks of the 20\% best models \ch{in terms of cross validation score} that have a FAP >0.05.
                The periods marked in red in Fig.~\ref{fig:l1perio} are represented by the blue dashed lines. }
        \label{fig:crossval}    
\end{figure}

Each model of $\mathrm{CV}_{20}$ might lead to different peaks selected with the FAP<0.05 criterion.  
The periods and  FAPs of the peaks selected are represented in Fig.~\ref{fig:crossval} by the yellow points. For a comparison with the best models, that is, the model with the maximal CV score, the periods appearing in the best model (red dots in Fig.~\ref{fig:l1perio}) are represented by the blue dashed lines. %The signals at 3.43, 5.19, 7.95, 12.0, 17.4, 366 and 1920 days appear in \ch{at least 90\%} of the $\mathrm{CV}_{20}$ models.
We represent the median of the FAPs for each of these periods in the $\mathrm{CV}_{20}$ and their FAP in the best models (purple and red, respectively). \ch{These values are also listed in Table~\ref{tab:cvresults}, where we also give the percentage of cases where a period corresponds to a FAP<0.05 in $CV_{20}$. }

%Note that the signals at 1.83 and 34.5 d are not included in the signals with FAP <0.05 for the $\mathrm{CV}_{20}$ models, except once for 34.5 d. Finally, we note that a signal at $\approx 640$ d  is included in 9\% of the $\mathrm{CV}_{20}$ models.

\begin{figure}
        \includegraphics[width=9.2cm]{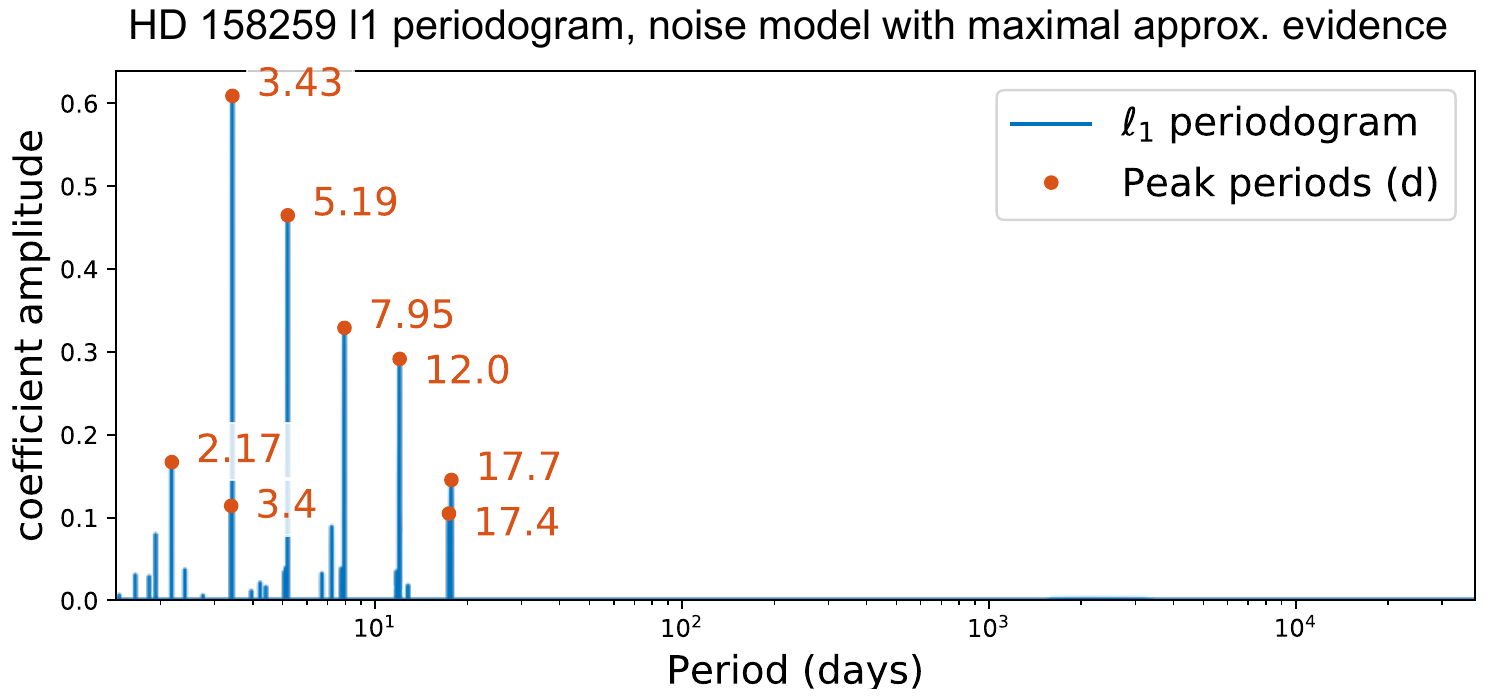}  
        \caption{$\ell_1$ periodogram corresponding to the highest approximate evidence. }
        \label{fig:l1perioevidence}
\end{figure}

\begin{figure} \centering
        \hspace{-0.2cm}
        \includegraphics[width=9 cm]{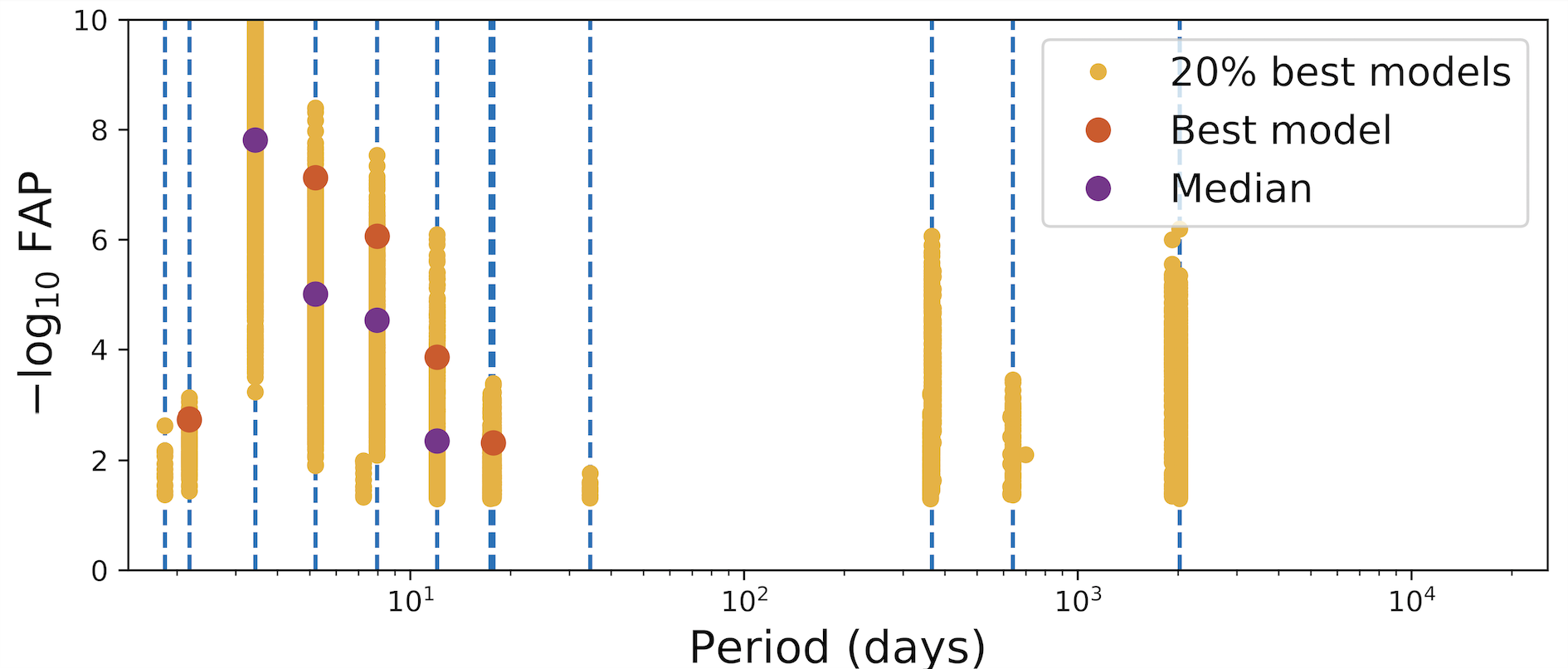}
        \caption{ FAPs of the peaks of the 20\% best models \ch{in terms of Laplace approximation of the evidence} that have a FAP >0.05.
                The periods marked in red in Fig.~\ref{fig:l1perio} are represented by the blue dashed lines. }
        \label{fig:evidence}    
\end{figure}

\ch{We plotted the same figure as~\ref{fig:crossval} and the same table as~\ref{tab:cvresults} for the 20\% best models in terms of evidence ($E_{20}$) (Fig.~\ref{fig:evidence} and table~\ref{tab:evresults}).         The $\ell_1$ periodogram corresponding to the highest ranked model is shown in Fig.~\ref{fig:l1perioevidence}). %The notable difference is that eow frequency signals and boosts the high frequencies. 
        Peaks at 3.4, 5.2, 7.9, and 12 d have similar behaviors with cross validation and approximate evidence. We see three notable differences when ranking a model with evidence: $\approx 10\%$ of the selected models display significant peaks at 1.84 or 2.17 d, 17.4 reaches a 5\% FAP in only 38\% of $E_{20}$, and 8\%  of the models  favor 17.4 over 17.7 d. Finally, the inclusion percentage and average FAP of long period signals significantly drops. 
        We have also ranked models with AIC~\citep{akaike1974} and BIC~\citep{schwarz1978}, which yield results very similar to approximate evidence. }    

When using a LASSO-type estimator~\citep{tibshirani1994}, such as the $\ell_1$ periodogram, for a fixed dictionary (here, a fixed noise model), it is common practice to select the model with cross-validation as the solution follows the so-called LASSO path. Here, this would constitute a viable alternative to selecting the peaks that have FAP <0.05. This was tested on a grid of parameters such as~\ref{tab:cv}, and it yields very similar conclusions. %\ch{One could also compute only evidence estimations , such a technique coulf}

\subsection{Long-term model}
\label{app:cv2}

It has been found that the $\log R'_{hk}$ has a strong long-term signal. This one can be included in the analysis with a Gaussian process analysis, similarly to~\cite{haywood2014}. 
We consider a simple covariance model for the $\log R'_{HK}$
\begin{align}
\begin{split}
V_{kl}(\theta ) & =  \delta_{k,l} \sigma_{W}^2 + \sigma_{C}^2 c(k,l) +   \sigma_{R}^2 \e^{-\frac{(t_k-t_l)^2}{2\tau_R^2}} 
\end{split} .
\label{eq:rhkkernel}
\end{align}

\begin{figure} \centering
        \hspace{-0.2cm}
        \includegraphics[width=9.4cm]{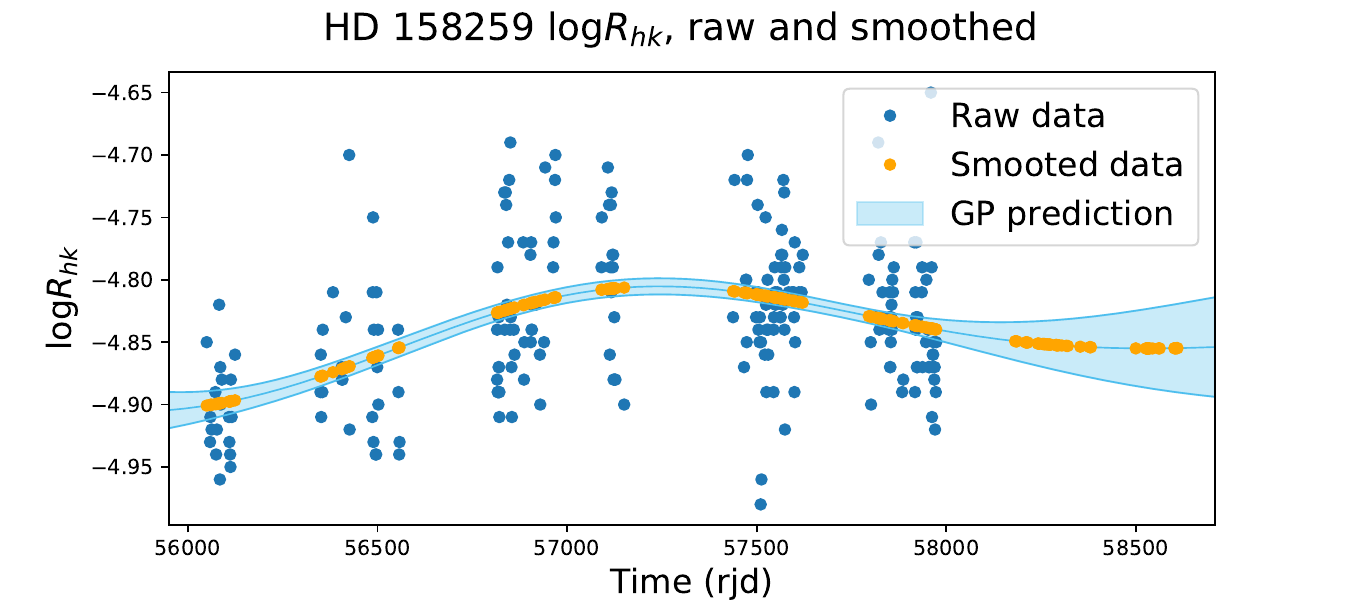}
        \label{fig:rhkgp}
        \caption{Raw and smoothed $\log R'_{HK} $ time series. The dark blue points represent the raw $\log R'_{HK} $ data used for the prediction. The light blue lines represent the Gaussian process prediction and its $\pm 1 \sigma$ error bars (see eq;~\eqref{eq:rhkkernel}). The orange points are the predicted values of the Gaussian process  at the radial velocity measurement times.}
\end{figure}

We assume that $\sigma_{R}$ is equal to the standard deviation of the $\log R'_{HK}$ data, and we fit the parameters $\sigma_{W}$ and $\tau$. We find $\sigma_{W} = 0.9 \sigma_{R}$ and $\tau = 770$ d. We then predicted the Gaussian process value and its covariance with formulae 2.23 and 2.24 in~\cite{rasmussen2005}. In Fig.~\ref{fig:rhkgp}, we represent the raw $\log R'_{HK}$ data on which the fitting is made. The Gaussian process and the one sigma standard deviation of the marginal distribution at $t$ are represented in light blue, and the prediction at the radial velocity measurement times is in orange. 

\begin{figure} \centering
        \hspace{-0.2cm}
        \includegraphics[width=9 cm]{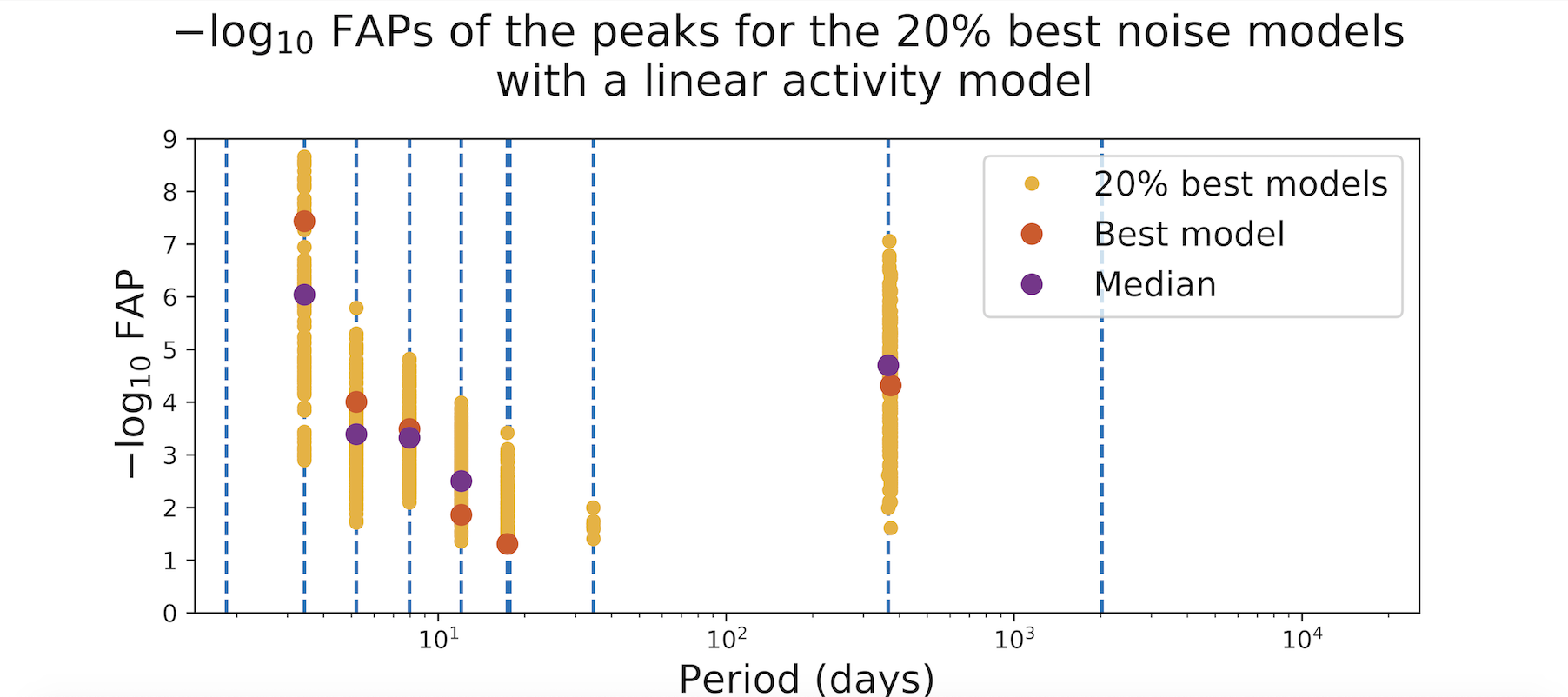}
        \caption{ FAPs of the peaks of the 20\% best models that have a FAP >0.05, when adding a linear activity model fitted as a Gaussian process.
                The periods marked in red in Fig.~\ref{fig:l1perio} are represented by the blue dashed lines. }
        \label{fig:crossvalgp}  
\end{figure}

We then performed the same analysis as in Section~\ref{app:cv1}, except that we include the smoothed  $\log R'_{HK}$ (orange points in Fig.~\ref{fig:rhkgp}) as a linear predictor in the model. The results are presented in Fig.~\ref{fig:crossvalgp}. These are almost identical to the results of Section~\ref{app:cv1}, except that the 2000 d \ch{and 640 d} signals disappear. %We point out that the cross validation score of the best model is -236, while the best cross validation scores including a fitted long term sinusoid is -228, pointing to a very slight advantage for the sinusoidal model of the long trend, but they cannot be distinguished with certainty.


\subsection{Discussion}
\label{app:cv_discuss}

The noise model has a strong impact on the false alarm probabilities. When instantiating the covariance matrix with an exponential kernel, such as Eq.~\ref{eq:rhkkernel}, the characteristic time $\tau$ can be seen as setting the threshold between ``short'' and ``long'' periods. The FAPs of the signals with periods lower than  $\tau$ are lower and higher, respectively, compared to the ones computed with a white noise model. Using  a correlated noise model reduces the chances of spurious detection claims at long periods, but on the other hand, it decreases the statistical power (see~\cite{delisle2019a} for a more detailed discussion on that point).
The  procedure of Section~\ref{app:cv1} aims to balance the two types of errors: If the white noise model cannot be excluded (here, this means being in the best 20\% noise models), then the short and long period signal  appear with lower and higher FAPs, respectively, resulting in a compromise value. We have found the result to be insensitive to the exact threshold on the models selected (choosing the best 10, 20, and 30 \% yields identical conclusions).  The interpretation of the differences between cross validation and evidence is that evidence appears to favor models with stronger correlated components, which damp the amplitude of long period signals and boost high frequency signals. 

We have found that signals at  3.43, 5.19, 7.95, 12.0, 17.4, and 366  d consistently appear in the models. The 1920 and $\ch{640}$  d signals appearing in Section~\ref{app:cv1} disappear when modeling the activity as in Section~\ref{app:cv2}, pointing to a stellar origin. \ch{The strength of the 17.4 d signal varies with the model ranking technique, which makes its detection less strong than the other signals. We, however, note that this one was put in competition with a quasi-periodic model at the same period, which is not the case of the other signals.} Overall, we claim that the 3.43, 5.19, 7.95, 12.0, 17.4, and 366  d periodicities are present in the signal \ch{and there is a candidate at 1.84/2.17 d. }
        %Note that the analysis of section~\ref{app:cv1} has also been made in a case where 23 d is replaced by 11.6 d in the array of noise models considered~\ref{tab:cv}, and the conclusions are identical. 
        
        As a remark, the noise model selection procedure is close to~\cite{jones2017}, which ranks noise models with cross validation, BIC, and AIC. The difference lies in the fact that instead of comparing noise models with free parameters, we compare models that are couples of the noise model with fixed parameters and planets with fixed periods.

        \begin{table}
                \caption{Value of the parameters used to define the grid of models tested. }
                \label{tab:cv}
                \centering
                \begin{tabular}{p{1.5cm}|p{3.1cm}|p{1.cm}|p{0.9cm}}
                        Parameter  & Values & Highest CV score & Highest evidence\\   \hline  \hline
                        $\sigma_W$ (m/s) & 0, 1 ,1.5, 2 ,2.5, 3 & 1  & 1  \\
                        $\sigma_{C}$ (m/s) & 0.75  & 0.75& 0.75 \\
                        $\sigma_{R}$ (m/s) & 0, 1 ,1.5, 2 ,2.5, 3, 3.5  & 2.5 & 2.5 \\ 
                        $\tau_R$ (d)&0,1,6  & 1  & 6\\
                        $\sigma_{QP}$ (m/s)& 1, 1.5, 2, ,2.5, 3  & 2   &3 \\ 
                        $\tau_{QP}$ (d)&0, 10, 20, 60 & 10 d& 60 \\
                        $P_\mathrm{act}$ (d) & 11.6, 17.4, 23.3, 34.0 &  11.6 & 34 \\
                \end{tabular}
        \end{table}
        
        \begin{table}   \caption{Periods appearing in the $\ell_1$ periodogram of the model with the highest CV score and their false alarm probabilities. The third columns show the percentage of models in the 20\% best CV score ($CV_{20}$ noise models) where the periodicity has a FAP<0.05, and the fourth column shows the median FAP of these periodocities in the $CV_{20}$ models.}
                \centering
                \label{tab:cvresults}
                \begin{tabular}{p{1cm}|p{1.5cm}|p{1.5cm}|p{1.5cm}} Period (d) & FAP (best fit) & Inclusion in the model & $\mathrm{CV}_{20}$ median FAP \\ \hline \hline 1.839& $1.00$ & 0.0\% &  - \\ 2.178& $1.00$ & 0.0\% &  - \\ 3.432& $1.97\cdot10^{-5}$ & 100.0\% & $3.03\cdot10^{-4}$\\ 5.197& $8.02\cdot10^{-3}$ & 100.0\% & $1.03\cdot10^{-3}$\\ 7.953& $1.83\cdot10^{-3}$ & 100.0\% & $1.63\cdot10^{-3}$\\ 12.03& $2.76\cdot10^{-3}$ & 100.0\% & $2.55\cdot10^{-4}$\\ 17.4& $2.26\cdot10^{-2}$ & 100.0\% & $4.75\cdot10^{-3}$\\ 17.74& $1.00$ & 0.0\% &  - \\ 34.59& $1.00$ & 0.0\% &  - \\ 365.7& $1.11\cdot10^{-6}$ & 100.0\% & $2.17\cdot10^{-7}$\\ 640.0& $1.35\cdot10^{-2}$ & 89.79\% & $1.39\cdot10^{-3}$\\ 1920.& $2.12\cdot10^{-2}$ & 100.0\% & $8.15\cdot10^{-4}$\\ \end{tabular}\end{table}

        \begin{table} \caption{Periods appearing in the $\ell_1$ periodogram of the model with the highest approximated evidence and their false alarm probabilities. The third columns show the percentage of models in the 20\% best evidence ($E_{20}$ noise models) where the periodicity has a FAP<0.05, and the fourth column shows the median FAP of these periodicities in the $E_{20}$ models. } 
                \label{tab:evresults}
                \centering\begin{tabular}{p{1cm}|p{1.5cm}|p{1.5cm}|p{1.5cm}} Period (d) & FAP (best fit) & Inclusion in the model & $\mathrm{E}_{20}$ median FAP \\ \hline \hline 1.839& $1.00$ & 0.714\% &  - \\ 2.178& $1.83\cdot10^{-3}$ & 9.486\% &  - \\ 3.432& $5.78\cdot10^{-12}$ & 100.0\% & $1.98\cdot10^{-8}$\\ 5.197& $7.49\cdot10^{-8}$ & 100.0\% & $1.23\cdot10^{-5}$\\ 7.953& $8.67\cdot10^{-7}$ & 100.0\% & $3.15\cdot10^{-5}$\\ 12.03& $1.38\cdot10^{-4}$ & 90.38\% & $4.35\cdot10^{-3}$\\ 17.4& $1.00$ & 38.07\% &  - \\ 17.74& $4.81\cdot10^{-3}$ & 8.771\% &  - \\ 34.59& $1.00$ & 0.0\% &  - \\ 365.7& $1.00$ & 56.07\% &  - \\ 640.0& $1.00$ & 8.966\% &  - \\ 1920.& $1.00$ & 39.44\% &  - \\ \end{tabular}\end{table}

        \section{Periodogram analysis }
        \label{app:periodogram}
        
        \subsection{Iterative search}
        
        In this section, we perform the period search in an iterative way. We follow a standard  planet search algorithm. We first compute at which frequencies the two maximum peaks of the spectral windows are attained (besides the peak in $\omega = 0$) and denote them by $\omega_{S_1}$ and $\omega_{S_2}$. Here, $\omega_{S_1} = 1.0027$ and $\omega_{S_2} = 0.999$ cycles/day.
        
         The planet count is initialized at 0. 
        We compute the generalized Lomb-Scargle periodogram, and we search at which frequency its maximum is attained $\omega_0$. We then fit a Keplerian signal initialized at $\omega_0$, but depending on the value of $\omega_0$, we might choose to fit $\omega_0- \omega_{S_i}$ instead, where $i=1,2$. We then add one to the planet count and compute the periodogram of the residuals. The frequency of the maximum peak is denoted by $\omega_1$; we then fit two Keplerian functions initialized at $\omega_0$ and $\omega_1$, compute the periodogram of the residuals, and so on until a non significant signal is found.
        
\begin{figure} \centering
                \includegraphics[width=9.2cm]{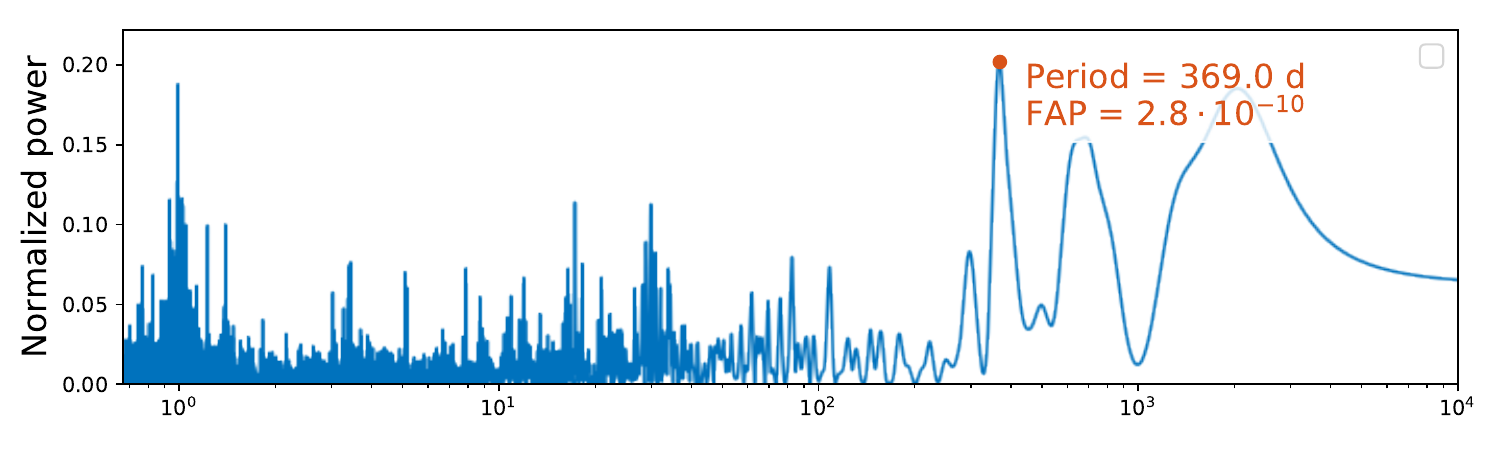}
                \includegraphics[width=9.2cm]{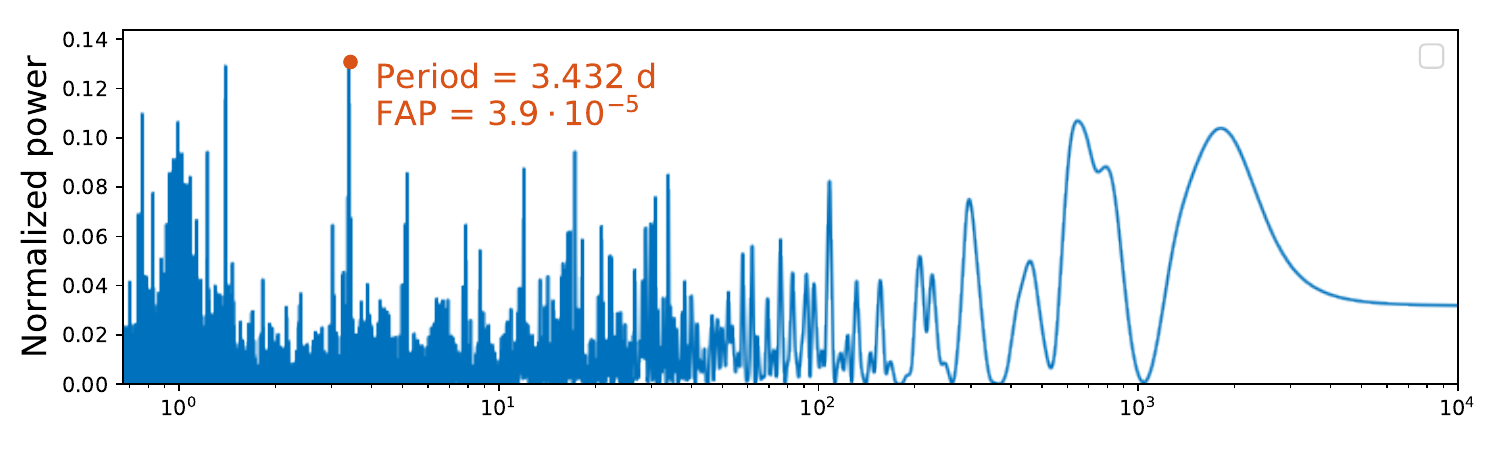}
                \includegraphics[width=9.2cm]{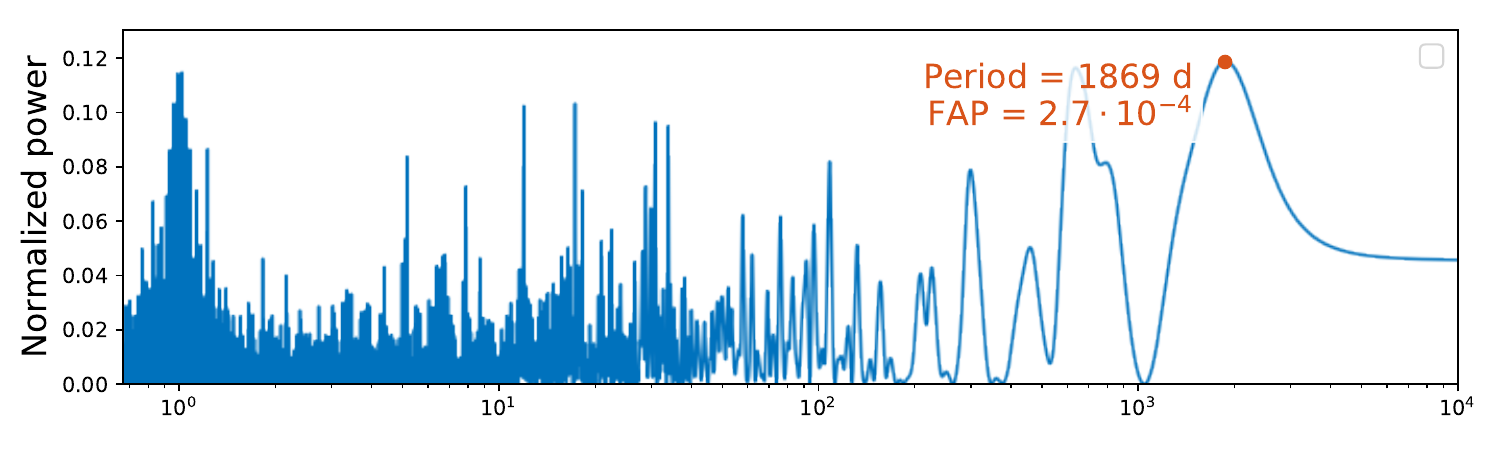}
                \includegraphics[width=9.2cm]{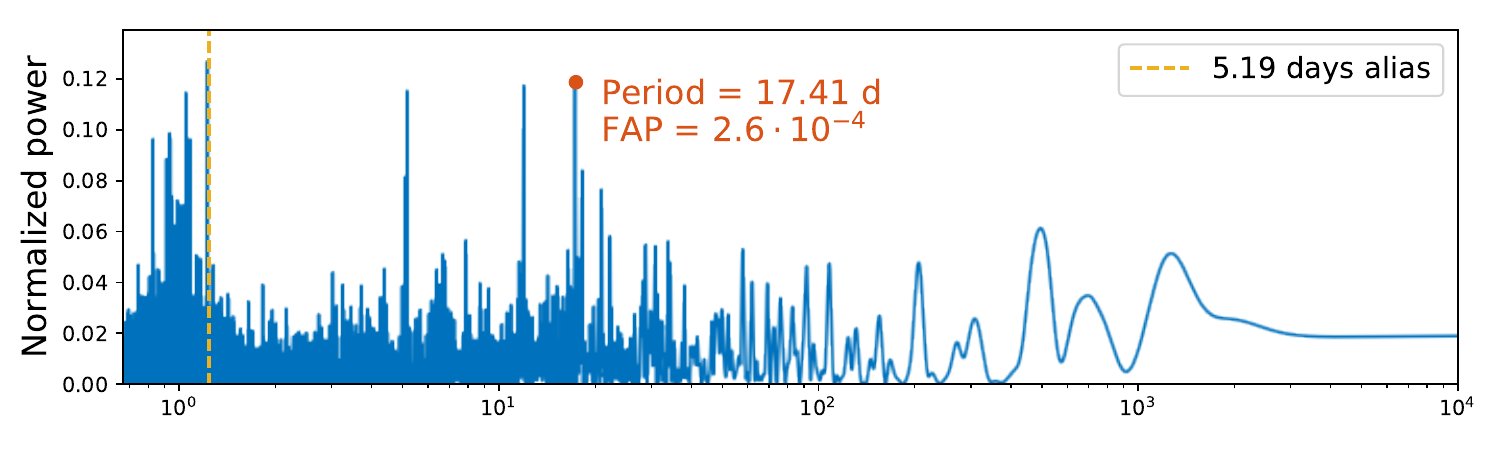}
                \includegraphics[width=9.2cm]{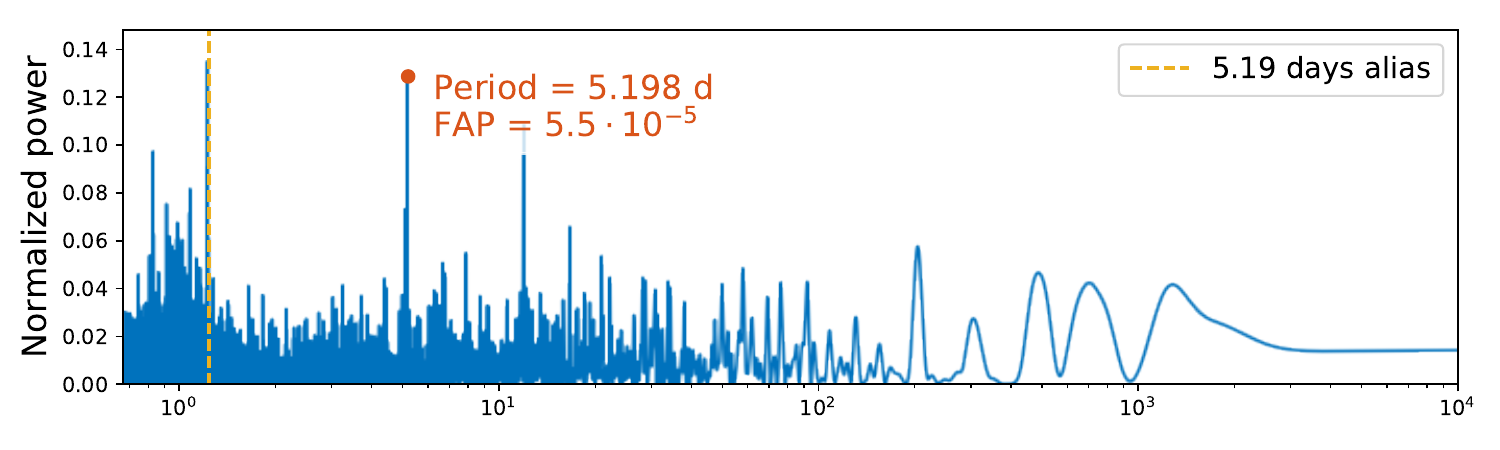}
                \includegraphics[width=9.2cm]{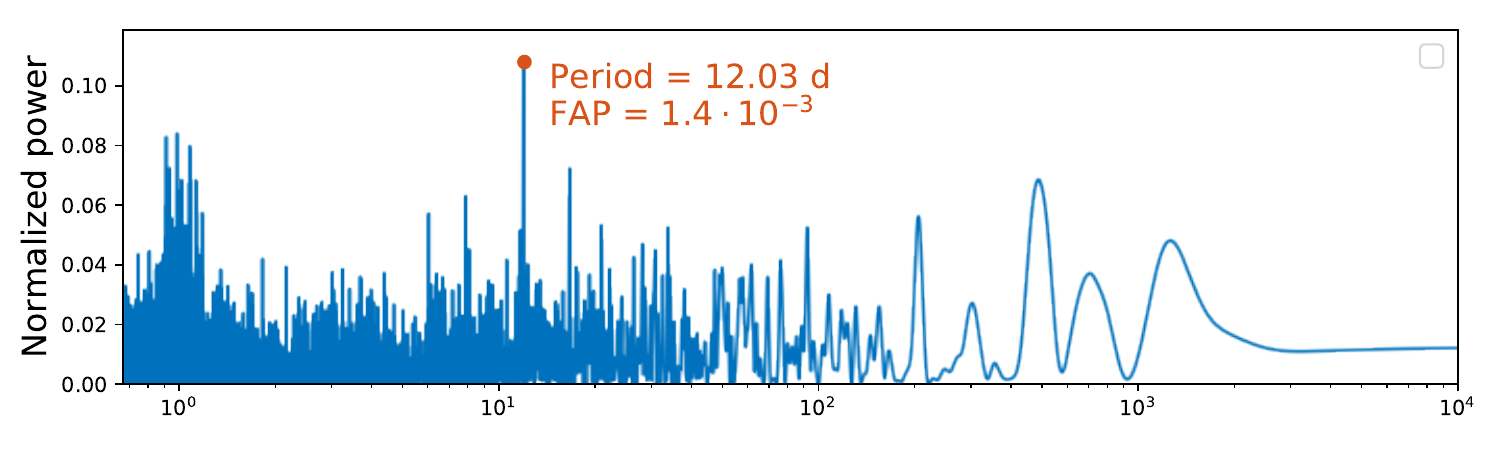}
                \includegraphics[width=9.2cm]{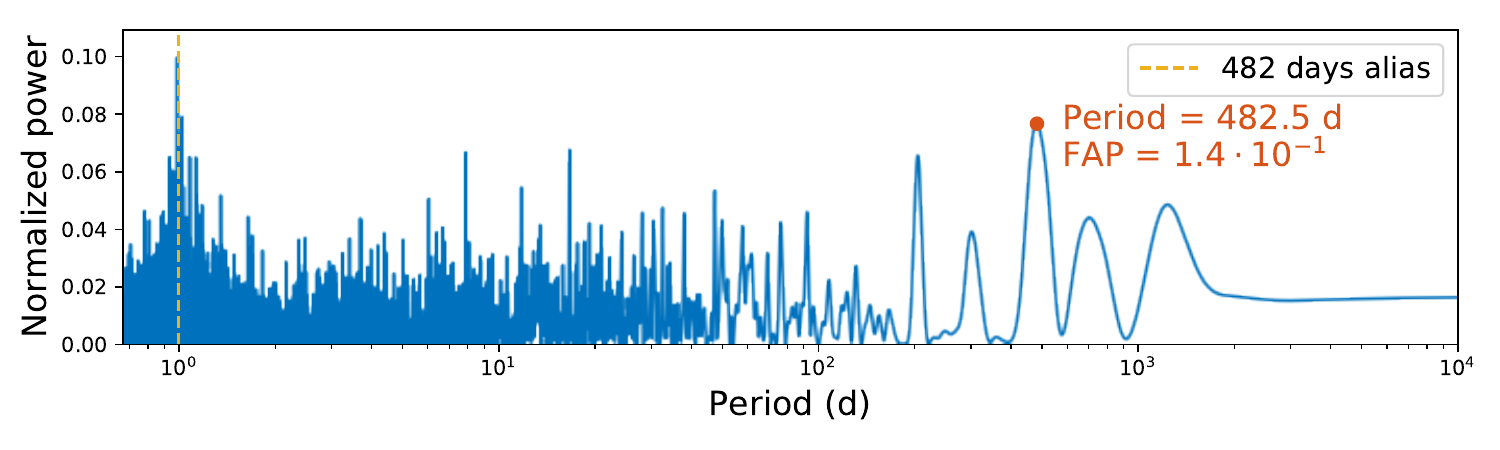} 
                \caption{Subsequent periodograms after a fit of a Keplerian orbit model initialized at the maximum peak of the periodogram }
                \label{fig:periodograms}
        \end{figure}

        In Fig.~\ref{fig:periodograms} we represent the subsequent periodograms, on a period range spanning from 0 to 1.5 cycles/day, which were computed iteratively. The periods selected at each iteration are marked in red, and their false alarm probability corresponding to their peaks is given. 
        It might happen that the maximum of the periodogram occurs at a period near 1 day, in which case the corresponding period is highlighted by a vertical yellow line. Since in each occurrence of this situation, there are peaks at a longer period which are aliases of these ones, the peaks at longer periods are preferred. 
        
        Overall, the periodogram results are similar to those of the $\ell_1$ periodogram, the most striking difference being that  in the classical approach, the 7.95 d-signal does not appear, while the 17.4 d-signal has a stronger significance. \ch{Furthermore, the 1.84/2.17 candidate does not appear. } In the classical approach, the 7.95 d-signal is absorbed in the Keplerian fit initialized at 17.4 d. When performing the iterative search with sinusoids only, the 7.95 d-signal appears at the fifth iteration.  The decreased significance of the 17.4 d-period in the $\ell_1$ periodogram compared to the classical periodogram stems from accounting for correlated noises in the $\ell_1$ periodogram. As discussed in Section~\ref{app:cv_discuss}, the signals with a period greater than the time-scale of the noise see their significance decrease with respect to a white noise model.

        \subsection{Aliases}
        \label{sec:aliases}
        
        In the previous section, we have seen that at the fourth iteration, the highest peak  in the periodogram occurs at 1.238 d, which, given the spectral window, is an alias of 5.198 d. Furthermore, when performing an iterative search with a circular model between 0 and 1.5 cycles/day, the subsequent maximum periodogram peaks include aliases of 3.4, 7.95,  17.4, and 2000 d. This raises the question of whether the periodicities claimed here might in fact  originate from shorter periods.

        It seems improbable that any of the planets would be at the aliases close to one day. Indeed, planet couples with period ratios near 1.52 are found to be common~\citep{lissauer_architecture_2011,fabrycky_architecture_2014,steffen2015}. If the chain found was spurious, the aliases would have to fall by chance such that a 3:2 chain would be compatible with the data. This cannot be completely excluded, but it seems improbable.

        % However, in the circular, iterative approach, after the six signals considered here, there   In the $\ell_1$ periodogram  

        %For a given parametrization of the noise, we compute the periodogram as defined in~\cite{baluev2008, delisle2019a}. This one is defined at a frequency $\nu$ as  
        %\begin{align}
        %P_V(\nu) =  \frac{\chi_\mathcal{H}^2 - \chi_\mathcal{K}^2(\nu)}{\chi_\mathcal{H}^2}. 
        %\label{eq:rgls}
        %\end{align}

        %where $\chi_\mathcal{H}$ and $\chi_\mathcal{K}$ are the $\chi^2$ of the residuals of a least square fit, respectively of a linear base model $\mathcal{H}$ and a linear base model plus a sinusoidal component at frequency $\nu$, $\mathcal{K}$. 
        
        %The period search has here been restricted to a grid between 0 and 0.66 cycles/day, to exclude the alias of the 3.43 day signal, which is the shortest apparent period (3.43 days). A similar analysis on a larger period range yields the detection of signals that are aliases of the 3.43, 5.19, 7.95, 12.0 periodicities close to one day. 
        
        %It seems however unlikely that the planet periods would be in the one day region, while their aliases are spaced 
        
        %\section{MCMC, supplementary information}
        
        \section{Phase and amplitude consistency}
        \label{app:phasecons}
        
        A property of planetary signals is that, aside from signatures of gravitational interaction between planets, their amplitude and phase do not vary with time. 
         To check the consistency in phase and the amplitude of the signals found, we performed two tests.

\begin{figure} \centering
                \includegraphics[width=8.3cm]{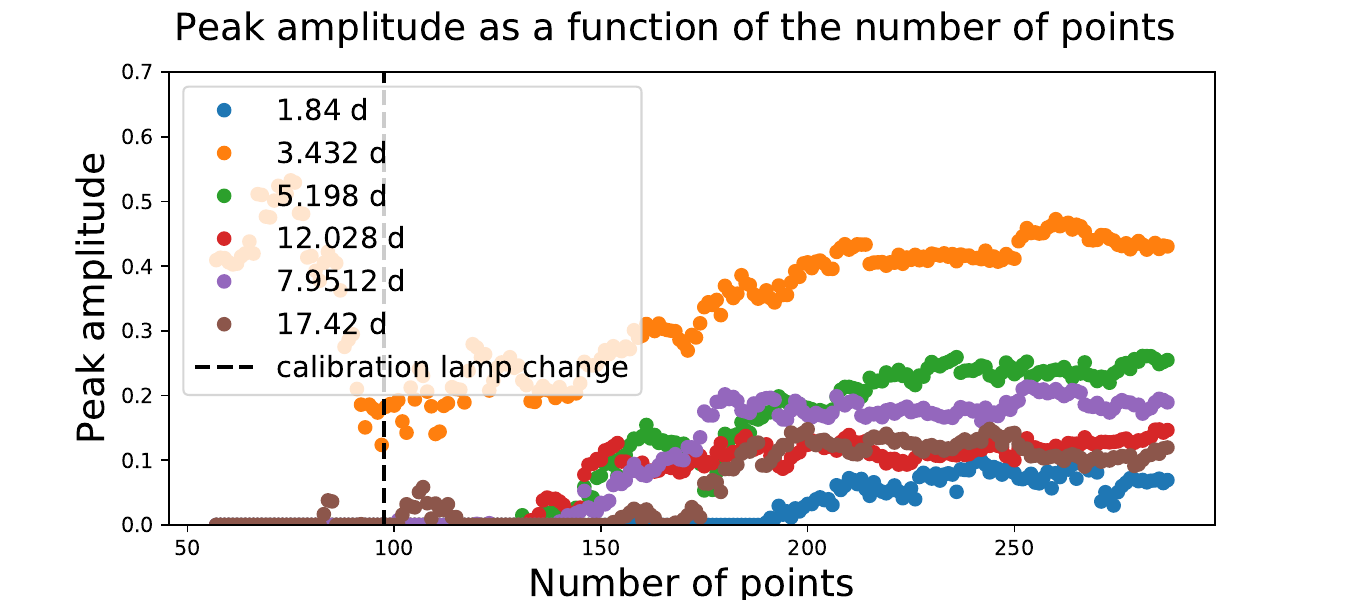}
                \includegraphics[width=8.3 cm]{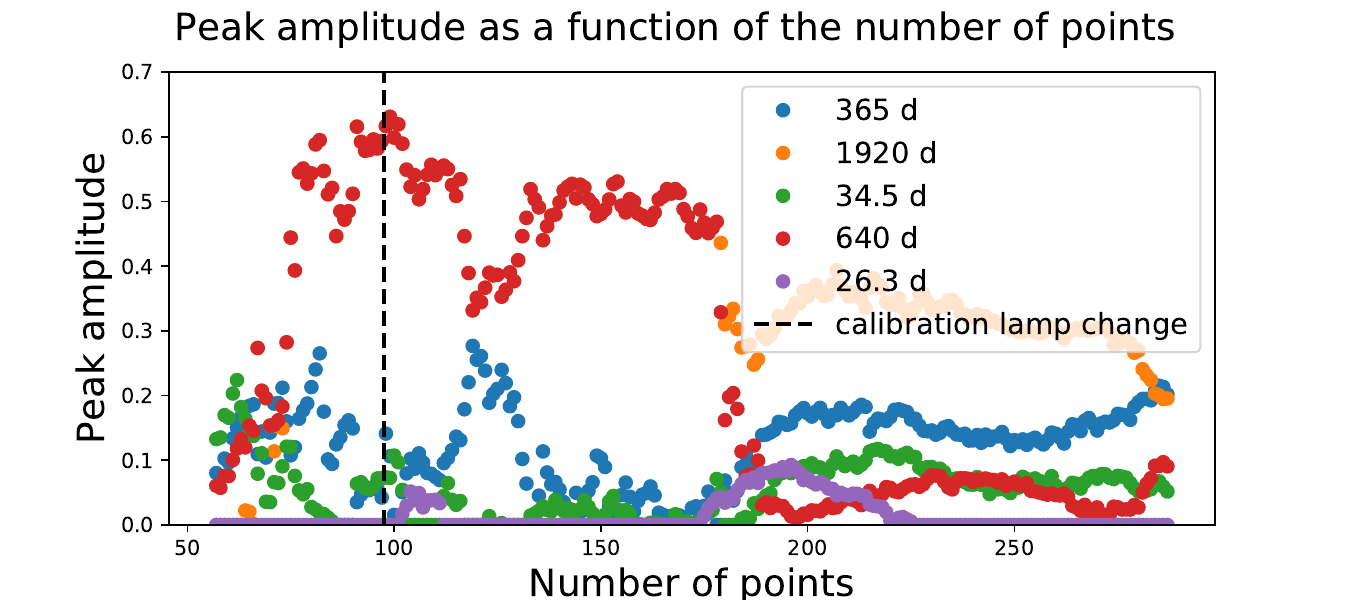}
                \caption{Evolution of the $\ell_1$ periodogram amplitudes of peaks at different periods. Top: Planets and planet candidates. Bottom: Other signals. The black dashed line indicates the transition from calibration with thorium-argon to Fabry-Perot. }
                \label{fig:peakevol}
        \end{figure}

                First, we computed the $\ell_1$ periodogram of the signal starting with the first 60 points and adding 1 point at a time up to 287, with the noise model corresponding to the best cross validation score, similarly to the procedures suggested in~\cite{schuster1898} and~\cite{mortier2017}. 
                From $\approx 180$ points, the signals appearing in the $\ell_1$ periodogram are identical to those from Table.~\ref{tab:fapsbody}. This result is also consistent when it is performed backwards, starting with the last 60 points. 
                In Fig.~\ref{fig:peakevol}, we represent the evolution of the peak amplitudes at several periods as a function of the number of points $n$ in the vicinity of periods of interest. By vicinity, we mean that the value plotted is the maximum of the peak occurring within $1/T_{\mathrm{obs}(n)}$ in frequency of the period of interest; $T_{\mathrm{obs}(n)}$ being the total observation time of the data set including $n$ points. It appears that the amplitude of the signals reported as planets and planetary candidates increases steadily (top panel), while other signals show more variability (bottom panel).  We note that before 180 points, the frequency resolution is insufficient in distinguishing signals at 640 and 1920 d. 
        
        It appears that most of the candidates appear from $\approx 130 $ points. This could be due to a phenomenon analogous to  phase transitions observed in sparse recovery with random matrices, where the number of signals appropriately recovered changes sharply in the vicinity of a critical number of observations~\citep{amelunxen2013}. 
                However, the $\ell_1$-periodogram of the first, middle, and last 150  points show important differences. The first 150 points fail to unveil most of the signals that are later confirmed, while they are seen in the $\ell_1$-periodogram of the middle and last 150 points.  %Incidentally, the change of the calibration lamp from thorium-argon to Fabry-Perot occurs at  BJD 2458181, after the 97\textsuperscript{th} measurement. This suggests that the improvement in calibration had a rôle in unveiling the system.

\begin{figure} \centering
                \includegraphics[width=4.3cm]{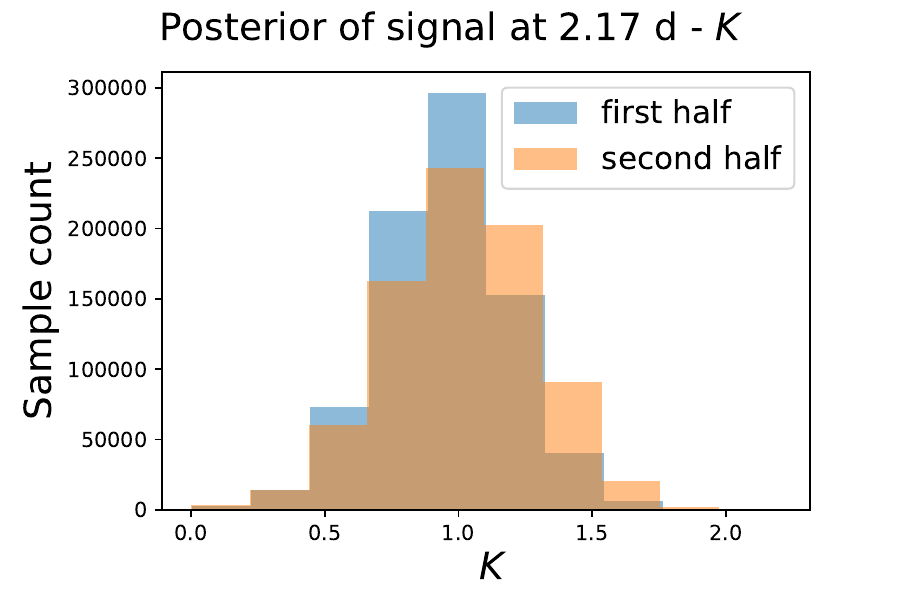}
                \includegraphics[width=4.3 cm]{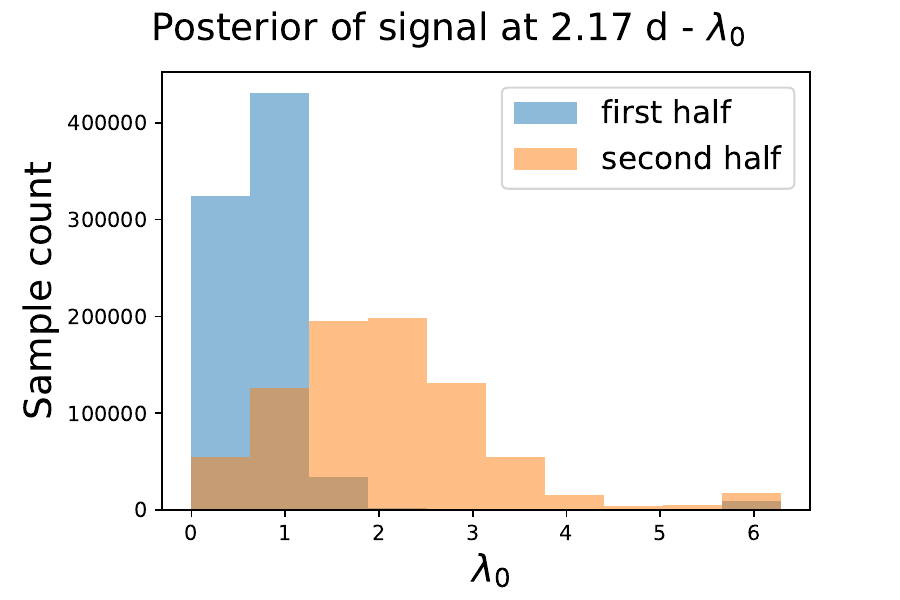}
                
                \includegraphics[width=4.3cm]{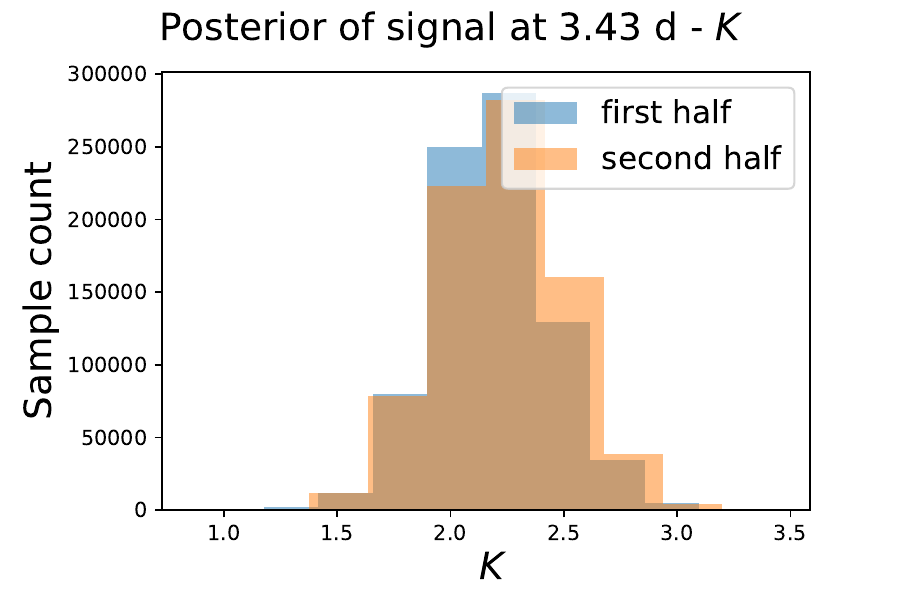}
                \includegraphics[width=4.3 cm]{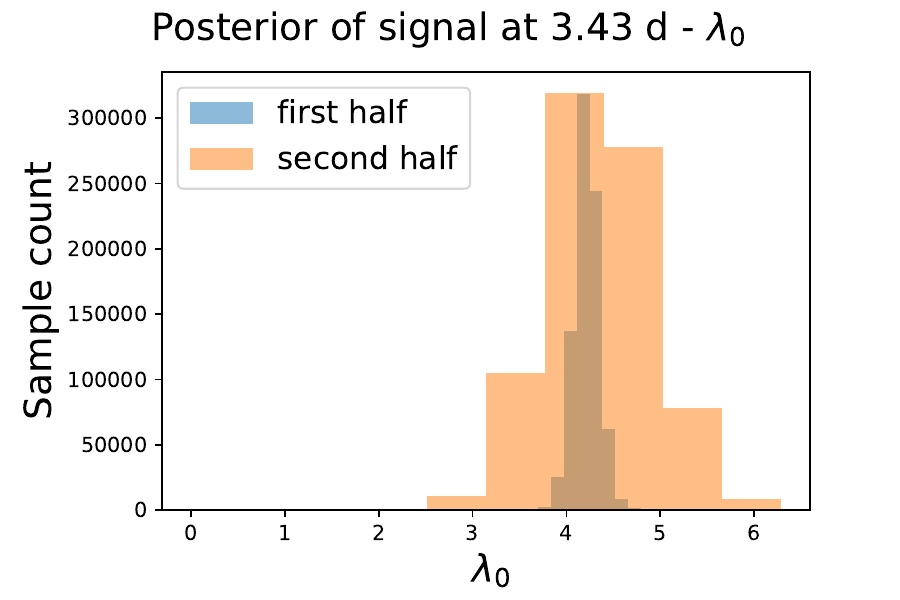}
                
                \includegraphics[width=4.3cm]{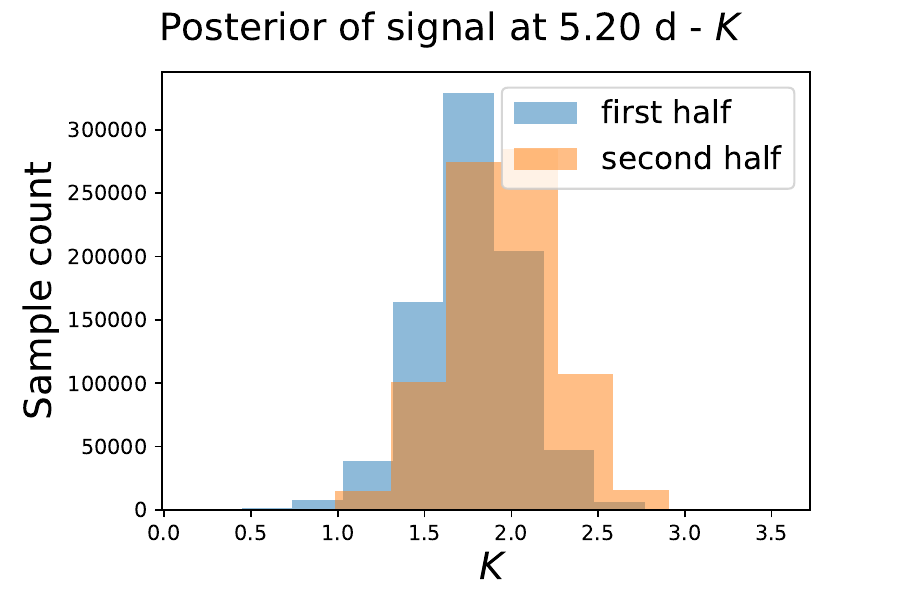}
                \includegraphics[width=4.3 cm]{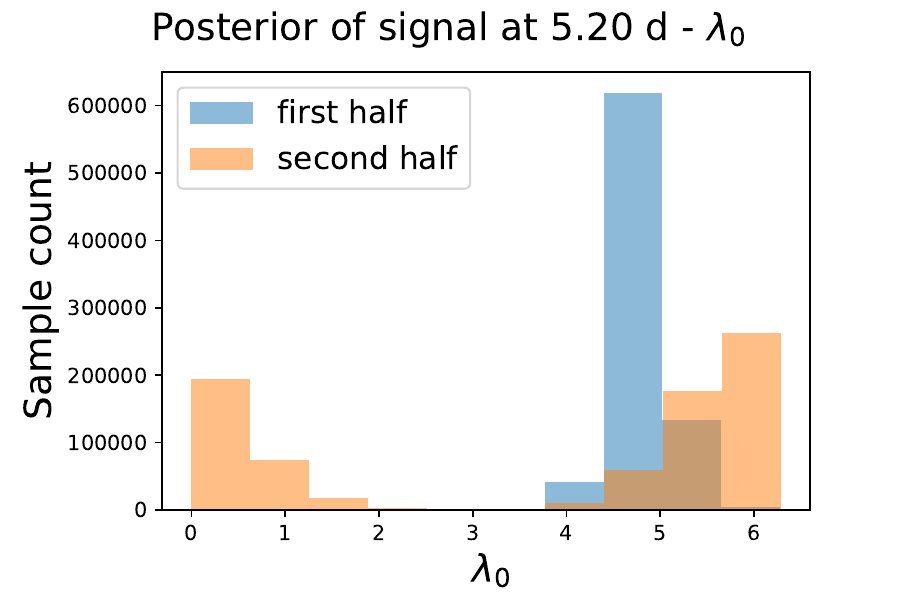}
                
                \includegraphics[width=4.3cm]{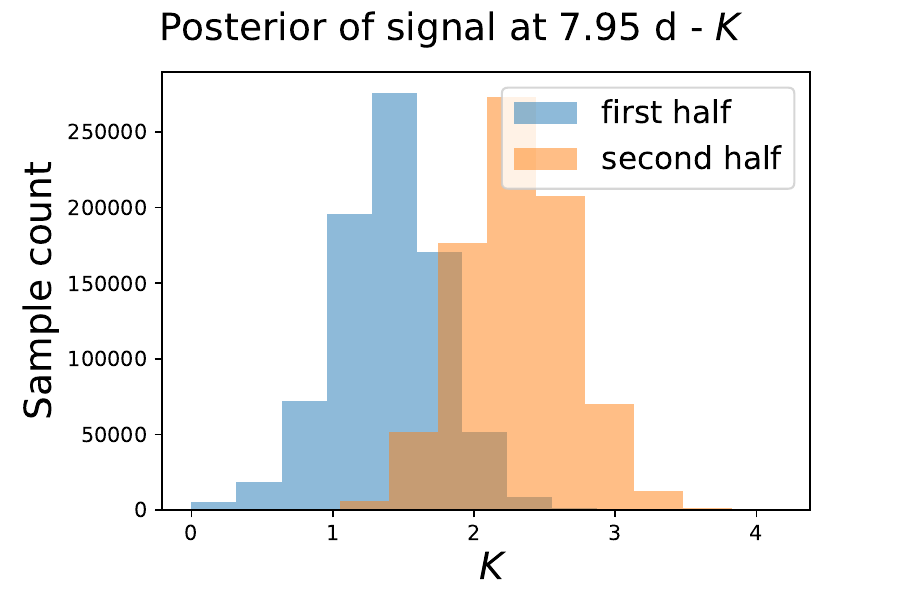}
                \includegraphics[width=4.3 cm]{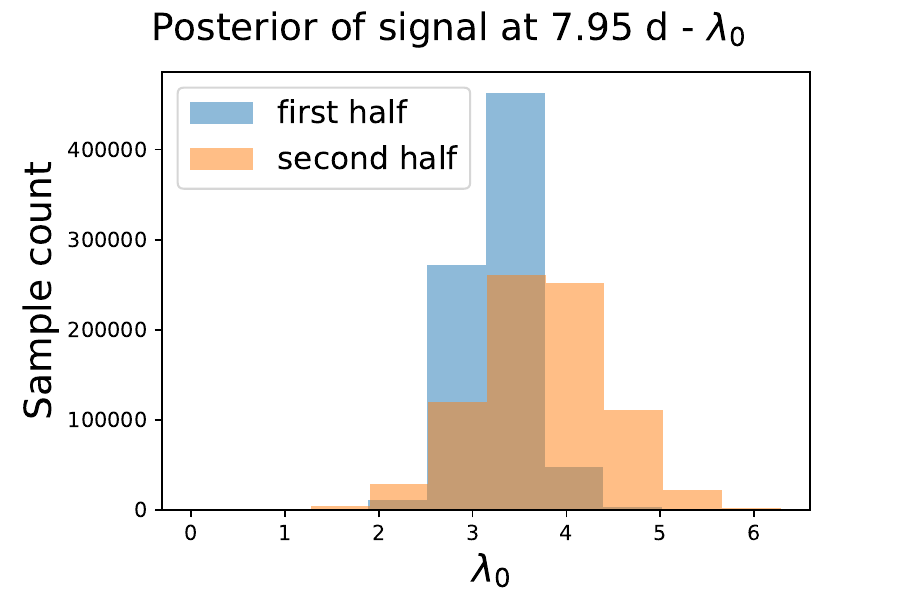}
                
                \includegraphics[width=4.3cm]{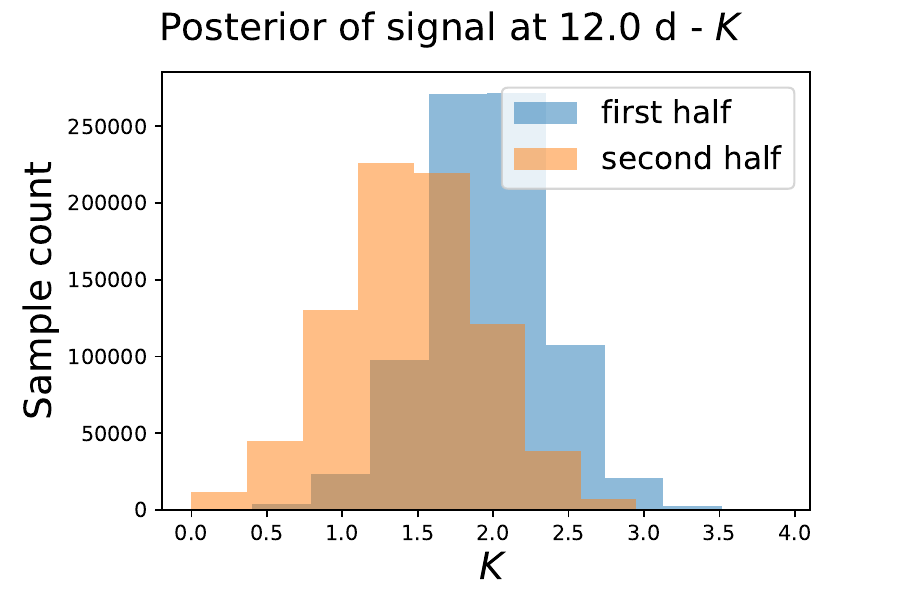}
                \includegraphics[width=4.3 cm]{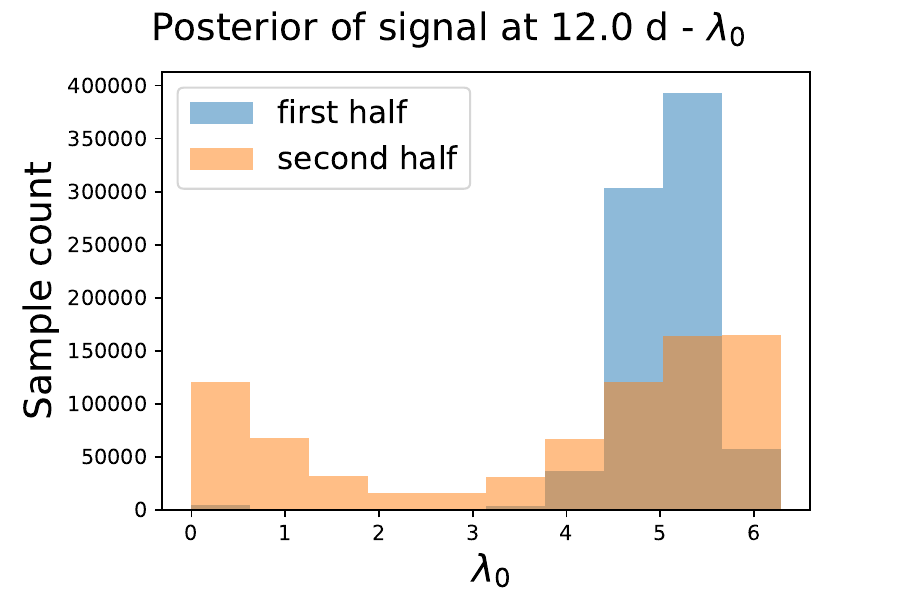}
                
                \includegraphics[width=4.3cm]{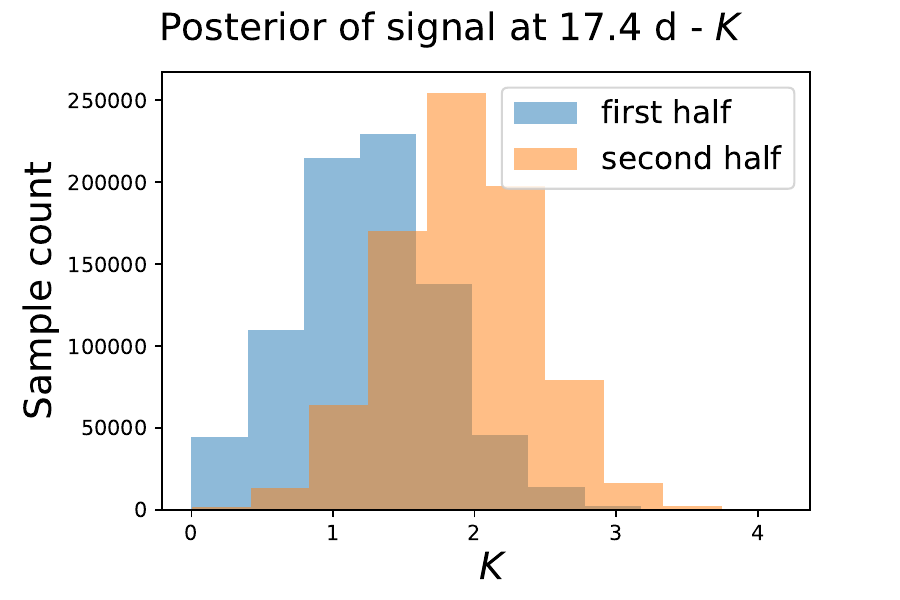}
                \includegraphics[width=4.3 cm]{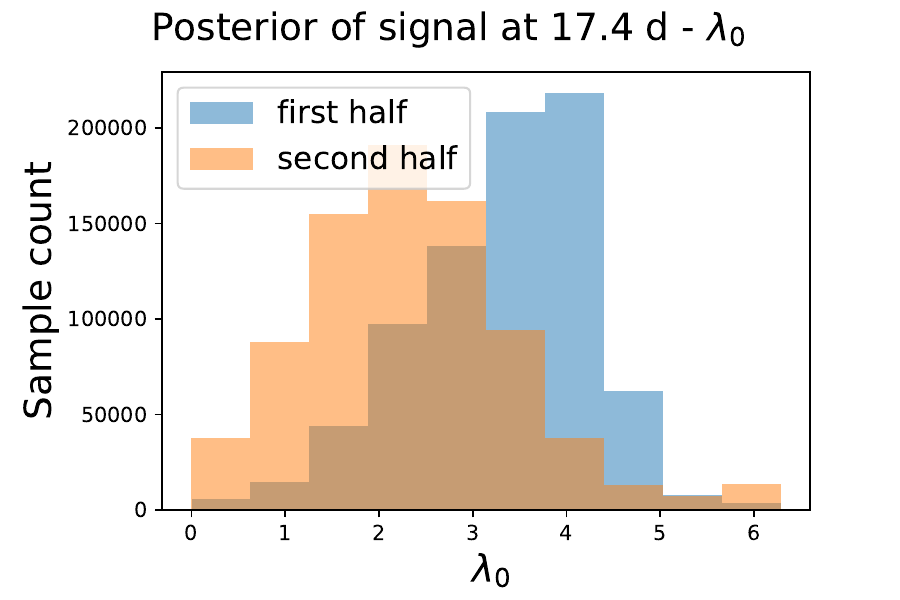}
                
                \includegraphics[width=4.3cm]{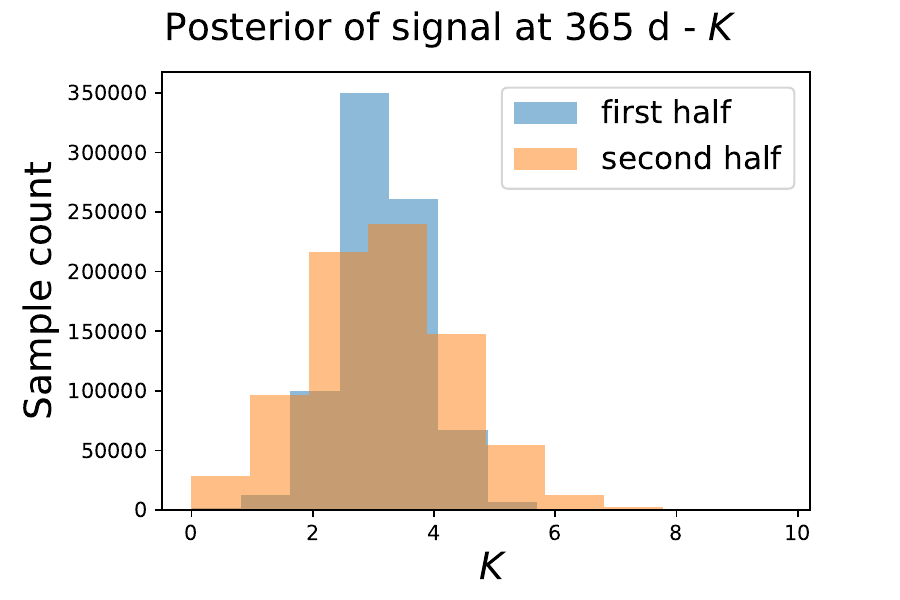}
                \includegraphics[width=4.3 cm]{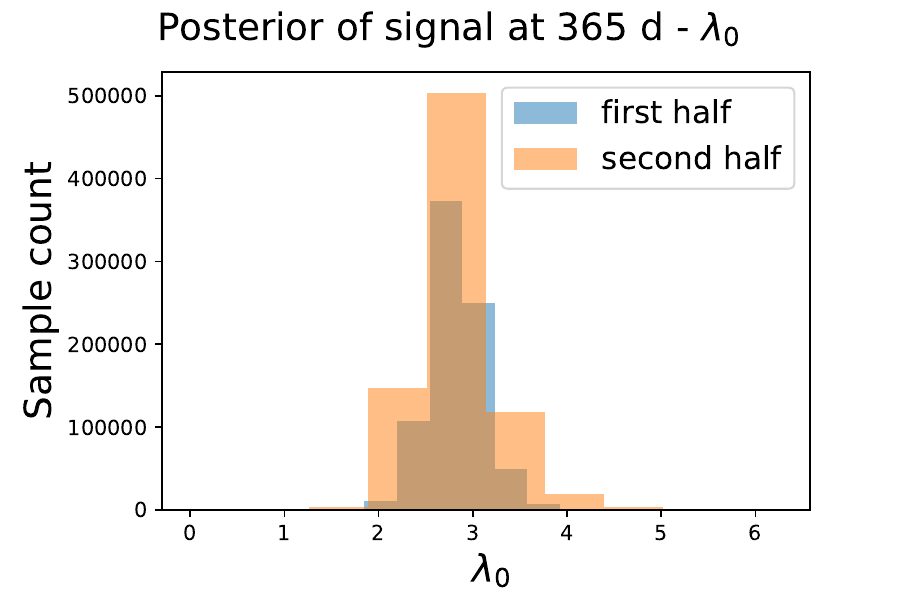}
                \caption{\ch{Amplitude ($K$) and  phase ($\lambda_0$) posteriors computed on the first and second half of the data for the signals.} }
                \label{fig:phaseamp}
        \end{figure}

                We now check the consistency in phase of the signals. We consider the points up to BJD 2457529.4867 (excluded) and after this date, so that we have two data sets of 144 and 143 points. The model described in Section~\ref{sec:orbitalelts} was fitted onto each half of the data. The likelihood and priors are identical, except that tight priors were set on the periods of the signals, that is, we set a Gaussian prior  as given in Table~\ref{tab:mcmcresults} with a standard deviation $ 1/T_{\mathrm{obs}}$ in frequency, where $T_{\mathrm{obs}}$ is the total observation time. %Indeed, the proper detection of all the signals necessitate 180 points, and the fit does not converge properly if priors on the periods are not set, 
                The posteriors of the first and second half of the data are represented in blue and red, respectively, in Fig.~\ref{fig:phaseamp}.The 1 sigma intervals of semi amplitude and longitude at reference time overlap in all cases, including the yearly signal. 
        
\begin{figure} \centering
                \includegraphics[width=4.3 cm]{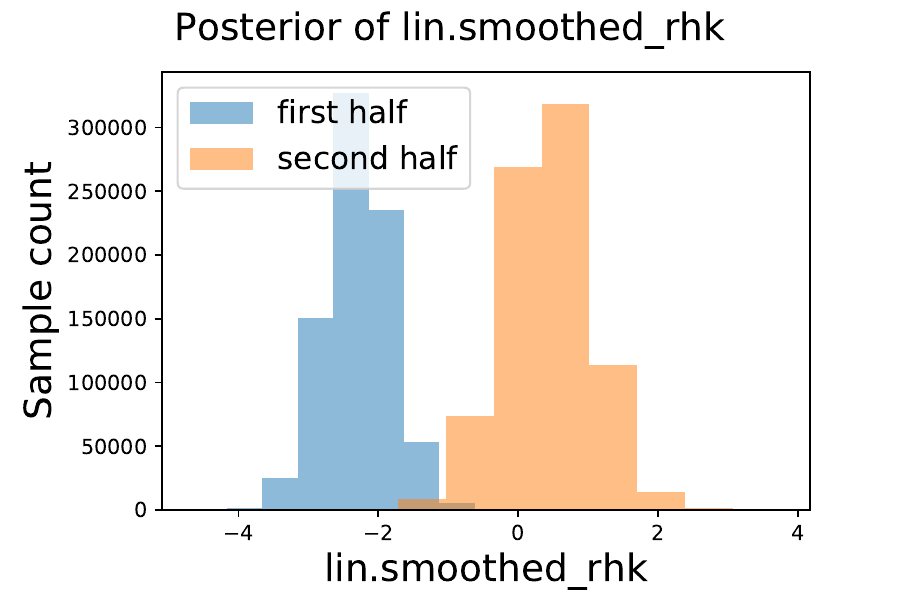}
                \includegraphics[width=4.3 cm]{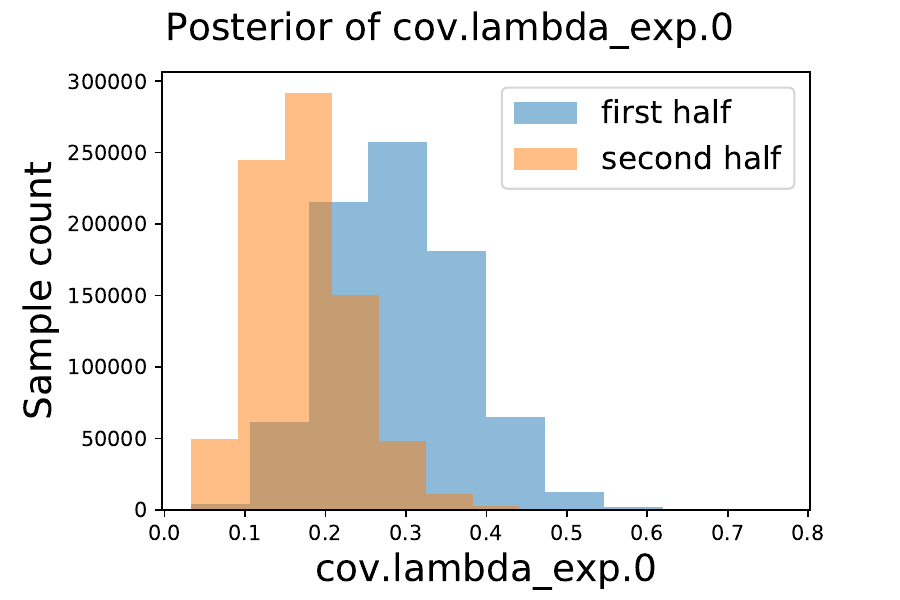}
                \caption{\ch{Amplitude ($K$) and  phase ($\lambda_0$) posteriors computed on the first and second half of the data for the signals.} }
                \label{fig:noise_halfhalf}
        \end{figure}

                The coefficient of correlation with the smoothed $\log R'_{HK}$ flips sign between the first and second part of the fit, and the noise seems to exhibit slightly different time scales. In Fig.~\ref{fig:noise_halfhalf}, we represent the distribution of the coefficient of the  smoothed $\log R'_{HK}$  and  inverse time-scale of the noise for the first and second halves of the data (left and right, respectively). Together with the test performed on the peak amplitudes, this suggests an evolution of the noise properties on the time-scale of the data, which is potentially due to instrumental and/or stellar features. 
        
        \section{Model parameters}
        \label{app:mcmc}
        
        \subsection{MCMC}
        
        To compute the uncertainties on the orbital elements, we performed a Monte Carlo Markov chain analysis. In this appendix, we define  the likelihood, priors, and convergence tests used. 
        
        We assume a Gaussian likelihood. We denote the time series of radial velocities  with $y$, the density of $y$ knowing the parameters is of the form
        \begin{align}
        p(y|\theta, \eta ) = \frac{1}{\sqrt{2\pi}^N | V (\eta)|} \mathrm{e}^{  \frac{1}{2} (y-f(\theta)) ^T V(\eta)^{-1}  (y-f(\theta)) }.
        \label{eq:likelihood}
        \end{align}
        Our signal model $f(\theta)$ includes Kelplerian models initialized at \ch{2.177}, 3.432, 5.198, 7.951, 12.03, 17.39, and 361 d, and a linear part, with an offset and the smoothed $\log R'_{HK}$ as defined in Appendix~\ref{app:cv2}. This one is centered and normalized by its standard deviation, so that its amplitude can be interpreted as a velocity.
        
        The noise model includes a free white noise jitter and an exponential decay term, so that the noise model is 
        \begin{align}
        V_{kl}(\eta ) & =  \delta_{k,l} (\sigma_{k}^2 + \sigma_{W}^2) + \sigma_{C}^2 c(k,l) +   \sigma_{R}^2 \e^{-\frac{|t_k-t_l|}{\tau_R}} 
        \label{eq:kernelmcmc}
        \end{align}
        where $\eta$ = ($\sigma_W$, $\sigma_{R}$, $\tau_R$) are free parameters and $\sigma_C$ is fixed to 1 m/s. In total, we have \ch{40} parameters; $\theta$ and $\eta$  have \ch{37} and 3 components, respectively.
        
        The prior distributions on the parameters are defined in Table~\ref{tab:mcmcparameters}.  For the eccentricity, we first ran the simulation with a looser prior (beta distribution with $\alpha=1$ and $\beta=4$).  Each MCMC sample was taken as an initial condition for the system. Its evolution was integrated 1 kyr in the future using the 15-th order N-body integrator IAS15~\citep{Rein2015}, from the package REBOUND~\citep{Rein2012}. General relativity was included via REBOUNDx, using the model of~\citet{Anderson1975}. We considered the system to be unstable if any two planets have an encounter below their mutual Hill radius. From the 36782 original samples, around 1700 passed the stability test. 
                
        To increase the number of effective samples and obtain reliable 3$\sigma$ intervals, we reran the MCMC. We redefined the prior on eccentricity for each planet $i$ with a beta distribution such that $\alpha_i=1$ and $\beta_i$ is such that the variance of the prior is equal to the variance of the empirical distribution of the eccentricities of the points that survived the integration. We note that  $\lambda_0$ is the mean longitude at the reference epoch 57500. For planet $b$, the prior $T_c$ was set in accordance with TESS data.

        The convergence was checked by computing the number of effective samples in each parameter chain  as in~\cite{delisle2018}. We find that each chain has at least 18,000 effective samples, which indicates convergence of the chain.
        To compute the uncertainties on values of $m sin i $, we  took the uncertainties on the stellar mass into
account and generated independent samples of Gaussian distributions with a mean and standard deviation of 1.08 and 0.1 $M_\odot $, which is in accordance with~\cite{chandler2016}.

        \begin{table}
                \centering
                \caption{Variables used for the computation of the MCMC analysis and their prior distributions. }
                \label{tab:mcmcparameters}
                \begin{tabular}{p{2cm}|p{6cm}}    Parameter & prior \\ \hline \hline                   
                        offset 1 & Uniform on [-200 , 200 ] km.s\textsuperscript{-1}  \\ 
                        smoothed rhk & Gaussian with mean 0 and $\sigma = 4$ m.s\textsuperscript{-1}  \\ 
                        \hline
                        $P$ & Uniform (no specified bounds)\\
                        $K$   & Uniform (no specified bounds) \\
                        $\lambda_0$ at BJD 2457500 & Uniform on [0, $2\pi$] (for $c,d,e,f,g$)\\ 
                        $T_c (b)$  & Gaussian of mean 2458766.049072 and $\sigma=$ 0.003708\\             
                        $\sqrt{e} \cos \omega$, $\sqrt{e} \sin \omega$&  beta prior on $e$ with $\alpha=1$ and $\beta_b = 11.6, \beta_c = 24.2, \beta_d = 17.9, \beta_e = 21.2, \beta_f = 21.4,  \beta_f = 20.9$, uniform prior on $\omega$ on [0, $2\pi$]
                        \\ \hline 
                        $\sigma_{W}^2$  & Truncated Gaussian with $\sigma = 4$ m\textsuperscript{-2}.s\textsuperscript{-2} \\ 
                        $\sigma_{R}^2$  & Truncated Gaussian with $\sigma = 4$ m\textsuperscript{-2}.s\textsuperscript{-2} \\     
                        $1/\tau$ & $\log$ uniform on [1/24, 30] d\textsuperscript{-1}  \\     
                \end{tabular}
        \end{table}

        \begin{table*}
                \caption{Estimates and credible intervals of the orbital parameters for circular orbital models and a noise model including a jitter and a red noise model with exponential decay.  We give three point estimates: the maximum likelihood, the posterior mean, and median. The credible intervals are given as the shortest intervals containing $x$\% of the sample with $x = 68.27$, $95.45,$ and $99.73$\%. }
                \label{tab:mcmcresults}
                
                \begin{tabular}{p{2cm}|p{1.3cm}|p{1.3cm}|p{1.3cm}|p{3.2cm}|p{3.2cm}|p{3.2cm}}
                        Parameter & ML fit & posterior mean & posterior median & 68.27\% confidence interval & 95.45\% confidence interval & 99.73\% confidence interval\\ \hline \hline
                        \multicolumn{7}{c}{Linear parameters} \\ \hline  
                        offset& 13536.45& 13536.91& 13536.92& [13536.30 , 13537.53]& [13535.63 , 13538.14]& [13534.93 , 13538.81]\\ smoothed rhk& -1.323185& -1.271674& -1.270171& [-1.676 , -0.836]& [-2.135 , -0.430]& [-2.62 , 0.05]\\ \hline  
                        \multicolumn{7}{c}{Noise parameters} \\ \hline  
                        
                        $\sigma_W$ (d)& 0.03232& 0.66044& 0.64209& [0.28 , 0.93]& [0.08 , 1.24]& [0.00 , 1.53]\\ $\sigma_R$ (d)& 11.40892& 3.57813& 3.56055& [3.28 , 3.82]& [3.04 , 4.15]& [2.82 , 4.54]\\ $\tau$ (d)& 4.28426& 5.39636& 5.17960& [3.84 , 6.23]& [3.06 , 8.21]& [2.521 , 11.38]\\ \hline
                        
                        \multicolumn{7}{c}{Planet $b$}\\ \hline$P$ (d)& 2.178& 2.178& 2.178 & [2.17790 , 2.17809]& [2.17780 , 2.17818]& [2.17771 , 2.17830]\\ $K$ (m/s) & 1.36626& 1.05103& 1.04994& [0.86 , 1.23]& [0.68 , 1.43]& [0.49 , 1.63]\\ $\lambda_0$ (rad) & 1.574033& 1.571026& 1.571012& [1.555 , 1.585]& [1.541 , 1.601]& [1.526 , 1.615]\\ $\sqrt{e} \cos \omega$ & 0.45803& 0.09082& 0.08526& [-0.13 , 0.29]& [-0.28 , 0.49]& [-0.43 , 0.62]\\ $\sqrt{e} \sin \omega$ & 0.40728& 0.02293& 0.02150& [-0.16 , 0.22]& [-0.36 , 0.39]& [-0.50 , 0.55]\\$m\sin i$ ($M_\oplus$)& 2.70& 2.22& 2.21& [1.77 , 2.61]& [1.38 , 3.05]& [0.98 , 3.48]\\  
                        Density ($\rho_\oplus$) & 1.33 &  1.11 & 1.09 &  [0.81 , 1.32] &  [0.62 , 1.66] & [0.42 , 2.05] \\ \hline  
                        
                        \multicolumn{7}{c}{Planet $c$}
                        \\ \hline$P$ (d)& 3.4322& 3.4320& 3.432& [3.43184 , 3.43230]& [3.43160 , 3.43253]& [3.4313 , 3.4328]\\ $K$ (m/s) & 2.23060& 2.26100& 2.26101& [2.06 , 2.45]& [1.87 , 2.66]& [1.65 , 2.84]\\ $\lambda_0$ (rad) & 1.31125& 1.36244& 1.36396& [1.18 , 1.54]& [0.99 , 1.73]& [0.78 , 1.92]\\ $\sqrt{e} \cos \omega$ & 0.04091& -0.01385& -0.01286& [-0.14 , 0.11]& [-0.25 , 0.22]& [-0.34 , 0.32]\\ $\sqrt{e} \sin \omega$ & 0.10077& -0.00090& -0.00090& [-0.13 , 0.12]& [-0.24 , 0.24]& [-0.34 , 0.35]\\ $m\sin i$ ($M_\oplus$)& 5.53& 5.60& 5.59& [5.01 , 6.20]& [4.45 , 6.85]& [3.94 , 7.48]\\  \hline

                        \multicolumn{7}{c}{Planet $d$}\\ \hline$P$ (d)& 5.1979112& 5.1980814& 5.1980803& [5.1972 , 5.1989]& [5.1964 , 5.1997]& [5.1955 , 5.2007]\\ $K$ (m/s) & 1.99999& 1.90714& 1.90756& [1.68 , 2.13]& [1.44 , 2.36]& [1.21 , 2.60]\\ $\lambda_0$ (rad) & 4.53784& 4.37415& 4.37424& [4.12 , 4.65]& [3.83 , 4.90]& [3.54 , 5.17]\\ $\sqrt{e} \cos \omega$ & 0.38891& 0.11966& 0.12116& [-0.04 , 0.29]& [-0.17 , 0.41]& [-0.30 , 0.51]\\ $\sqrt{e} \sin \omega$ & -0.08914& -0.03209& -0.03217& [-0.18 , 0.12]& [-0.32 , 0.24]& [-0.42 , 0.36]\\ $m\sin i$ ($M_\oplus$)& 5.62& 5.41& 5.39& [4.70 , 6.15]& [3.99 , 6.89]& [3.35 , 7.76]\\ \hline  
                        
                        \multicolumn{7}{c}{Planet $e$}\\ \hline$P$ (d)& 7.950& 7.951& 7.951& [7.9489 , 7.9532]& [7.9468 , 7.9554]& [7.944 , 7.957]\\ $K$ (m/s) & 2.01711& 1.85725& 1.85832& [1.57 , 2.12]& [1.27 , 2.39]& [1.02 , 2.71]\\ $\lambda_0$ (rad) & 2.09952& 2.05759& 2.05296& [1.75 , 2.35]& [1.46 , 2.67]& [1.13 , 3.00]\\ $\sqrt{e} \cos \omega$ & -0.27297& -0.07014& -0.06581& [-0.20 , 0.09]& [-0.35 , 0.20]& [-0.47 , 0.31]\\ $\sqrt{e} \sin \omega$ & 0.26144& 0.02982& 0.02877& [-0.11 , 0.17]& [-0.24 , 0.30]& [-0.35 , 0.42]\\$m\sin i$ ($M_\oplus$)& 6.55& 6.08& 6.06& [5.05 , 7.02]& [4.07 , 8.01]& [3.14 , 9.17]\\  \hline 
                        
                        \multicolumn{7}{c}{Planet $f$}\\ \hline$P$ (d)& 12.022& 12.028& 12.028& [12.019 , 12.037]& [12.008 , 12.046]& [11.999 , 12.058]\\ $K$ (m/s) & 1.97486& 1.63327& 1.63426& [1.29 , 1.98]& [0.94 , 2.30]& [0.63 , 2.71]\\ $\lambda_0$ (rad) & 3.35529& 3.16016& 3.15748& [2.63 , 3.67]& [2.05 , 4.21]& [1.48 , 4.92]\\ $\sqrt{e} \cos \omega$ & -0.07334& 0.00713& 0.00620& [-0.13 , 0.15]& [-0.27 , 0.28]& [-0.38 , 0.41]\\ $\sqrt{e} \sin \omega$ & 0.19106& 0.00505& 0.00402& [-0.14 , 0.14]& [-0.27 , 0.28]& [-0.40 , 0.40]\\ $m\sin i$ ($M_\oplus$)& 7.43& 6.14& 6.12& [4.77 , 7.45]& [3.46 , 8.84]& [2.206 , 10.38]\\ \hline  
                        
                        \multicolumn{7}{c}{\cht{Strong planet candidate} $g$}\\ \hline$P$ (d)& 17.46& 17.42& 17.42& [17.396 , 17.455]& [17.36 , 17.48]& [17.33 , 17.51]\\ $K$ (m/s) & 1.34260& 1.62597& 1.63026& [1.23 , 2.02]& [0.82 , 2.42]& [0.37 , 2.85]\\ $\lambda_0$ (rad) & 0.61538& 1.64808& 1.56750& [0.79 , 2.21]& [0.11 , 2.99]& [8.61 , 6.22]\\ $\sqrt{e} \cos \omega$ & -0.01637& -0.00227& -0.00157& [-0.15 , 0.14]& [-0.29 , 0.28]& [-0.40 , 0.42]\\ $\sqrt{e} \sin \omega$ & -0.06039& -0.01280& -0.01112& [-0.15 , 0.14]& [-0.30 , 0.27]& [-0.43 , 0.39]\\$m\sin i$ ($M_\oplus$)& 5.73& 6.91& 6.91& [5.15 , 8.64]& [3.473 , 10.52]& [1.55 , 12.3]\\  \hline 
                        
                        \multicolumn{7}{c}{Yearly signal (probably systematic)}\\ \hline$P$ (d)& 364.4& 365.0& 365.0& [364.06 , 366.03]& [363.08 , 367.03]& [362.13 , 367.93]\\ $K$ (m/s) & 3.53665& 3.50129& 3.47300& [2.66 , 4.35]& [1.79 , 5.26]& [0.83 , 6.46]\\ $\lambda_0$ (rad) & 2.71212& 2.93677& 2.91814& [2.69 , 3.12]& [2.48 , 3.40]& [2.27 , 3.82]\\ $\sqrt{e} \cos \omega$ & -0.03186& -0.08448& -0.08278& [-0.30 , 0.12]& [-0.48 , 0.31]& [-0.61 , 0.47]\\ $\sqrt{e} \sin \omega$ & -0.01764& -0.02944& -0.02962& [-0.23 , 0.16]& [-0.39 , 0.34]& [-0.54 , 0.48]\\   $m\sin i$ ($M_\oplus$)& 41.5& 40.8& 40.3& [29.9 , 50.0]& [20.4 , 61.6]& [11.2 , 76.0]\\  \hline  
                        
                \end{tabular}
                
        \end{table*}

        %$5.5^{+0.4}_{-0.6}$
        %$5.1^{+0.5}_{-0.6}$
        %$5.8^{+0.8}_{-0.8}$
        %$6.5^{+1.0}_{-1.3}$
        %$6.7^{+1.8}_{-1.6}$
        %$40.^{+8.0}_{-8.7}$
        %$2.1^{+0.3}_{-0.3}$
        
        \subsection{Model consistency}
        
        It might happen that the parametrized model chosen (eq.~\ref{eq:likelihood} and~\ref{eq:kernelmcmc}) is such that \ch{no model of this class accurately represents the data. In that case,} the orbital elements are not reliable as they are computed with an incorrect model. 
        
        To check whether the model is consistent with the data, we study the residuals. Following~\cite{hara2019ecc}, we define
        \begin{align}
        \label{eq:rw}
        r_w := W(\widehat{\eta})^{1/2} (y - f(\widehat{\theta}))
        \end{align}
        where $y$ is the data,  $W$ is the inverse of the covariance matrix, $f$ is the signal model, and $\widehat{\theta},\widehat{\eta}$ are the maximum likelihood values of the parameters, as given in Table~\ref{tab:mcmcresults}. If the mode is consistent, then $r_w $ should be approximately behaved as a Gaussian variable of mean 0 and variance 1. 
        
\begin{figure} \centering
                \includegraphics[width=9cm]{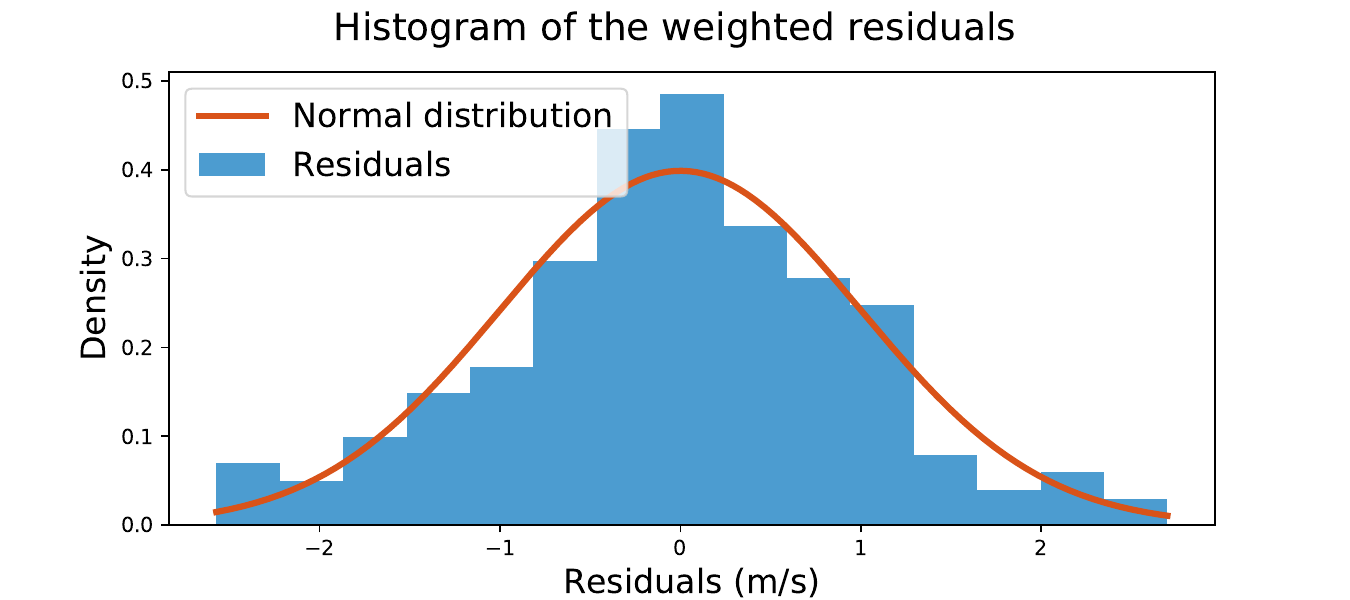}
                \caption{Histogram of the weighted residuals (blue) and probability density function of a normal variable (red).}
                \label{fig:rwhist}
        \end{figure}

\begin{figure} \centering
                \includegraphics[width=9cm]{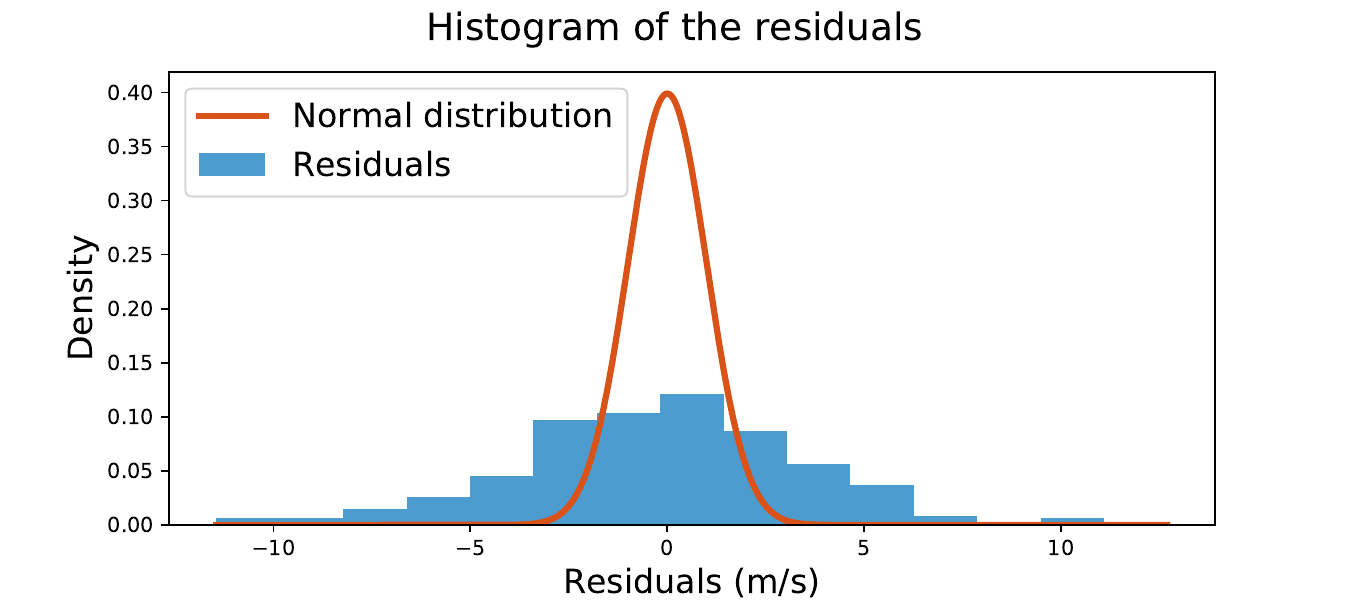}
                \caption{Histogram of the residuals (blue) and probability density function of a normal variable (red).}
                \label{fig:rhist}
        \end{figure}

        In Fig.~\ref{fig:rwhist}, we represent the normalized histogram of $r_W$ (blue) and the probability density function of a normal variable (red).  The histogram shows moderate asymmetry, which is inconclusive. We performed a Shapiro-Wilk normality test~\citep{shapirowilk1965}, yielding a $p$-value of 0.1, which means that the behavior of the residuals is consistent with a normal distribution. For comparison purposes, we plotted the distribution of the non normalized residuals in Fig.~\ref{fig:rhist}.

\begin{figure} \centering
                \includegraphics[width=9cm]{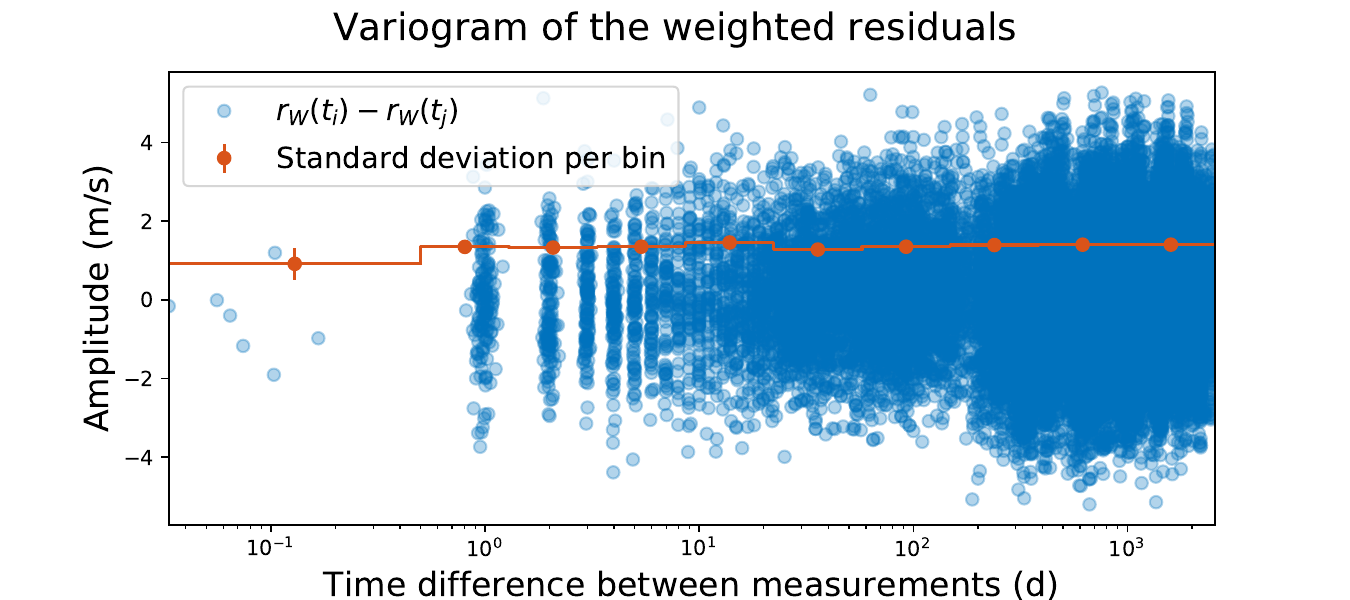}
                \caption{Difference between the weighted residuals as a function of the time interval between them (blue). The red stair curves represents the standard deviation of the residuals difference in each time bin.}
                \label{fig:rwvar}
        \end{figure}

\begin{figure} \centering
                \includegraphics[width=9cm]{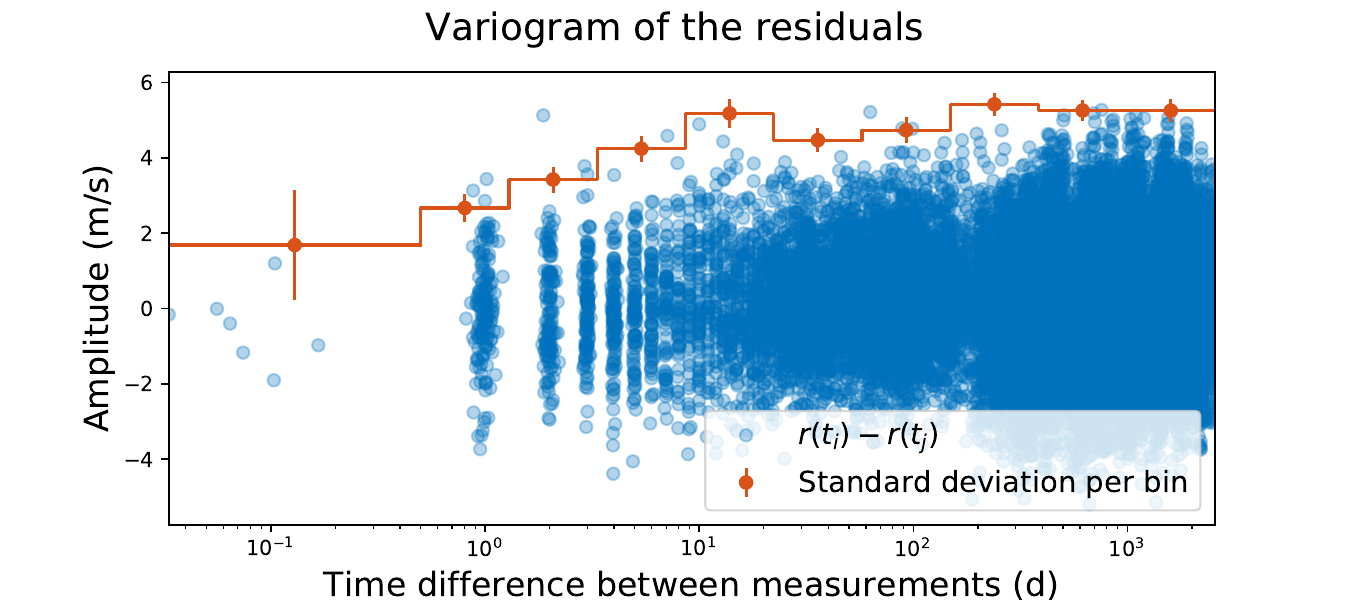}
                \caption{Difference between the residuals (not weighted) as a function of the time interval between them (blue). The red stair curves represent the standard deviation of the residuals difference in each time bin.}
                \label{fig:resvar}
        \end{figure}

        Secondly, we searched for potential correlations in the weighted residuals. For all combinations of measurement times $t_i>t_j$, we represent $d_{ij} := r_W(t_j) - r_W(t_i)$ as a function of $t_j - t_i$. We then computed the standard deviation of the $d_{ij}$ such that $t_j - t_i$ is in a certain time bin. Ten such intervals are considered, with a constant length in $\log$ scale. The results are represented in Fig.~\ref{fig:rwvar}, where it is apparent that no significant correlations remain in the residuals. For comparison purposes, in Fig.~\ref{fig:resvar}, we show the same plots for the nonweighted residuals of the model, where it appears that the dispersion of $d_{ij}$ increase as a function of  $t_j - t_i$. 
        
        In conclusion, there is no sign of missed variance and temporal correlations in the residuals, such that the signal model of Eqs.~\ref{eq:likelihood} and~\ref{eq:kernelmcmc} seems appropriate.
        
\end{document}